\documentclass[10pt]{report}
\usepackage{amsxtra,amssymb,amsthm,amsmath,latexsym}

\theoremstyle{plain}
\newtheorem{theorem}{Theorem}[section]
\newtheorem{definition}[theorem]{Definition}
\newtheorem{proposition}[theorem]{Proposition}
\newtheorem{claim}[theorem]{Claim}
\newtheorem*{claimN1}{Claim N1: The basic integral equation}
\newtheorem{lemma}[theorem]{Lemma}

\newtheorem{remark}[theorem]{Remark}
\numberwithin{equation}{section}

\begin{document}

A.G.Ramm, One-dimensional inverse scattering and spectral problems, 

\newcommand{\res}{\operatorname{Res}}

\newcommand{\refT}[1]{Theorem~\ref{T:#1}}
\newcommand{\refS}[1]{Section~\ref{S:#1}}
\newcommand{\refU}[1]{Subsection~\ref{U:#1}}
\newcommand{\refL}[1]{Lemma~\ref{L:#1}}
\newcommand{\refC}[1]{Chapter~\ref{C:#1}}
\newcommand{\refCL}[1]{Claim~\ref{CL:#1}}
\newcommand{\refP}[1]{Proposition~\ref{P:#1}}
\newcommand{\refD}[1]{Definition~\ref{D:#1}}
\newcommand{\refR}[1]{Remark~\ref{R:#1}}

\def\ds{\displaystyle}
\def\nd{\noindent}

\def\ra{\rightarrow}
\def\lra{\longrightarrow}
\def\R{{\mathbb R}}
\def\+R{{(\mathbb R_+)}}
\def\xR{{(\mathbb R_x)}}
\def\NR{{(\mathbb R_N)}}
\def\N{{\mathbb N}}
\def\Z{{\mathbb Z}}
\def\C{{\mathbb C}}
\def\oH{\buildrel\circ\over H}
\def\oH1{\buildrel\circ\over H\kern-.02in{}^1}
\def\qed{{\hfill $\Box$}}
\def\l{\ell}
\def\dotf{\dot{f}}
\def\tildea{\widetilde a}
\def\tildeb{\widetilde b}
\def\tildeg{\widetilde g}
\def\tildeh{\widetilde h}
\def\tildev{\widetilde v}
\def\tildeA{\widetilde A}
\def\tildeF{\widetilde F}
\def\tildeH{\widetilde H}
\def\tildeL{\widetilde L}
\def\tildeR{\widetilde R}
\def\ind{\hbox{\,ind\,}}
\def\arg{\hbox{\,arg\,}}

\def\notint{\int\hskip-.075in\smallsetminus}

\def\cop{\bot\hskip-.075in\bot}
\def\bysame{\rule{.5in}{.005in},\ }
\def\Im{\hbox{\,Im\,}}
\def\supp{\hbox{\,supp\,}}
\def\const{\hbox{\,const\,}}
\def\barq{\overline{q}}
\def\calJ{\mathcal J}
\def\calS{\mathcal S}
\def\calA{\mathcal A}
\def\calD{\mathcal D}
\def\calP{\mathcal P}
\def\calL{\mathcal L}
\def\calH{\mathcal H}
\def\calG{\mathcal G}

\def\be{\begin{equation}}\def\ee{\end{equation}}


\def\vl{\varphi_\ell}
\def\iro{\int^r_0}
\def\k{K(r,\rho)}
\def\ul{u_\ell}
\def\dl{\{\delta_\ell\}}


\title{
One-dimensional inverse scattering and spectral problems
   \thanks{
   key words: property C for ODE, inverse spectral and scattering problems, inverse problems for PDE and ODE,  spectral and scattering theory.   
   }
   \thanks{
Math subject classification: 35R30, 34B25, 34A55, 81F05, 81F15     }
}
\author{
A.G. Ramm\\
 Mathematics Department, Kansas State University, \\
 Manhattan, KS 66506-2602, USA\\
ramm@math.ksu.edu\\
http://www.math.ksu.edu/\,$\widetilde{\ }$\,ramm}

\date{}

\maketitle\thispagestyle{empty}

\begin{abstract}

Inverse scattering and spectral one-dimensional problems are discussed
systematically in a self-contained way. Many novel results due to the
author are presented. The classical results are often presented in a
new way. Several highlights of the new results include:

\begin{description}

\item[1)] Analysis of the invertibility of the steps in the 
Gel'fand-Levitan
and Marchenko inversion procedures,

\item[2)] Theory of the inverse problem with $I\hbox{-function}$ as the data
and its applications;

\item[3)] Proof of the property C for ordinary differential operators,
numerous applications of property C;

\item[4)] Inverse problems with ``incomplete" data;

\item[5)] Spherically symmetric inverse scattering problem with 
fixed-energy data: analysis of the Newton-Sabatier (NS) scheme for
inversion of fixed-energy phase shifts is given. This analysis shows that 
the NS scheme is fundamentally wrong, and is not a valid inversion method.

\item[6)] Complete presentation of the Krein inverse scattering theory
is given. Consistency of this theory is proved.

\item[7)] Quarkonium systems;



\item[8)] A study of the properties of $I\hbox{-function}$;

\item[9)] Some new inverse problems for the heat and wave equations are 
studied.

\item[10)] A study of inverse scattering problem for an inhomogeneous
Schr\"odinger equation;


\end{description}

\end{abstract}

\tableofcontents


\chapter{Introduction}\label{C:1}

\section{Why is this paper written?}\label{S:1.0}

There are excellent books \cite{M} and \cite{L},
where inverse spectral and scattering problems are discussed in detail.
The author decided to write this paper for the following reasons:
1) He gives a new approach to the uniqueness of the solutions to these problems.
This approach is based on {\it property C} for Sturm-Liouville operators;
2) the inverse problem with $I\hbox{-function}$ as the data is studied
and applied to many inverse problems;
3) a detailed analysis of the invertibility of the steps in Marchenko
and Gel'fand-Levitan (GL) inversion procedures is given;
4) inverse problems with ``incomplete" data are studied;
5) a detailed presentation of Krein's inversion method with proofs is 
given apparently for the first time;
6) a number of new results for various inverse problems are presented.
These include, in particular,
a) analysis of the Newton-Sabatier (NS) inversion scheme for finding a
potential given the corresponding fixed-energy phase shifts:
it is proved that the NS scheme is fundamentally wrong and is not an inversion 
method;
b) a method for finding confining potential
(a quarkonium system) from a few experimental data;
c) solution of several new inverse problems for the heat- and
wave equations;
d) a uniqueness theorem for finding a potential $q$ from a part of the
corresponding fixed-energy phase shifts;
and many other results which are taken from \cite{R}, \cite 
{R1}-\cite{R29}. 

Due to the space limitations, several important questions are not 
discussed: inverse scattering on the full line, iterative methods
for finding potential $q$: a) from two spectra
\cite{R},\cite{R5}, b) from $S-$matrix alone 
when $q$ is compactly supported \cite{R9}, approximate methods for finding
$q$ from fixed-energy phase shifts \cite{R14},\cite{R15}, property of 
resonances \cite{R}, \cite{R29}, inverse
scattering for systems of equations, etc.

\section{Auxiliary results}\label{S:1.1}
Let $q(x)\in L_{1,1}$,
$L_{1,m}=\{q: q(x)=\overline{q(x)},
\ \int^\infty_0 (1+x)^m |q(x)|dx <\infty,
\hbox{\ and\ } q\in L^2_{loc} \+R \}$,
where
$L^2_{loc}$ $(\R_+)$ consists of functions belonging to
$L^2(0,a)$ for any  $a<\infty$,
and overline stands for complex conjugate.

Consider the differential expression
$\l u=-u''+q(x)u$ with domain of definition
$D(l_0)=\{u:u(0)=0,\ u\in C^2_0(0,\infty) \}$,
where $C^2_0(0,\infty)$ is the set of $C^2(\R_+)\hbox{-functions}$
vanishing in a neighborhood of infinity, $\R_+:=[0,\infty)$.
If $H$ is the Hilbert space $L^2(\R_+)$, then
$\l_0$ is densely defined symmetric linear operator in $H$,
essentially self-adjoint, that is, the closure $\l$ of $\l_0$ in $H$
is selfadjoint. It is possible to construct a selfadjoint operator
$\l$ without assuming that $q\in L^2_{loc}(\R_+)$. Such a theory is
technically more difficult, because it is not even obvious a priori
that the set
$D(\l_0)u:=\{u:u\in C^2_0(\R_+),\ \l u\in L^2(\R_+)\}$ is dense in $H$
(in fact, it is dense). Such a theory is presented in \cite{Nai}.
If one drops the assumption $q\in L^2_{loc}$, then $D(\l_0)$
is not a domain of definition of $\l$ since there are functions
$u\in \calD(\l_0)$ for which $\l u\notin L^2(\R_+)$.
In the future we mean by $\l$ a self-adjoint operator generated by the
differential expression $\l$ and the boundary condition $u(0)=0$.

This operator has absolutely continuous spectrum, which fills
$(0,\infty)$, and discrete, finite, negative spectrum
$\{-k^2_j\}_{1\leq j\leq J}$, where $-k^2_j$ are the
eigenvalues of $\l$,
all of them are simple,
\begin{equation}\label{e1.1.0}
 \l\varphi_j:=-\varphi^{\prime\prime}_j+q\varphi_j =-k^2_j\varphi_j,
 \quad \varphi_j(0)=0, \quad \varphi'_j(0)=1,
 \end{equation}
where $\varphi_j$ are corresponding eigenfunctions which are real-valued
functions, and 
\begin{equation}\label{e1.1.1}
 \frac{1}{c_j}:=\int^\infty_0 \varphi^2_j\,dx.
 \end{equation}

The functions $\varphi(x,k)$ and $\theta(x,k)$ are defined as the unique
solutions to the problems:
\begin{equation}\label{e1.1.2}
 \l\varphi=k^2\varphi,\ x>0;\quad \varphi(0,k)=0,\ \varphi'(0,k)=1,
 \end{equation}
\begin{equation}\label{e1.1.3}
 \l\theta=k^2\theta,\ x>0;\quad \theta(0,k)=1, \quad \theta'(0,k)=0.
 \end{equation}
These functions are well defined for any $q(x)\in L^1_{loc}(\R_+)$.
Their existence and uniqueness can be proved by using the Volterra
equations for $\varphi$ and $\theta$. If $q\in L_{1,1}$, then the Jost
solution $f(x,k)$ exists and is unique. This solution is defined
by the problem:
\begin{equation}\label{e1.1.4}
 \l f:=-f''+qf=k^2f, \  f(x,k)=\exp (ikx)+o(1) \hbox{\ as\ }
 x\to +\infty; \  f(0,k):=f(k).
 \end{equation}
Existence and uniqueness of $f$ is proved by means of the
Volterra equation:
\begin{equation}\label{e1.1.5}
 f(x,k)=exp(ikx)+\int^\infty_x \frac{\sin[k(t-x)]}{k}\, q(t)f(t,k)dt.
 \end{equation}
If $q\in L_{1,1}$ then this equation implies that $f(x,k)$ is an
analytic function of $k$ in $\C_+=\{k: Imk>0\}$, 
$f(x,k)=\overline{f(x,-k)}$
for $k>0$.
The Jost function is defined as $f(k):=f(0,k)$. It has exactly $J$ simple
roots $ik_j$, $k_j>0$, where $-k^2_j$, $1\leq j\leq J$, are the
negative eigenvalues of $\l$.
The number $k=0$ can be a zero of $f(k)$. If $f(0)=0$,
then $\dotf(0)\not=0$, where $\dotf(k):=\frac{df}{dk}$.
Existence of $\dotf(0)$ is a fine result under the only assumption
$q\in L_{1,1}$ ( see Theorem 3.1.3 below, and \cite{R})
and an easy one if
$q\in L_{1,2}:=\{q:q=\barq,\ \int^\infty_0(1+x^2)|q(x)|dx<\infty\}$.
The phase shift $\delta(k)$ is defined by the formula
\begin{equation}\label{e1.1.6}
 f(k)=|f(k)|exp(-i\delta(k)), \qquad \delta(\infty)=0,
 \qquad f(\infty)=1,
 \end{equation}
where the last equation in \eqref{e1.1.16} follows from \eqref{e1.1.5}.
Because $q(x)=\overline{q(x)}$, one has $\delta(-k)=-\delta(k)$ for
$k\in\R$.
One defines the $S\hbox{-matrix}$ by the formula
\begin{equation}\label{e1.1.7}
 S(k):=\frac{f(-k)}{f(k)}, \qquad k\in\R.
 \end{equation}
The function $S(k)$ is not defined for complex $k$ if $q\in L_{1,1}$,
but if $|q(x)|\leq c_1 exp(-c_2|x|^\gamma)$, $\gamma >1$,
then $f(k)$ is an entire function of $k$ and $S(k)$
is meromorphic in $\C$. If $q(x)=0$ for $x>a$, then $f(k)$ is an
entire function of exponential type $\leq 2a$ (see \refS{5.0}).

If $q\in L_{1,1}$, then at $k^2=-k^2_j,k_j>0$, the Jost solution
$f_j(x):=f(x,ik_j)$ is proportional to $\varphi_j(x):=\varphi(x,ik_j)$,
$f_j$ and $\varphi_j$ both belong to $L^2(\R_+)$.
The integral equation for $\varphi$ is:
\begin{equation}\label{e1.1.8}
 \varphi(x,k)=\frac{\sin(kx)}{k} +\int^x_0 \frac{\sin k(x-s)}{k}
 q(s)\varphi(s,k)ds.
 \end{equation}
One has:
\begin{equation}\label{e1.1.9}
 \varphi(x,k)=\frac{f(x,k)f(-k)-f(x,-k)f(k)}{2ik},
 \end{equation}
because the right-hand side of \eqref{e1.1.9} solves equation
\eqref{e1.1.4} and satisfies conditions  \eqref{e1.1.2} at $x=0$.
The first condition \eqref{e1.1.2} is obvious, and the second one
follows from the Wronskian formula:
\begin{equation}\label{e1.1.10}
 f'(0,k)f(-k)-f'(0,-k)f(k)=2ik.
 \end{equation}
If $k=ik_j$ then $f_j(x)\in L^2(\R_+)$, as one can derive easily
from equation \eqref{e1.1.5}.
In fact, $|f_j(x)|\leq ce^{-k_jx}$, $x\geq 0$.
If $k>0$, then $f(x,-k)=f(x,k)$.
If $q=\barq$ then $f_j(x)$ is a real-valued function.
The function $f(x,k)$ is analytic in $\C_+$ but is, in general,
not defined for $k\in\C_-:=\{k:Imk<0\}$.
In particular, \eqref{e1.1.10}, in general, is valid on the real axis only.
However, if $|q(x)|\leq c_1 exp(-c_2|x|^\gamma)$, $\gamma>1$, then
$f(k)$ is defined on the whole complex plane of $k$, as was mentioned
above. Let us denote $f(x,k):=f_+(x,k)$ for $k\in\C_+$ and let
$f_-(x,k)$ be the second, linearly independent, solution to equation
\eqref{e1.1.4} for $k\in\C_+$.
If $f_+\in L^2(\R_+)$, then $f_-\notin L^2(\R_+)$.
One can write a formula, similar to \eqref{e1.1.9},
for $k\in\C_+$:
\begin{equation}\label{e1.1.11}
 \varphi(x,k)=c(k)[f_-(0,k) f(x,k)-f(0,k)f_-(x,k)],
 \end{equation}
where $c(k)=\const\not= 0$. For $\varphi(x,ik_j)\in L^2(\R_+)$,
it is necessary and sufficient that $f(ik_j)=0$.
In fact
\begin{equation}\label{e1.1.12}
 f(ik_j)=0, \quad \dotf(ik_j)\not= 0, \quad 1\leq j\leq J,
 \end{equation}
where $\dotf=\frac{df}{dk}$. To prove the second relation in
\eqref{e1.1.12}, one differentiates \eqref{e1.1.4} with respect to $k$ and 
gets
\begin{equation} \label{e1.1.12a}
 \dotf''+k^2\dotf-q\dotf=-2kf.
 \end{equation}
Existence of the derivative $\dot f$ with respect to $k$ in $\C_+$ follows 
easily from equation \eqref{e1.1.5}.
Multiply \eqref{e1.1.12a} by $f$ and \eqref{e1.1.4} by $\dotf$, subtract and
integrate over $\R_+$, then by parts, put $k=ik_j$, and get:
$$
 -2ik_j \int^\infty_0 f^2_j dx=(f\dotf'-f'\dotf)|_0^\infty
 = f'(0,ik_j)\dotf(ik_j).\notag
$$
Thus
\begin{equation} \label{e1.1.13}
  \int^\infty_0 f^2_j dx= \frac{f'(0,ik_j)\dotf(ik_j)}{-2ik_j}
  :=\frac{1}{s_j}>0.
 \end{equation}
It follows from \eqref{e1.1.13} that $\dotf(ik_j)\not= 0$.
The numbers $s_j>0$ are called the norming constants:
\begin{equation} \label{e1.1.14}
 s_j=-\frac{2ik_j}{f'(0,ik_j)\dotf(ik_j)},
 \quad 1\leq j\leq J.
 \end{equation}

\begin{definition}\label{D:1.1.1}
Scattering data is the triple:
\begin{equation}\label{e1.1.15}
 \calS:=\{S(k),k_j,s_j,\ 1\leq j\leq J\},
 \quad S(k):=\frac{f(-k)}{f(k)}, \quad k_j>0, \quad s_j>0.
 \end{equation}
\end{definition}

The Jost function $f(k)$ may vanish at $k=0$.
If $f(0)=0$, then the point $k=0$ is called a resonance.
If $|q(x)|\leq c_1 exp(-c_2|x|^\gamma)$, $\gamma>1$,
then the zeros of $f(k)$ in $\C_-$ are called resonances.
As we have seen above, there are finitely many zeros of $f(k)$ in $\C_+$,
these zeros are simple, their number $J$ is the number of negative
eigenvalues $-k^2_j$, $1\leq j\leq J$, of the selfadjoint Dirichlet
operator $\ell$. If $q\in L_{1,1}$ then the negative
spectrum of $\ell$ is finite  \cite {M}.

The phase shift $\delta(k)$, defined in \eqref{e1.1.16}, is related to
$S(k)$:
\begin{equation}\label{e1.1.16}
 S(k)= e^{2i\delta(k)},
 \end{equation}
so that $S(k)$ and $\delta(k)$ are interchangeable in the scattering data.
One has $f_j(x)=f'(0,ik_j)\varphi_j(x)$, because
$\frac{f(x,ik_j)}{f'(0,ik_j)}$ solves \eqref{e1.1.2}.
Therefore
\begin{equation}\label{e1.1.17}
 \int^\infty_0 \varphi^2_j dx=\frac{1}{s_j[f'(0,ik_j)]^2} :=\frac{1}{c_j}.
 \end{equation}
Thus
\begin{equation}\label{e1.1.18}
 c_j=-\frac{2ik_j f'(0,ik_j)}{\dotf(ik_j)},
 \quad 1\leq j\leq J.
 \end{equation}
In \refS{4.0} the notion of spectral function $\rho(\lambda)$ is
defined. It will be proved in \refS{5.0} for $q\in L_{1,1}$ that
the formula for the spectral function is:
\begin{equation}\label{e1.1.19}
 d\rho(\lambda)=
 \left\{
 \begin{aligned}
    \frac{\sqrt{\lambda} d\lambda} {\pi|f(\sqrt{\lambda})|^2},
       & \quad \lambda\geq 0,\\
    \sum^J_{j=1} c_j \delta(\lambda+k^2_j)d\lambda, &\quad \lambda<0,
 \end{aligned}
 \right.
 \end{equation}
where $c_j$ are defined in \eqref{e1.1.17}-\eqref{e1.1.18}.
The spectral function is defined in \refS{4.0}
for any $q\in L^1_{loc}(\R_+)$, $q=\barq$. Such a $q$ may grow at 
infinity.
On the other hand, the scattering theory is constructed for $q\in L_{1,1}$.

Let us define the index  of $S(k)$:
\begin{equation} \label{e1.1.20}
{\mathcal J}:=
\ind S(k):=\frac{1}{2\pi i} \Delta_\R \arg S(k) =\frac{1}{2\pi i}
 \int^\infty_{-\infty} d \ln S(k).
 \end{equation}
This definition implies that $\ind S(k)=\ind f(-k)-\ind f(k)=-2\ind f(k)$.
Therefore:
\begin{equation}\label{e1.1.21}
 \ind S(k)=\left\{
 \begin{aligned} -2J & \hbox{\ if\ } f(0)\not=0,\\
               -2J-1 & \hbox{\ if\ } f(0)=0,\end{aligned}
 \right.
 \end{equation}
because a simple zero $k=0$ contributes $\frac{1}{2}$ to the index,
and the index of an analytic in $\C_+$ function $f(k)$, such that
$f(\infty)=1$, equals to the  number of zeros of $f(k)$ in $\C_+$ plus
half of the number of its zeros on the real axis, provided that
all the zeros are simple. This follows from the argument principle.

In \refS{4.1} and \refS{5.1} the existence and uniqueness of the
transformation (transmutation) operators will be proved.
Namely,
\begin{equation}\label{e1.1.22}
 \varphi(x,k)=\varphi_0(x,k)+\int^x_0 K(x,y) \varphi_0(y)dy:=(I+K)\varphi_0,
 \quad \varphi_0:=\frac{\sin(kx)}{k},
 \end{equation}
and
\begin{equation}\label{e1.1.23}
 f(x,k)=e^{ikx} +\int^\infty_x A(x,y)e^{iky}dy
 :=(I+A)f_0, \quad f_0:=e^{ikx},
 \end{equation}
and the properties of the kernels $A(x,y)$ and $K(x,y)$ are discussed
in \refS{5.1} and \refS{4.1} correspondingly. The transformation operator
$I+K$ transforms the solution $\varphi_0$ to the equation \eqref{e1.1.2}
with $q=0$ into the solution $\varphi$ of \eqref{e1.1.2}, satisfying the
same as $\varphi_0$ boundary conditions at $x=0$.
The transformation operator $I+A$ transforms the solution $f_0$
to equation \eqref{e1.1.4} with $q=0$ into the solution $f$
of \eqref{e1.1.4} satisfying the same as $f_0$ ``boundary conditions at
infinity".

One can prove (see [M] and Sec. 5.7) the following estimates
\begin{equation} \label{e1.1.24}
 |A(x,y)|\leq c\sigma\left( \frac{x+y}{2} \right), \quad c=\const>0,
 \quad \sigma(x):=\int^\infty_x |q(t)|dt,
 \end{equation}
\begin{equation}\label{e1.1.25}
 \bigg|A_y(x,y)+\frac{1}{4}q\left(\frac{x+y}{2}\right)\bigg|+
 \bigg|A_x(x,y)+\frac{1}{4}q\left(\frac{x+y}{2}\right)\bigg|
 \leq c\sigma(x)\sigma\left(\frac{x+y}{2}\right),
 \end{equation}
and $A(x,y)$ solves the equation:
\begin{equation}\label{e1.1.26}
 A(x,y)=\frac{1}{2} \int^\infty_{\frac{x+y}{2}} q ds +
 \int^\infty_{\frac{x+y}{2}} ds
 \int^{\frac{y-x}{2}}_0 dt q(s-t) A(s-t,s+t).
 \end{equation}
By $H^m=H^m(\R_+)$ we denote Sobolov spaces $W^{m,2}$.
The kernel $A(x,y)$ is the unique solution to \eqref{e1.1.26},
and also of the problem \eqref{e5.0.1}-\eqref{e5.0.3}.

\section{Statement of the inverse scattering 
and inverse spectral problems.}\label{S:1.2}

ISP: {\textit Inverse Scattering problem (ISP) consists of finding
$q\in L_{1,1}$ from the corresponding scattering data  $\calS$
(see \eqref{e1.1.5}).}

A study of ISP consists of the following:

\begin{description}

\item[1)] One proves that ISP has at most one solution (see Theorem 
5.2.1).

\item[2)] One finds necessary and sufficient conditions for
$\calS$ to be scattering data corresponding to a $q\in L_{1,1}$
(characterization of the scattering data problem).

\item[3)] One gives a reconstruction method for calculating $q\in L_{1,1}$
from the corresponding $\calS$.

\end{description}

In \refC{5} these three problems are solved.

ISpP: {\textit Inverse spectral problem consists of finding $q$ from
the corresponding spectral function.}

A study of ISpP consists of the similar steps:

\begin{description}
\item[1)] One proves that ISpP has at most one solution in an
appropriate class of $q$: if $q_1$ and $q_2$ from this class generate
the same $\rho(\lambda)$, then $q_1=q_2$.

\item[2)] One finds necessary and sufficient conditions on
$\rho(\lambda)$ which guarantee that $\rho(\lambda)$ is a spectral
function corresponding to some $q$ from the above class.

\item[3)] One gives a reconstruction method for finding $q(x)$ from
the corresponding $\rho(\lambda)$.

\end{description}

\section{Property C for ODE.} \label{S:1.3}

Denote by $\l_m$ operators $\l$ corresponding to potentials
$q_m\in L_{1,1}$, and by $f_m(x,k)$ the corresponding Jost solutions, 
$m=1,2$.

\begin{definition}\label{D:1.3.1}
We say that a pair $\{\l_1,\l_2\}$ has property $C_+$ iff the set
$\{f_1(x,k)f_2(x,k)\}_{\forall k>0}$ is complete (total)
in $L^1(\R_+)$.
\end{definition}

This means that if $h\in L^1(\R_+)$ then
\begin{equation}\label{e1.3.1}
 \left\{ \int^\infty_0 h(x) f_1(x,k) f_2(x,k)dx=0
 \quad \forall k>0 \right\} \Rightarrow h=0.
 \end{equation}

We prove in \refS{2.1} that a pair $\{\l_1,\l_2\}$ does have property
$C_+$ if $q_m\in L_{1,1}$.
Let $\l \varphi :=-\varphi''+q(x)\varphi$,
and let $\varphi_j$ correspond to $q=q_j$,
\begin{equation}\label{e1.3.2}
 \l \varphi-k^2\varphi=0, \quad \varphi(0,k)=0;
 \quad \varphi'(0,k)=1; \quad \l\theta-k^2\theta=0,
 \quad \theta(0,k)=1, \quad \theta'(0,k)=0.
 \end{equation}

\begin{definition}\label{D:1.3.2}
We say that a pair $\{\l_1,\l_2\}$ has property $C_+$
iff the set $\{\varphi_1({\mathbf\cdot},k)\varphi_2({\mathbf\cdot},k)\}$
is complete in $L^1(0,b)$
for any $b>0$, $b<\infty$.
\end{definition}

This means that if $h\in L^1(0,b)$, then:
\begin{equation}\label{e1.3.3}
 \left\{  \int^b_0 h(x)\varphi_1(x,k)\varphi_2(x,k) dx\right\}_{\forall k>0}
 \Rightarrow h=0.
 \end{equation}

In \refT{2.2.2}
we prove that there is a $h\not= 0$ for which
$$
 \int^\infty_0 h(x)\varphi_1(x,k)\varphi_2(x,k)dx=0
 \qquad \forall k>0
$$
for a suitable $q_1\not=q_2$, $q_1,q_2\in L_{1,1}$.
Therefore Property $C_\varphi$ with $b=\infty$ does not hold, in general.

Property $C_\theta$ is defined similarly to Property $C_\varphi$,
with functions $\theta_j(x,k)$ replacing $\varphi_j(x,k)$.

In \refC{2}  we prove that properties $C_+$, $C_\varphi$ and $C_\theta$
hold, and give many applications of these properties throughout this
work.

\section{A brief description of the basic results.}\label{S:1.4}

The basic results of this work include:

\begin{description}
\item[1)] Proof of properties $C_+$, $C_\varphi$ and $C_\theta$.
Demonstration of many applications of these properties.

\item[2)] Analysis of the invertibility of the steps in the inversion
procedures of Gel'fand-Levitan (GL) for solving inverse spectral problem:

\end{description}

\begin{equation}\label{e1.4.1}
 \rho\Rightarrow L\Rightarrow K\Rightarrow q,
 \end{equation}
where
\begin{equation}\label{e1.4.2}
 \quad q=2\frac{d K(x,x)}{dx},
 \end{equation}
the kernel $L=L(x,y)$ is:
\begin{equation}\label{e1.4.3}
 L(x,y)=\int^\infty_{-\infty}
 \varphi_0 (x,\lambda) \varphi_0 (y,\lambda) d\sigma(\lambda),
 \quad d\sigma(\lambda):= d[\rho(\lambda)-\rho_0(\lambda)],
 \end{equation}
\begin{equation}
 d\rho_0=\left\{
   \begin{aligned}
     \frac{\sqrt{\lambda}d\lambda}{\pi},
         & \quad \lambda\geq 0,\\
      0, & \quad \lambda < 0.
   \end{aligned} \right.
  \notag
  \end{equation}
$\rho_0=\frac{2\lambda^{3/2}}{3\pi}$, $\rho_0$ is the spectral
function of $\l$ with $q=0$, and $K$ solves the Gel'fand-Levitan
equation
\begin{equation}\label{e1.4.4}
 K(x,y)+\int^x_0 K(x,s)L(s,y) ds+L(x,y)=0,
 \quad 0\leq y\leq x.
 \end{equation}

Our basic result is a proof of the invertibility of all the steps in
\eqref{e1.4.1}:
\begin{equation}\label{e1.4.5}
 \rho\Leftrightarrow L\Leftrightarrow K\Leftrightarrow q,
 \end{equation}
which holds under a weak assumption on $\rho$. Namely, assume that
\begin{equation}\label{e1.4.6}
 \rho \in \calG,
 \end{equation}
where $\calG$ is the set of nondecreasing functions $\rho$ of
bounded variation on every interval $(-\infty,b)$, $b<\infty$,
such that the following two assumptions, $A_1$) and $A_2$) hold.

Denote $L^2_0(\R_+)$ the set of $L^2(\R_+)$ functions vanishing
in a neighborhood of infinity. Let $h\in L^2_0(\R_+)$
and $H(\lambda):=\int^\infty_0 h(x) \varphi_0(x,\lambda)dx$.

{\bf Assumption $A_1)$} is:
\begin{equation}\label{e1.4.7}
 \hbox{If\ }h\in L^2_0(\R_+)\hbox{\ and\ }
 \int^\infty_{-\infty} H^2(\lambda)d\rho(\lambda)=0,
 \hbox{\ then\ } h=0.
 \end{equation}
Let
\begin{equation}\label{e1.4.8}
 \calH :=\{ H(\lambda):h\in C^\infty_0(\R_+)\},\quad 
H(\lambda):=\int^\infty_0 h(x) \varphi_0(x,\lambda)dx,
 \end{equation}

$\rho_1$ and $\rho_2$ belong to $\calP$, and 
$\nu:=\rho_1-\rho_2$ (see Section 4.2).

{\bf Assumption $A_2$)} is:
\begin{equation}\label{e1.4.9}
 \hbox{If\ } \int^\infty_{-\infty} H^2(\lambda)d\nu=0
 \quad \forall H\in\calH, \hbox{\ then\ }\nu=0.
 \end{equation}

In order to insure the one-to-one correspondence between spectral
functions $\rho$ and selfadjoint operators $\l$,
we assume that $q$ is such that the corresponding $\l$ is ``in the limit
point at infinity case".
This means that the equation ($\l-z)u=0$, $\Im z>0$ has exactly one
nontrivial solution in $L^2(\R_+)$, $\l u=-u''+q(x)u$.
If $q\in L_{1,1}$ then $\l$ is ``in the limit point at infinity case".

\begin{description}

\item[3)] Analysis of the invertibility of the Marchenko inversion
procedure for solving ISP:

\end{description}
\begin{equation}\label{e1.4.10}
 \calS\Rightarrow F\Rightarrow A\Rightarrow q,
 \end{equation}
where
\begin{equation}\label{e1.4.11}
 F(x):=\frac{1}{2\pi} \int^\infty_{-\infty} [1-S(k)]e^{ikx}dx+
 \sum^J_{j=1} s_j e^{-k_jx}:= F_s(x)+F_d(x),
 \end{equation}
\begin{equation}\label{e1.4.12}
 q(x)=-2 \frac{dA(x,x)}{dx},
 \end{equation}
and $A(x,y)$ solves the Marchenko equation
\begin{equation}\label{e1.4.13}
 A(x,y)+\int^\infty_x A(x,s) F(s+y)ds+F(x+y)=0,
 \quad 0\leq x\leq y<\infty.
 \end{equation}

Our basic result is a proof of the invertibility of the steps in
\eqref{e1.4.10}:
\begin{equation}\label{e1.4.14}
 \calS\Leftrightarrow F\Leftrightarrow A \Leftrightarrow q
 \end{equation}
under the assumption $q\in L_{1,1}$.
We also derive a {\it new equation} for
\begin{equation}
 A(y):=\left\{
 \begin{aligned}A(0,y), \quad & y\geq 0,\\
                       0, \quad & y<0.
 \end{aligned}\right.      
 \notag
 \end{equation}
This equation is:
\begin{equation}\label{e1.4.15}
 F(y)+A(y)+\int^\infty_0 A(s)F(s+y)ds=A(-y),
 \quad -\infty< y<\infty.
 \end{equation}
The function $A(y)$ is of interest because
\begin{equation}\label{e1.4.16}
 f(k)=1+\int^\infty_0 A(y) e^{iky}dy:=1+\widetilde A(k).
 \end{equation}
Therefore the knowledge of $A(y)$ is equivalent to the knowledge
of $f(k)$.

In \refS{5.4} we give necessary and sufficient conditions
  for $\calS$ to be the scattering data corresponding to
$q\in L_{1,1}$.
We also prove that if
\begin{equation}\label{e1.4.17}
 |q(x)|\leq c_1 \exp(-c_2|x|^\gamma), \quad \gamma>1,
 \end{equation}
and, in particular, if
\begin{equation}\label{e1.4.18}
 q(x)=0 \hbox{\ for\ } x>a,
 \end{equation}
then $S(k)$ alone determines $q(x)$ uniquely, because it determines
$k_j$, $s_j$ and $J$ uniquely under the assumption \eqref{e1.4.17}
or \eqref{e1.4.18}.

\begin{description}

\item[4)] 
We give a very short and simple proof of the uniqueness theorem which says
that the $I\hbox{-function}$,
\begin{equation}\label{e1.4.19}
 I(k):=\frac{f'(0,k)}{f(k)}, \quad \forall k>0,
 \end{equation}
determines $q\in L_{1,1}$ uniquely. The $I\hbox{-function}$
is equal to Weyl's $m\hbox{-function}$ if $q\in L_{1,1}$.

\end{description}

We give many applications of the above uniqueness theorem.
In particular, we give short and simple proofs of the uniqueness
theorems of Marchenko which say that $\calS$ determines $q\in L_{1,1}$
uniquely, and $\rho(\lambda)$ determines $q$ uniquely.
We prove that if \eqref{e1.4.18} (or \eqref{e1.4.17}) holds,
then either of the four functions $S(k)$, $\delta(k)$, $f(k)$,
$f'(0,k)$, determines $q(x)$ uniquely.
This result is applied in \refC{10} to the heat and wave equations.
It allows one to study some {\it new inverse problems}. For example, let
\begin{equation}\label{e1.4.20}
 u_{tt}=u_{xx}-q(x)u, \quad x>0, \quad t>0,
 \end{equation}
\begin{equation}\label{e1.4.21}
 u=u_t=0 \hbox{\ at\ }t=0. 
 \end{equation}
\begin{equation}\label{e1.4.22}
 u(0,t)=\delta(t) \hbox{\ or\ }u'(0,t)=\delta(t).
 \end{equation}
Assume
\begin{equation}\label{e1.4.23}
 q=0 \hbox{\ for\ } x>1, \quad q=\barq, \quad q\in L^1(0,1),
 \end{equation}
and let the extra data (measured data) be
\begin{equation}\label{e1.4.24}
 u(1,t)=a(t) \quad \forall t>0.
 \end{equation}

{\it The inverse problem is: given these data, find $q(x)$.}

Another example:
Let
\begin{equation}\label{e1.4.25}
 u_t=u_{xx}-q(x)u, \quad 0\leq x\leq 1, \quad t>0, \quad
 q\in L^1[0,1],
 \end{equation}
\begin{equation}\label{e1.4.26}
 u(x,0)=0
 \end{equation}
\begin{equation}\label{e1.4.27}
 u(0,t)=0, \quad u(1,t)=a(t), \quad a(t)\in L^1(\R_+),
 \quad a\not= 0.
 \end{equation}
The extra data are
\begin{equation}\label{e1.4.28}
 u_x(1,t)=b(t)\qquad \forall t>0.
 \end{equation}

{\it The inverse problem is: given these data, find $q(x)$.}

Using the above uniqueness results, we prove that these two inverse
problems have at most one solution. The proof gives also a
constructive procedure for finding $q$.

\begin{description}
\item[5)] We have already mentioned uniqueness theorems for some
inverse problems with ``incomplete data". ``Incomplete data" means the 
data
which are a proper subset of the classical data,
but ``incompleteness" of the data is compensated by the additional 
assumptions on $q$.
For example, the classical scattering data are the triple \eqref{e1.1.15},
but if \eqref{e1.4.18} or \eqref{e1.4.17} is assumed, then the
``incomplete data" alone, such as $S(k)$, or $\delta(k)$, or $f(k)$,
or $f'(0,k)$, $\forall k>0$, determine $q$ uniquely.
Another general result of this nature, that we prove in \refC{7},
is the following one.

\end{description}

Consider, for example, the problem
\begin{equation}\label{e1.4.29}
 \l \varphi_j=\lambda_j\varphi_j, \quad 0\leq x\leq 1;
 \quad \varphi_j(0)=\varphi_j(1)=0.
 \end{equation}
Other boundary conditions can also be considered.

Assume that the following data are given.
\begin{equation}\label{e1.4.30}
 \{\lambda_{m(j)}\forall j;\ q(x),\ b\leq x\leq 1\},
 \quad q(x)\in L^1[0,1], q=\barq,
 \end{equation}
where $0<b<1$, and
\begin{equation}\label{e1.4.31}
 m(j)=\frac{j}{\sigma} (1+\epsilon_j), \  |\epsilon_j|<1,
 \  \epsilon_j\rightarrow 0\hbox{\ as\ }j\to\infty,
 \  \sigma=\const, \  0<\sigma\leq 2.
 \end{equation}
Assume also
\begin{equation}\label{e1.4.32}
 \sum^\infty_{j=1} |\epsilon_j|<\infty.
 \end{equation}
We prove

\begin{theorem}\label{T:1.4.1}
Data \eqref{e1.4.30}-\eqref{e1.4.31} determine uniquely
$q(x)$ on the interval $0\leq x\leq b$ if $\sigma>2b$. If \eqref{e1.4.32} 
is assumed
additionally, then $q$ is uniquely determined if $\sigma\geq 2b$.
\end{theorem}

The $\sigma$ gives the ``part of the spectra" sufficient for the unique
recovery of $q$ on $[0,b]$. For example,
if $b=\frac{1}{2}$ and \eqref{e1.4.32} holds, then $\sigma=1$,
so ``one spectrum" determines uniquely $q$ on $[0,\frac{1}{2}]$.
If $b=\frac{1}{4}$, then $\sigma=\frac{1}{2}$, so ``half of the spectrum"
determines uniquely $q$ on $[0,\frac{1}{4}]$.
If $b=\frac{1}{5}$, then ``$\frac{2}{5}$ of the spectrum"
determine uniquely $q$ on $[0,\frac{1}{5}]$.
If $b=1$, then $\sigma=2$, and ``two spectra" determine $q$ uniquely
on the whole interval $[0,1]$.
The last result belongs to Borg \cite{B}.
By ``two spectra" one means $\{\lambda_j\}\bigcup\{\mu_j\}$,
where $\mu_j$ are the eigenvalues of the problem:
\begin{equation}\label{e1.4.33}
 \l u_j=\mu_j u_j, \quad u_j(0)=0, \quad u_j'(1)+h u_j(1)=0.
 \end{equation}
In fact, two spectra determine not only $q$ but the boundary
conditions as well \cite{M}.

\begin{description}

\item[6)]
Our basic results on the spherically symmetric inverse scattering problem 
with fixed-energy
data are the following.

\end{description}

{\it The first result:} If $q=q(r)=0$ for $r>a$, $a>0$ is an arbitrary
large fixed number, $r:=|x|$, $x\in\R^3$, $q=\barq$, and
$\int^a_0 r^2|q(r)|^2 dr<\infty$, then the data
$\{\delta_\l\}_{\forall \l \in\calL}$ determine $q(r)$ uniquely. Here
$\delta_\l$ is the phase shift at a fixed energy $k^2>0$,
$\l$ is the angular momentum, and $\calL$ is any fixed set of
positive integers such that
\begin{equation}\label{e1.4.34}
 \sum_{\l\in\calL} \frac{1}{\l}=\infty.
 \end{equation}

{\it  The second result is:}
If $q=q(x)$, $x\in\R^3$, $q=0$ for $|x|>a$, $q\in L^2(B_a)$,
where $B_a:=\{x:|x|\leq a\}$, then the knowledge of the scattering
amplitude $A(\alpha',\alpha)$ at a fixed energy $k^2>0$ and all
$\alpha'\in \widetilde{S}^2_j$
determine $q(x)$ uniquely \cite{R}, \cite{R7}. Here $\widetilde{S}^2_j$, 
$j=1,2$,
are arbitrary small open subsets in $S^2$
and $S^2$ is the unit sphere in $\R^3$.
The scattering amplitude is defined in \refS{6.0}.

{\it  The third result is:}
The Newton-Sabatier inversion procedure (see \cite{CS}, \cite{N})
is {\it fundamentally wrong.}

\begin{description}

\item[7)] Following \cite{R16} we present, apparently for the first time,
a detailed exposition (with proofs) of the Krein inversion theory for 
solving inverse
scattering problem and prove the consistency of this theory.

\item[8)] We give a method for recovery of a quarkonium system
(a confining potential) from a few experimental measurements.



\item[9)] We study various properties of the $I$-function.

\item[10)] We study an inverse scattering problem for inhomogeneous
Schr\"odinger equation.


\end{description}

\chapter{Property C for ODE} \label{C:2}

\section{Property $C_+$}\label{S:2.1}

By ODE in this section, the equation
\begin{equation}\label{e2.1.1}
 (\l-k^2)u:=-u''+q(x)u-k^2u=0
 \end{equation}
is meant. Assume $q\in L_{1,1}$. Then the Jost solution $f(x,k)$
is uniquely defined. In \refS{1.3} \refD{1.3.1}, property $C_+$
is explained. Let us prove

\begin{theorem}\label{T:2.1.1}
If $q\in L_{1,1}$, $j=1,2$ then property $C_+$ holds.
\end{theorem}

\begin{proof}
We use \eqref{e1.1.23} and \eqref{e1.1.24}.
Denote $A(x,y):=A_1(x,y)+A_2(x,s)$.
Let 
\begin{equation}\label{e2.1.2}
  \begin{aligned}
  0 & =\int^\infty_0
  dx h(x) f_1(x,k)f_2(x,k)
 \\ 
    & =\int^\infty_0 dx h(x)
   \left[ e^{2ikx} + \int^\infty_x A(x,y) e^{iky} dy \right.
 \\ 
   &\left.
   \quad +\int^\infty_x \int^\infty_x dy dz A_1(x,y)A_2(x,z) e^{ik(y+z)}
 \right]
 \end{aligned}
 \end{equation}
for some $h\in L^1(\R_+)$. Set $y+z=s$, $y-z=\sigma$ and get
\begin{equation}\label{e2.1.3}
 \int^\infty_x\int^\infty_x A_1(x,y)A_2(x,z) e^{ik(y+z)}
 dy dz=\int^\infty_{2x} T(x,s) e^{iks}ds,
 \end{equation}
where
\begin{equation}\label{e2.1.4}
 T(x,s)=\frac{1}{2} \int^{s-2x}_{-(s-2x)}
 A_1 \left( x,\frac{s+\sigma}{2} \right)
 A_2 \left( x,\frac{s-\sigma}{2} \right) d\sigma.
 \end{equation}

Thus, $f_1f_2=(I+V^*)e^{2ikx}$, where $V^*$ is the adjoint to a Volterra 
operator, $V^*f:=2\int^\infty_{x/2}A(x,2s)f(s)ds +2\int^\infty_{x}
T(x,2s)f(s)ds$.

Using \eqref{e2.1.3} and \eqref{e2.1.4} one rewrites \eqref{e2.1.2} as
\begin{multline}\label{e2.1.5}
 0 =\int^\infty_0  ds e^{2iks}
     \left[ h(s)+2 \int^s_0 A(x,2s-x) h(x) dx+2
     \int^s_0 T(x,2s) h(x) dx \right], 
    \forall k>0.
\end{multline}
The right-hand side is an analytic function of $k$ in $C_+$ vanishing
for all $k>0$. Thus, it vanishes identically in $C_+$ and,
consequently, for $k<0$. Therefore
\begin{equation}\label{e2.1.6}
 h(s)+2 \int^s_0 A(x,2s-x) h(x) dx+2 \int^s_0 T(x,2s) h(x)dx=0,
 \forall s>0.
 \end{equation}
Since $A(x,y)$ and $T(x,y)$ are bounded continuous functions,
the Volterra equation \eqref{e2.1.6} has only the trivial solution $h=0.$
\end{proof}

Define functions $g_\pm$ and $f_\pm$ as the solutions to equation
\eqref{e1.1.4} with the following asymptotics:
\begin{equation}\label{e2.1.7}
 g_\pm =\exp (\pm ikx)+o(1), \quad x\to -\infty,
 \end{equation}
\begin{equation}\label{e2.1.8}
 f_\pm =\exp (\pm ikx)+o(1), \quad x\to +\infty,
 \end{equation}
Let us denote $f_+=f$ and $g_+=g$.

\begin{definition} \label{d2.1.2}
The pair $\{\l_1,\l_2\}$ has property $C_-$ iff the set
$\{g_1g_2\}_{\forall k>0}$ is complete in $L^1(\R_-)$.
\end{definition}

Similar definition can be given with $(g_{-,j})$ replacing $g_j$, $j=1,2$.

As above, one proves:
\begin{theorem}\label{T:2.1.3}
If $q_j\in L_{1,1}(\R_-), j=1,2$, then property $C_-$ holds for
$\{\l_1,\l_2\}$.
\end{theorem}
By $L_{1,1}(\R)$ we mean the set
\begin{equation}\label{e2.1.9}
 L_{1,1}(\R):=\{q:q=\barq,\int^\infty_{-\infty} (1+|x|) |q(x)| dx<\infty\}.
 \end{equation}

\section{Properties $C_\varphi$ and $C_\theta$.}\label{S:2.2}
We prove only property $C_\varphi$. Property $C_\theta$ is proved
similarly. Property $C_\varphi$ is defined in \refS{1.3}.

\begin{theorem}\label{T:2.2.1}
If $q_j\in L_{1,1}, j=1,2$, then property $C_\varphi$ holds for
$\{\l_1,\l_2\}$.
\end{theorem}

\begin{proof}
Our proof is similar to the proof of \refT{2.1.1}. Using
\eqref{e1.1.22} and denoting $\phi=k\varphi$, $K:=K_1+K_2$, one writes
\begin{equation}\label{e2.2.1}
 \begin{aligned}
 \phi_1 \phi_2=
 &
  \sin^2(kx)+ \int^x_0 K(x,y) \sin(kx) \sin(ky) dy
 \\ &
 +\frac{1}{2} \int^x_0 \int^x_0 K_1(x,y) K_2(x,s)
 \{\cos [k(y-s)]-\cos[k(y+s)] \} dyds.
 \end{aligned}
 \end{equation}
Assume:
\begin{equation}\label{e2.2.2}
 0=\int^b_0 h(x) \phi_1(x,k) \phi_2(x,k) dx \quad \forall k>0.
 \end{equation}
Then
\begin{equation}\label{e2.2.3} 
  \begin{aligned}
  0  & =\int_0^bdx h(x) -\int_0^bdx h(x)\cos(2kx) 
  \\ &
   + \int_0^bds  \cos(ks)\int_s^bdx h(x) K(x,x-s) 
  \\ &
  -\int_0^{2b}ds\cos(ks)\int_{\frac{s}{2}}^{\min (b,s)}dx h(x)K(x,s-x) +I,
 \end{aligned}
 \end{equation}
where
\begin{equation*}
 I:=\int_0^bdx h(x) \int_0^x\int_0^xK_1(x,y)K_2(x,s)
 \{\cos[k(y-s)]-\cos[k(y+s)]\}dyds.
\end{equation*}
Let $y-s:=t$, $y+s:=v.$ Then
$$
 \int_0^x\int_0^x K_1K_2 \cos[k(y-s)]dyds=\int_0^xds\cos(ks)B_1(x,s),
$$
where
\begin{multline*}
  B_1 (x,s):=  \frac{1}{2} \int_{|s|}^{2x-|s|}
    \left[ K_1(x,\frac{s+v}{2})K_2 \left( x,\frac{v-s}{2} \right)
    +K_1 \left( x,\frac{v-s}{2} \right)
    K_2 \left( x,\frac{v+s}{2} \right) \right]dv,
\end{multline*}
$$  \int_0^x\int_0^x    K_1K_2\cos[k(y+s)]dyds
     =\int_0^{2x}B_2(x,s)\cos(ks)ds,
$$
$$  B_2(x,s):= \frac{1}{2} \int_{-\omega(s)}^{\omega(s)} K_1
   \left( x,\frac{t+s}{2} \right)
   K_2 \left( x,\frac{s-t}{2} \right) dt,
$$
and $\omega=s$ if $0\leq s \leq x$; $\omega=2x-s$ if $x\leq s \leq 2x$.

Therefore
\begin{equation}\label{e2.2.4}  
 I=\int_0^{b}ds \cos(ks)\int_s^bdx h(x)B_1(x,s) -\int_0^{2b}ds\cos(ks)
 \int_{\frac{s}{2}}^bdx h(x)B_2(x,s).
 \end{equation}
From \eqref{e2.2.3} 
and \eqref{e2.2.4}, 
taking $k\to \infty$, one gets:
\begin{equation*}
 \int_0^b h(x)dx=0,\notag
\end{equation*}
and (using completeness of the system $\cos (ks)$, $0<k<\infty$,
in $L^2(0,b)$) the following equation:
\begin{equation}\label{e2.2.5} 
 \begin{aligned}
  0=-\frac{h(\frac{s}{2})}{2} +\int_s^b
  & K(x,x-s) h(x)dx -\int_{s/2}^{\min(b,s)}dx h(x)K(x,s-x)
 \\
   & +\int_s^bdx h(x)B_1(x,s)
 -\int_{s/2}^bdx h(x)B_2(x,s).
 \end{aligned}
 \end{equation}
The kernels $K, B_1,$ and $B_2$ are bounded and continuous functions.
Therefore, if $b<\infty$ and $h(x)=0$ for $x>b$,
\eqref{e2.2.5} 
implies: 
\begin{equation*}
 |h(y)|\leq c\int_{2y}^b|h(x)|dx +c\int_{y}^b|h(x)|dx,
\end{equation*}
where $c>0$ is a constant which bounds the kernels $2K$, $2B_1$ and $2B_2$ 
from above and $2y=s$. From the above inequality one gets
\begin{equation}\label{e2.2.6}
 \max_{b-\epsilon \leq y \leq b} |h(y)|
 \leq c \epsilon \max_{b-\epsilon \leq y \leq b}|h(y)|,
 \end{equation}
where $\epsilon$, $0<\epsilon <b$, 
is sufficiently small so that $c\epsilon <1$
and $b-\epsilon< 2b-2\epsilon$. Then inequality
\eqref{e2.2.6} 
implies
$h(x)=0$ if $b-\epsilon <x<b$. Repeating this argument, one proves,
in finitely many steps, that $h(x)=0$,  $0<x<b$.  

\refT{2.2.1} is proved. 
\end{proof}

The proof of \refT{2.2.1} is not valid if $b=\infty$.  
{\it The result is not valid either if $b=\infty$.}
Let us give a counterexample.


\begin{theorem}\label{T:2.2.2}
There exists $q_1,q_2\in L_{1,1}$ and an $h\not=0$, such that
\begin{equation}\label{e2.2.7}
 \int^\infty_0 h(x)\varphi_1(x,k) \varphi_2(x,k)dx=0 \quad
 \forall k>0.
 \end{equation}
\end{theorem}

\begin{proof}
Let $q_1$ and $q_2$ are two potenetials in $L_{1,1}$ such that
$S_1(k)=S_2(k)$ $\forall k>0$, $\l_1$ and $\l_2$ have one negative
eigenvalue $-k^2_1$, which is the same for $\l_1$ and $\l_2$,
but $s_1\not= s_2$, so that $q_1\not= q_2$.
Let $h:=q_2-q_1$. Let us prove that \eqref{e2.2.7} holds.
One has $\l_1 \varphi_1=k^2\varphi_1$,  $\l_2 \varphi_2=k^2\varphi_2$.
subtract from the first equation the second and get:
\begin{equation}\label{e2.2.8}
 -\varphi''-k^2\varphi +q_1\varphi=h\varphi_2,
 \quad \varphi:=\varphi_1-\varphi_2,
 \quad \varphi(0,k)=\varphi'(0,k)=0.
 \end{equation}
Multiply \eqref{e2.2.8} by $\varphi_1$, integrate over $(0,\infty)$ and then
by parts to get
\begin{equation}\label{e2.2.9}
 \int^\infty_0 h\varphi_2\varphi_1 dx=
 (\varphi\varphi'_1-\varphi'\varphi_1)\big\vert^\infty_0=0,
 \quad \forall k>0.
 \end{equation}
At $x=0$ we use conditions \eqref{e2.2.8}, and at $x=\infty$
the phase shifts corresponding to $q_1$ and $q_2$ are the same
(because $S_1(k)=S_2(k)$) and therefore the right-hand side of
\eqref{e2.2.9} vanishes. \refT{2.2.2} is proved.

\end{proof}

\chapter{Inverse problem with $I$-function as the data} \label{C:3}
\section{Uniqueness theorem}\label{S:3.1}
Consider equation \eqref{e1.1.4} and assume $q\in L_{1,1}$
Then $f(x,k)$ is analytic in $\C_+$.
Define the $I$-function:
\begin{equation}\label{e3.1.1}
 I(k)=\frac{f'(0,k)}{f(k)}.
 \end{equation}
From \eqref{e3.1.1} it follows that $I(k)$ is meromorphic in $\C_+$ with
the finitely many simple poles $ik_j$, $1\leq j\leq J$.
Indeed, $ik_j$ are simple zeros of $f(k)$ and $f'(0,ik_j)\not=0$
as follows from \eqref{e1.1.3}.
Using \eqref{e1.1.18}, one gets
\begin{equation}\label{e3.1.2}
 a_j:=\res_{k=ik_j}
 I(k)= \frac{f'(0,ik_j)}{\dot f(ik_j)}
 =-\frac{c_j}{2ik_j},
 \quad k_j>0;
 \quad a_0=\frac{f'(0,0)}{\dot f(0)},
 \end{equation}
where $\Im a_j>0$, $1\leq j\leq J$, and $\Im a_0\geq 0$,
$a_0\not= 0$ iff $f(0)=0$.
We prove that if $q\in L_{1,1}$ and $f(0)=0$ then
$\dot f(0)$ exists and $\dot f(0)\not= 0$ (\refT{3.1.3} below).
This is a fine result.

\begin{lemma}\label{L:3.1.1}
The $I(k)$ equals to the Weyl function $m(k)$.
\end{lemma}

\begin{proof}
The $m(k)$ is a function such that
$\theta (x,k)+m(k)\varphi(x,k)\in L^2(\R_+)$ if $\Im k>0$.
Clearly
$$f(x,k)=c(k) [\theta(x,k)+m(k)\varphi(x,k)],$$
where $c(k)\not= 0$ , $\Im k>0$.
Thus
$I(k)=
\frac{\theta'(0,k)+m(k) \varphi'(0,k)}{\theta(0,k)+m(k) \varphi'(0,k)}
=m(k)$
because of \eqref{e1.1.2} and \eqref{e1.1.3}.
\end{proof}

Our basic uniqueness theorem is:

\begin{theorem}\label{T:3.1.2}
If $q_j\in L_{1,1}$, $j=1,2$, generate the same $I(k)$, then
$q_1=q_2$.
\end{theorem}

\begin{proof}
Let $p:=q_2-q_1$, $f_j$ be the Jost solution \eqref{e1.1.4} corresponding
to $q_j$, $w:=f_1-f_2$. Then one has
\begin{equation}\label{e3.1.3}
 -w''+q_1w-k^2w=pf_2,
 \quad |w|+|w'|=o(1),
 \quad x\to +\infty.
 \end{equation}
Multiply \eqref{e3.1.3} by $f_1$, integrate over $\R_+$, then by parts,
using \eqref{e3.1.3}, and get
\begin{multline}\label{e3.1.4}
 \int^\infty_0 pf_2f_1dx
  =  w'(0) f_1(0)-w(0)f'_1(0)
 \\ =f'_1(0,k)f_2(k)-f'_2(0,k)f_1(k)
  = f_1(k)f_2(k)[I_1(k)-I_2(k)]=0.
\end{multline}
By property $C_+$ (\refT{2.1.1}), $p(x)=0$.
\end{proof}

{\bf Remark:} If $q_j\in L_{1,1}, \,j=1,2,$ and  (*) $|I_1(k)-I_2(k)|\leq
ce^{-2a\Im k}$, where $k=|k|e^{i\arg k},\,\forall |k|>0,\, 0<\arg 
k<\pi$, then $q_1(x)=q_2(x)$ for almost all $x\in (0,a)$. This 
result is proved in \cite{GS1} for $q_j\in L^1_{loc}$. 
Our proof is based on \eqref{e3.1.4}, from which, using (*), one gets
(**) $ \int^\infty_0 pf_2f_1dx=O(e^{-2a\Im k})$. Note that
$f_1f_2=(I+V^*)e^{2ikx}$, where $V^*$ is the adjoint to a Volterra 
operator (see the formula below (2.1.4)). Thus, (**) can be written as
 (***) $ \int^\infty_0 p_1 e^{2ikx}dx=O(e^{-2a\Im k})$, where 
$p_1:=(I+V)p$. Formulas (6.5.6) (see Chapter 6 below) and (***) imply 
$p_1=0$
for almost all $x\in (0,a)$. Since $V$ is a Volterra operator,
it follows that $p=0$ for almost all $x\in (0,a)$, as claimed.

\begin{theorem}\label{T:3.1.3}
If $q\in L_{1,1}$ and $f(0)=0$, then $\dotf (0)$ exists and $\dotf(0)\not=0$.
\end{theorem}

\begin{proof}
Let us prove that $f(k)=ik\tildeA_1(k)$,  $\tildeA_1(0)\not=0$,
$\tildeA_1:=\int^\infty_0$ $e^{ikt} A_1(t)dt$,
and $A_1\in L^1(\R_+)$.
Let $A_1(t):=\int^\infty_t A(s)ds$,
 $\tildeA:=f(k)-1$, and $A(y)=A(0,y)$,
where $A(x,y)$ is defined in \eqref{e1.1.23} and $A(y)\in L^1(\R_+)$
by \eqref{e1.1.24}.
Integrating by parts, one gets
$\tildeA(k)=-exp(ikt)A_1(t)\vert^\infty_0+ik\tildeA_1=ik\tildeA_1-1$.
Thus $f(k)=ik\tildeA_1$. The basic difficulty is to prove that
$A_1\in L^1(\R_+)$.
If this is done, then $\lim_{k\to 0}\frac{f(k)}{k}=\dotf(0)$
exists and $\dotf(0)=i\tildeA_1(0)$.
To prove that $\dotf(0)\not=0$, one uses the Wronskian formula
\eqref{e3.2.2} with $x=0$:  $f(-k)f'(0,k)-f(k)f'(0,-k)=2ik$.
Divide by $k$ and let $k\to 0$. Since existence of $\dotf(0)$ is proved,
one gets $-\dotf(0)f'(0,0)=i$, so $\dotf(0)\not= 0$.
We have used here the existence of the limit
$\lim_{k\to 0}f'(0,k)=f'(0,0)$.
The existence of it follows from \eqref{e1.1.23}:
\begin{equation}\label{e3.1.5}
 f'(0,k)=ik-A(0,0)+\int^\infty_0 A_x(0,y) e^{iky}dy,
 \end{equation}
and
\begin{equation}\label{e3.1.6}
 f'(0,0)=-A(0,0)+\int^\infty_0 A_x(0,y)dy.
 \end{equation}
From \eqref{e1.1.25} one sees that $A_x(0,y)\in L^1(\R_+)$.
Thus, to complete the proof, one has to prove $A_1\in L^1(\R_+)$.
To prove this, use \eqref{e1.4.13} with $x=0$ and \eqref{e1.4.11}.
Since $f(ik_j)=0$, one has $\tildeA(ik_j)=-1$. Therefore \eqref{e1.4.13}
with $x=0$ yields:
\begin{equation}\label{e3.1.7}
 A(y)+\int^\infty_0 A(t) F_s(t+y)dt+ F_s(y)=0, \quad y\geq 0.
 \end{equation}
Integrate \eqref{e3.1.7} over $(x,\infty)$ to get:
\begin{equation}\label{e3.1.8}
 A_1(x)+ \int^\infty_0 A(t) \int^\infty_x F_s(t+y)dy\,dt
  +\int^\infty_x F_s(y)dy=0,\quad x\geq 0,
 \end{equation}
where $F_s(y)\in L^1 \+R $.
Integrating by parts yields:
\begin{equation}\label{e3.1.9}
 \int^\infty_0 A(t) \int^\infty_x F_s(t+y) dy\,dt
 = A_1(0) \int^\infty_x F_s(y)dy
 -\int^\infty_x A_1(t)F_s(x+t)dt.
 \end{equation}
Because $0=f(0)=1+\int^\infty_0 A(y)dy$,
one has $A_1(0)=-1$.
Therefore \eqref{e3.1.8} and \eqref{e3.1.9} imply:
\begin{equation}\label{e3.1.10}
 A_1(x)-\int^\infty_0 A_1(t) F_s(x+t) dt=0, \quad x\geq 0.
 \end{equation}
From this equation and from the inclusion $F_s(t)\in L^1(\R_+)$,
one derives $A_1\in L^1(\R_+)$ as follows.
Choose a $T(t)\in C^\infty_0(\R_+)$ such that
$\Vert F_s-T\Vert_{L^1(\R_+)}\leq 0.5$,
and let $Q:=F_s-T$. Then \eqref{e3.1.10} can be written as:
\begin{equation}\label{e3.1.11}
 A_1(x)-\int^\infty_0 Q(x+t) A_1(t)dt
 =a(x):= \int^\infty_0 T(x+t) A_1(t)dt,
 \quad x\geq 0.
 \end{equation}
Since $T\in C^\infty_0(\R_+)$ and $A\in L^1(\R_+)$,
it follows that $A_1$ is bounded.
Thus $a\in L^1(\R_+)$.
The operator $QA_1:=\int^\infty_0 q(x+t) A_1(t)dt$
has norm $\Vert Q\Vert_{L^1(\R_+)\to L^1(\R_+)}$ $\leq 0.5$.
Therefore equation \eqref{e3.1.3} is uniquely solvable in $L^1(\R_+)$ and
$A_1\in L^1(\R_+)$. \refT{3.1.3} is proved.

\end{proof}

\section{Characterization of the $I$-functions}\label{S:3.2}

One has                              
\begin{equation}\label{e3.2.1}
  \Im I(k)=\frac{1}{2i}
  \left ( \frac{f'(0,k)}{f(k)}
    - \frac{\overline{f'(0,k)}}{\overline{f(k)}}\right)
  =\frac{k}{|f(k)|^2},
 \end{equation}
where the Wronskian formula was used with $x=0$:
\begin{equation}\label{e3.2.2}
 \overline{f(x,k)} f'(x,k)-f(x,k) \overline{f'(x,k)}=2ik.
 \end{equation}
From \eqref{e1.1.19} and \eqref{e3.2.1} with $k=\sqrt{\lambda}$,
one gets:
\begin{equation}\label{e3.2.3}
 \frac{1}{\pi} \Im I(\sqrt{\lambda})d\lambda=d\rho,
 \quad \lambda\geq 0.
 \end{equation}
The $I(k)$ determines uniquely the points $ik_j$, $1\leq j\leq J$,
as the (simple) poles of $I(k)$ on the imaginary axis, and the numbers
$c_j$ by \eqref{e3.1.2}.
Therefore $I(k)$ determines uniquely the spectral function $\rho(\lambda)$
by formula \eqref{e1.1.19}. The characterization of the class of
spectral functions
$\rho(\lambda)$, given in \refS{4.5} induces a characterization
of the class of $I$-functions.

The other characterization of the $I$-functions one obtains by 
establishing
a one-to-one correspondence between the $I$-function and the scattering
data $\calS$ \eqref{e1.1.15}.
Namely, the numbers $k_j$ and $J$, $1\leq j\leq J$, are obtained
from $I(k)$ since $ik_j$ are the only poles if $I(k)$ in $\C_+$,
the numbers $s_j$ are obtained by the formula
(see \eqref{e1.1.14} and \eqref{e3.1.2}):
\begin{equation}\label{e3.2.4}
 s_j=-\frac{2ik_j}{a_j[\dotf(ik_j)]^2}, 
 \end{equation}
if $f(k)$ is found from $I(k)$. Finally, $f(k)$ can be uniquely
recovered from $I(k)$ by solving a Riemann problem. To derive
this problem, define
\begin{equation}\label{e3.2.5}
 w(k):= \prod^J_{j=1} \frac{k-ik_j}{k+ik_j}
 \quad \hbox{\ if\ }\,\,\, I(0)<\infty,
 \end{equation}
and
\begin{equation}\label{e3.2.6}
 w_0(k):= \frac{k}{k+i \kappa} w(k),
 \quad \hbox{\ if\ }I(0)=\infty,
 \quad \kappa\not=k_j \quad \forall j.
 \end{equation}
Assumption \eqref{e3.2.5}, means that $f(0)\not= 0$, and
\eqref{e3.2.6} means $f(0)=0$.

Define
\begin{equation}\label{e3.2.7}
 h(k):=w^{-1}(k)f(k), \quad I(0)<\infty
 \end{equation}
\begin{equation}\label{e3.2.8}
 h_0(k):= w^{-1}_0 (k)f(k), \quad I(0)=\infty.
 \end{equation}
Write \eqref{e3.2.1} as $f(k)=\frac{k}{\Im I(k)} \frac{1}{f(-k)}$,
or
\begin{equation}\label{e3.2.9}
 h_+(k)=g(k)h_-(k),
 \quad -\infty <k <\infty,
 \end{equation}
where $h_+(k):=h(k)$ is analytic in $\C_+$,
$h_+(k)\not=0$ in $\C_+$,
the closure of $\C_+$,
$h(\infty)=1$ in $\overline{\C_+}$, $h_-(k):=h(-k)$
has similar properties in $\C_-$,
\begin{equation}\label{e3.2.10}
 g(k)=\frac{k}{\Im I(k)}\,\, \hbox{\ if\ } \,\, I(0)<\infty,
 \quad g(k)=\frac{k}{\Im I(k)} \frac{k^2+1}{k^2}\,\,
 \hbox{\ if\ }\,\, I(0)=\infty,
 \end{equation}
$g(k)>0$ for $k>0$, $g(k)$ is bounded in a neighborhood of $k=0$ and has a
finite limit at $k=0$. From \eqref{e3.2.9} and the properties
of $h$, one gets:
\begin{equation}\label{e3.2.11}
 h(k)=\exp
 \left( \frac{1}{2\pi i} \int^\infty_{-\infty}
 \frac{\ln g(t)}{t-k}dt\right),
 \end{equation}
and
\begin{equation}\label{e3.2.12}
 f(k)=w(k)h(k), \quad \Im k\geq 0.
 \end{equation}
In \refS{3.4} we prove:
\begin{equation}\label{e3.2.13}
 \frac{1}{2\pi} \int^\infty_{-\infty}[I(k)-ik] e^{-ikt}dk=
 -\frac{r_0}{2}- \sum^J_{j=1} r_j e^{k_jt},
 \quad t<0,
 \end{equation}
where $r_j=-ia_j$. Taking $t\to -\infty$ in \eqref{e3.2.13},
one finds step by step all the numbers $r_j$, $k_j$ and $J$.
If $I(0)<\infty$, then $r_0=0$. Thus the data \eqref{e1.1.15} are
algorithmically recovered from $I(k)$ known for all $k>0$.

A characterization of $\calS$ is given in \refS{5.4},  
and thus an implicit characterization of $I(k)$ is also given.

\section{Inversion procedures.}\label{S:3.3}
Both procedures in \refS{3.2}, which allow one to construct either
$\rho(\lambda)$ or $\calS$ from $I(k)$ can be considered as inversion
procedures $I\Rightarrow q$ because in \refC{4} and \refC{5} 
reconstruction procedures
are given for recovery of $q(x)$ from either $\rho(\lambda)$
or $\calS$. All three data, $I(k)$, $\rho(\lambda)$ and $\calS$
are equivalent.
Thus, our inversion schemes are:
\begin{equation}\label{e3.3.1}
 I(k)\Rightarrow \rho(\lambda) \Rightarrow q(x),
 \end{equation}
\begin{equation}\label{e3.3.2}
 I(k)\Rightarrow \calS \Rightarrow q(x),
 \end{equation}
where \eqref{e1.4.1} gives the details of the step
$\rho(\lambda)\Rightarrow q(x)$,
and \eqref{e1.4.10} gives the details of the step
$\calS\Rightarrow q(x)$.

\section{Properties of $I(k)$}\label{S:3.4}
In this section, we derive the following formula for $I(k)$:

\begin{theorem}\label{T:3.4.1}
One has
\begin{equation}\label{e3.4.1}
 I(k)=ik+\sum^J_{j=0} \frac{a_j}{k-ik_j} +\widetilde{a}(k),
 \quad \widetilde{a}(k)=\int^\infty_0 a(t) e^{ikt}dt,
 \end{equation}
where $k_0$, $\Im a_0>0$ if and only if $f(0)=0$, $a_j$ are the
constants defined in \eqref{e3.1.2}, $\Im a_j>0$, $1\leq j\leq J$,
$ a(t)\in L^1(\R_+)$
if $f(0)\not= 0$ and $q\in L_{1,1}$,
$a(t)\in L^1(\R_+)$ if $f(0)=0$ and $q\in L_{1,3}(\R_+)$.
\end{theorem}

We prove this result in several steps which are formulated as lemmas.
Using \eqref{e1.1.23} one gets
\begin{equation}\label{e3.4.2}
 \begin{aligned}
 I(k)= &\frac{ik-A(0,0)+\int_0^\infty A_x(0,y)e^{iky}dy}{1+\tildeA(k)},\\
       &  A(y):=A(0,y), 
          \ \tildeA(k):=\int^\infty_0 A(y) e^{iky} dy.
 \end{aligned}
 \end{equation}

One has (cf. \eqref{e3.2.5})
\begin{equation}\label{e3.4.3}
 f(k)=1+\tildeA(k):=f_0(k) w(k)\frac{k}{k+i \kappa},\,\,\,
 w(k):= \prod^J_{j=1} \frac{k-ik_j}{k+ik_j},
 \kappa \not= k_j\ \quad \forall j
 \end{equation}
\begin{equation}\label{e3.4.4}
 f_0(k)\not= 0 \hbox{\ in\ } \C_+, \quad f_0(\infty)=1,
 \end{equation}
$f_0(k)$ is analytic in $\C_+$, the factor $\frac{k}{k+i}$
in \eqref{e3.4.3} is present if and only if $f(0)=0$, and
$w(k)\frac{k}{k+i\kappa}:= w_0(k)$.

\begin{lemma}\label{L:3.4.2}

If $f(0)\not=0$ and $q\in L_{1,1}(\R_+)$ then
\begin{equation}\label{e3.4.5}
  f_0(k)=1+\tilde{b}_0(k), \quad b_0(x)\in W^{1,1}\+R, \quad
||b_0||_{W^{1,1}\+R}:=\int^\infty_0 (|b_0|+|b'_0|) dx<\infty.
\end{equation}
\end{lemma}


\begin{proof}
It is sufficient to prove that, for any $1 \leq j \leq J$, the function
\begin{equation}  \label{e3.4.6}
  \frac{k+ik_j}{k-ik_j} f(k) = 1+ \int^\infty_0 g_j (t) e^{ikt}\, dt, \quad
  g_j \in W^{1,1} (\R_+).
  \end{equation}
Since
$\frac{k+ik_j}{k-ik_j} = 1+ \frac{2ik_j}{k-ik_j}$, and since
$A(y) \in W^{1,1} (\R_+)$ provided that $q \in L_{1,1}(\R_+)$
(see \eqref{e1.1.24}--\eqref{e1.1.25}),
it is sufficient to check that
\begin{equation}  \label{e3.4.7}
  \frac{f(k)}{k-ik_j} = \int^\infty_0 g(t)e^{ikt}\,dt,
  \quad g \in W^{1,1} (\R_+).
  \end{equation}
Note that
\begin{equation}\label{e3.4.7'}
 \frac{k-ik_j}{k+ik_j} =\int^\infty_{-\infty} e^{ikt}
 \left[ \delta(t)-2k_je^{-k_jt} \theta(t) \right] dt,
 \quad \theta(t):= \left\{
 \begin{array}{ll}1,&t\geq 0\\0,&t<0.\end{array}
 \right.\tag{$3.4.7'$}
\end{equation}
One has $f(ik_j)=0$, thus
\begin{equation*}
 \begin{aligned}
 \frac{f(k)}{k-ik_j}& = \frac{f(k) -f(ik)}{k-ik_j}
    = \int^\infty_0 dy A(y)             
 \frac{e^{i(k-ik_j)y}-1}{k-ik_j} e^{-k_jy}\,dy  \\
   & = \int^\infty_0 A(y)e^{-k_jy} i\int^y_0 e^{i(k-ik_j)s}\, ds
   = \int^\infty_0 e^{iks} h_j(s)\,ds,
   \end{aligned}
\end{equation*}
where
\begin{equation}  \label{e3.4.8}
  h_j(s) :=
  i \int^\infty_s A(y)e^{-k_j(y-s)}\,dy=i\int^\infty_0 
A(t+s)e^{-k_jt}\,dt.
  \end{equation}
From \eqref{e3.4.8} one obtains \eqref{e3.4.7}
since $A(y) \in W^{1,1}(\R_+)$.

\refL{3.4.2} is proved.
\end{proof}

\begin{lemma}\label{L:3.4.3}
If $f(0)=0$ and $q\in L_{1,2} (\R_+)$, then \eqref{e3.4.5} holds.
\end{lemma}

\begin{proof}
The proof goes as above with one difference:
if $f(0)=0$ then $k_0 =0$ is
present in formula \eqref{e3.4.1} and in
formula \eqref{e3.4.8} with $k_0 =0$
one has
\begin{equation}  \label{e3.4.9}
  h_0 (s) = i \int^\infty_0 A(t+s)\, dt.
  \end{equation}
Thus, using \eqref{e1.1.24}, one gets
\begin{equation*}
  \begin{aligned}
     \int^\infty_0 & | h_0(s)|\,ds \leq c\int^\infty_0\,ds \int^\infty_0\, dt
    \int^\infty_{\frac{t+s}{2}} |q(u)|\, du
    \\
  &\quad = 2c\int^\infty_0\, ds
    \int^\infty_{\frac{s}{2}}\, dv
    \int^\infty_v |q(u)|\, du
  \leq 2c \int^\infty_0\, ds
    \int^\infty_{\frac{s}{2}} | q(u)| u\,du
    \\
  &\quad = 4c\int^\infty_0 u^2 |q(u)|\, du < \infty
  \quad \hbox{\ if\ } \quad q \in L_{1,2}(\R_+), 
  \end{aligned}
\end{equation*}
where $c>0$ is a constant.
Similarly one checks that $h_0^\prime (s) \in L^1(\R_+)$ if 
$q \in L_{1,2}(\R_+)$.

\refL{3.4.3} is proved.
\end{proof}

\begin{lemma}\label{L:3.4.4}
Formula \eqref{e3.4.1} holds.
\end{lemma}

\begin{proof}
Write
\begin{equation*}
  \frac{1}{f(k)} = \frac{\frac{k+i}{k}
  \prod^J_{j=1}\frac{k+ik_j}{k-ik_j}} {f_0(k)}.
  \end{equation*}
Clearly
\begin{equation*}
  \frac{k+i}{k} \prod^J_{j=1} \frac{k+ik_j}{k-ik_j}
  = 1 + \sum^J_{j=0} \frac{c_j}{k-ik_j},\quad k_0 :=0, \quad k_j>0.
  \end{equation*}

By the Wiener-Levy theorem \cite[\S 17]{GRS}, one has
\begin{equation}  \label{e3.4.5'}\tag{$3.4.5'$}
  \frac{1}{f_0(k)} = 1 + \int^\infty_0 b(t)e^{ikt}\,dt,
  \quad b(t) \in W^{1,1} (\R_+),
  \end{equation}
where $f_0(k)$ is defined in (3.4.3).
Actually, the Wiener-Levy theorem yields $b(t) \in L^1(\R_+)$.
{\it However, since $b_0 \in W^{1,1}(\R_+)$,
one can prove that $b(t) \in W^{1,1} (\R_+)$.}
Indeed, $\widetilde{b}$ and $\widetilde b_0$ are related by the equation:
\begin{equation*}
  (1 + \widetilde b_0)(1+ \widetilde{b}) = 1, \quad \forall k\in \R,
  \end{equation*}
which implies
\begin{equation*}
  \widetilde{b} = -\widetilde b_0 -\widetilde b_0 \widetilde{b},
  \end{equation*}
or
\begin{equation}  \label{e3.4.10}
  b(t) = -b_0(t) - \int^t_0 b_0(t-s) b(s)\, ds := -b_0 -b_0 \ast  b,
  \end{equation}
where $\ast$  is the convolution operation.

Since $b^\prime_0 \in L^1(\R_+)$ and $b\in L^1(\R_+)$ the convolution
$b_0^\prime  \ast  b \in L^1 (\R_+)$.
So, differentiating \eqref{e3.4.10} one sees that
$b^\prime  \in L^1 (\R_+)$, as claimed.

From the above formulas one gets:
\begin{equation}  \label{e3.4.11}
  I(k) = (ik-A(0) + \widetilde A_1) (1+ \widetilde{b})
  \left(1 + \sum^J_{j=0}\frac{c_j}{k-ik_j}\right)
  = ik+c + \sum^J_{j=0} \frac{a_j}{k-ik_j} + \widetilde{a},
  \end{equation}
where $c$ is a constant defined in \eqref{e3.4.13} below, the constants
$a_j$ are defined in \eqref{e3.4.14} and the function $\widetilde{a}$
is defined in \eqref{e3.4.15}.
We will prove that $c=0$ (see \eqref{e3.4.17}).

To derive \eqref{e3.4.11}, we have used the formula:
\begin{equation*}
  ik\widetilde{b} = ik
  \left[ \frac{e^{ikt}}{ik}b(t) \bigg|^\infty_0 - \frac{1}{ik}
  \int^\infty_0 e^{ikt} b^\prime(t) dt\right] =-b(0) -\widetilde b^\prime, 
  \end{equation*}
and made the following transformations:
  \begin{align}         \label{e3.4.12}
    & I(k) = ik-A(0)-b(0)
    -\widetilde b^\prime+\widetilde A_1-A(0)\widetilde{b}
    +\widetilde A_1 \widetilde{b}
   \sum^J_{j=0} \frac{c_j ik}{k-ik_j}
   \notag \\
  &-\sum^J_{j=0} \frac{c_j[A(0) + b(0)]}{k-ik_j} + \sum^J_{j=0}
    \frac{\widetilde{g}(k) - \widetilde{g} (ik_j)}{k-ik_j} c_j
    + \sum^J_{j=0}
    \frac{\widetilde{g}(ik_j)c_j}{k-ik_j},
  \end{align}
where
\begin{equation}
  \widetilde{g}(k) := -\widetilde b^\prime  + \widetilde A_1 - A(0)
  \widetilde{b} + \widetilde A_1\widetilde{b}.
  \notag
  \end{equation}
Comparing \eqref{e3.4.12} and \eqref{e3.4.11} one concludes that
\begin{equation}  \label{e3.4.13}
  c := -A(0) -b(0) +i\sum^J_{j=0} c_j,
  \end{equation}
\begin{equation}  \label{e3.4.14}
  a_j := -c_j\left[k_j + A(0) + b(0) -\widetilde{g} (ik_j)\right],
  \end{equation}
\begin{equation}  \label{e3.4.15}
  \widetilde{a}(k) := \widetilde{g}(k) + \sum^J_{j=0} 
  \frac{\widetilde{g} (k)-\widetilde{g} (ik_j)}{k-ik_j} c_j.
  \end{equation}

To complete the proof of \refL{3.4.4}
one has to prove that $c=0$, where $c$
is defined in \eqref{e3.4.13}.
This follows from the asymptotics of $I(k)$ as
$k \to \infty$. Namely, one has:
\begin{equation}
  \label{e3.4.16}
  \widetilde{A}(k) = -\frac{A(0)}{ik} - \frac{1}{ik} \widetilde A^\prime. 
  \end{equation}
From \eqref{e3.4.16} and \eqref{e3.4.2} one gets:
  \begin{align}     \label{e3.4.17}
    I(k) &= (ik-A(0) + \widetilde A_1)
    \left[1-\frac{A(0)}{ik} + o\left(\frac{1}{k}\right)\right]^{-1}
    \notag\\
  & =(ik-A(0) + \widetilde A_1)
  \left(1+ \frac{A(0)}{ik} + o\left(\frac{1}{k}\right)\right)
    =ik+o(1), \quad k \to +\infty.
     \end{align}
From \eqref{e3.4.17} and \eqref{e3.4.11} it follows that $c=0$.
\refL{3.4.4} is proved.
\end{proof}

\begin{lemma}\label{L:3.4.5}
One has $a_j=ir_j$, $r_j>0$, $1 \leq j \leq J$, and $r_0=0 $ if
$f(0) \neq 0$, and $r_0>0$ if $f(0)=0$.
\end{lemma}

\begin{proof}
From \eqref{e3.1.2} one gets:
\begin{equation}  \label{e3.4.18}
  a_j= -\frac{c_j}{2ik_j} = i \frac{c_j}{2k_j} := ir_j ,
  \quad r_j :=\frac{c_j}{2k_j} >0, \quad j>0.
  \end{equation}
If $j=0$, then
\begin{equation}  \label{e3.4.19}
  a_0= \operatorname*{Res}_{k=0} I(k)
  := \frac{f^\prime (0,0)}{\dotf(0)}.
  \end{equation}
Here by $\operatorname*{Res}_{k=0}I(k)$ we mean the right-hand side of
\eqref{e3.4.19} since $I(k)$
is, in general, not analytic in a disc centered at $k=0$, it is analytic in 
$\C_+$ and, in general, cannot be continued analytically into $\C_-$.
By \refT{3.1.3} the right-hand side of \eqref{e3.4.19}
is well defined and
\begin{equation}  \label{e3.4.20}
  a_0= -\frac{i}{\left[ \dotf(0) \right]^2} = ir_0,
  \quad r_0 := -\frac{1}{[\dotf(0)]^2}.
  \end{equation}
From \eqref{e1.1.23} one gets:
\begin{equation} \label{e3.4.21}
 \dotf(0) = i \int^\infty_0 A(y) \,y \,dy.
 \end{equation}

Since $A(y)$ is a real-valued function if $q(x)$ is real-valued
(this follows from the integral equation \eqref{e1.1.26},
formula \eqref{e3.4.21} shows that
\begin{equation}  \label{e3.4.22}
  \left[\dotf(0) \right]^2 <0,
  \end{equation}
and \eqref{e3.4.20} implies
\begin{equation}  \label{e3.4.23}
  r_0 >0.
  \end{equation}

\refL{3.4.5} is proved.
\end{proof}

One may be interested in the properties of function $a(t)$ in
\eqref{e3.4.1}. These
can be obtained from \eqref{e3.4.15} and \eqref{e3.4.5}
as in the proof of \refL{3.4.2} and \refL{3.4.3}.

In particular, the statements of \refT{3.4.1} are obtained.

\begin{remark}\label{R:3.4.6}
Even if $q(x) \not\equiv 0$ is compactly supported, one cannot
claim that $a(t)$ is compactly supported.
\end{remark}

\begin{proof}
Assume for simplicity that $J=0$ and $f(0) \neq 0$. In this case, if 
$a(t)$ is
compactly supported then $I(k)$ is an entire function of exponential type.
It is proved in
\cite[p.278]{R}
that if $q(x) \not\equiv 0$
is compactly supported,
$q \in L^1(\R_+)$, then $f(k)$ has infinitely many zeros in $\C$. The
function $f^\prime  (0,z) \neq 0$ if $f(z) = 0$.
Indeed, if $f(z) =0$ and $f^\prime (0,z) =0$
then $f(x,z) \equiv 0$ by the uniqueness of the solution of
the Cauchy problem for equation \eqref{e1.1.4} with $k=z$.
Since $f(x,z) \not\equiv 0$, one has a contradiction,
which proves that $f^\prime (0,z) \neq 0$
if $f(z) =0$. Thus $I(k)$ cannot be an entire function if
$q(x) \not\equiv 0$, $q(x) \in L^1 (\R_+)$ and $q(x)$ is compactly supported.
\end{proof}

Let us consider the following question:

{\it What are the potentials for which $a(t) =0$ in \eqref{e3.4.1}?}

In other words, let us assume
\begin{equation}  \label{e3.4.24}
  I(k) = ik + \sum^J_{j=0} \frac{ir_j}{k-ik_j},
  \end{equation}
and find $q(x)$ corresponding to $I$-function \eqref{e3.4.24},
and describe the decay
properties of $q(x)$ as $x \to +\infty$.

We give two approaches to this problem. The first one is as follows.

By definition
\begin{equation}  \label{e3.4.25}
  f^\prime (0,k) = I(k) f(k), \quad f^\prime (0,-k) = I(-k)f(-k),
  \quad k \in \R.
  \end{equation}
Using \eqref{e3.4.25} and \eqref{e1.1.10} one gets
$  [I(k) -I(-k)] f(k)f(-k) = 2ik$, or
\begin{equation}  \label{e3.4.26}
  f(k)f(-k) = \frac{k}{Im I(k)}, \quad \forall k \in \R.
  \end{equation}

By \eqref{e3.4.18}  one can write (see \eqref{e1.1.19})
the spectral function
corresponding to the  $I$-function \eqref{e3.4.24}
$(\sqrt{\lambda} = k)$:
\begin{equation}  \label{e3.4.27}
 d\rho(\lambda)=
 \begin{cases}
 \frac{Im\,I(\lambda)}{\pi}\, d\lambda, & \lambda\geq 0,\\
 \sum^J_{j=1} 2k_jr_j \delta(\lambda+k^2_j)\,d\lambda, & \lambda<0,
 \end{cases}
  \end{equation}
where $\delta(\lambda)$ is the delta-function.

Knowing $d\rho (\lambda)$ one can recover $q(x)$ algorithmically by the
scheme \eqref{e1.4.1}.

Consider an example. Suppose $f(0) \neq 0, \quad J=1$,
\begin{equation}  \label{e3.4.28}
  I(k) = ik + \frac{ir_1}{k-ik_1} = ik + \frac{ir_1(k+ik_1)}{k^2+k_1^2}=
  i \left(k+ \frac{r_1k}{k^2+k^2_1}\right) - \frac{r_1k_1}{k^2+k^2_1}.
  \end{equation}
Then \eqref{e3.4.27} yields:
\begin{equation}  \label{e3.4.29}
  d\rho (\lambda) =
  \begin{cases}
  \frac{d\lambda}{\pi} \left( \sqrt{\lambda}
    + \frac{r_1\sqrt{\lambda}}{\lambda+k^2_1} \right), & \lambda>0, \\
  2k_1r_1\delta(\lambda +k^2_1)\,d\lambda, & \lambda<0.
  \end{cases}
  \end{equation}
Thus \eqref{e1.4.3} yields:
\begin{equation}  \label{e3.4.30}
  L(x,y)=\frac{1}{\pi} \int^\infty_0 d\lambda \frac{r_1 \sqrt{\lambda}}
  {\lambda + k^2_1} \frac{\sin \sqrt{\lambda}x}{\sqrt{\lambda}}
  \frac{\sin \sqrt{\lambda} y}{\sqrt{\lambda}}
  + 2k_1r_1\frac{sh(k_1x)}{k_1}  \frac{sh(k_1y)}{k_1},
  \end{equation}
and, setting $\lambda = k^2$ and taking for simplicity $2k_1r_1=1$,
one finds:
\begin{equation}\label{e3.4.31}
  \begin{aligned}
  L_0(x,y)
  & :=   \frac{2r_1}{\pi} \int^\infty_0 \frac{dk k^2}{k^2 + k^2_1}
  \frac{\sin (kx) \sin (ky)}{k^2}
    \\  
  &= \frac{2r_1}{\pi} \int^\infty_0
   \frac{dk \sin (kx) \sin (ky)}{k^2 + k^2_1}
  \\
  & = \frac{r_1}{\pi} \int^\infty_0
  \frac{dk[\cos k (x-y) - \cos k(x+y)]}{k^2 +k^2_1}
   \\
  & = \frac{r_1}{2k_1} \left(e^{-k_1|x-y|} -e^{-k_1(x+y)}\right),
  \quad k_1 >0,
  \end{aligned}
\end{equation}
where the known formula was used:
\begin{equation} \label{e3.4.32}
 \frac{1}{\pi}\int^\infty_0 \frac{\cos kx}{k^2+a^2}\,dk
 =\frac{1}{2a}\, e^{-a|x|},\qquad a>0, \qquad x\in \R.
 \end{equation}
Thus
\begin{equation}  \label{e3.4.33}
  L(x,y)=\frac{r_1}{2k_1} \left[ e^{-k_1|x-y|} -e^{-k_1(x+y)} \right]
  + \frac{sh (k_1x)}{k_1}\ \frac{sh(k_1y)}{k_1}.
  \end{equation}

Equation \eqref{e1.4.4} with kernel \eqref{e3.4.33}
is not an integral equation with degenerate kernel:
  \begin{align}  \label{e3.4.34}
  K(x,y) &+\int^x_0 K(x,t)
    \left[ \frac{ e^{-k_1|t-y|} -e^{-k_1(t+y)}}{2k_1/r_1}
    + \frac{sh(k_1t)}{k_1} \frac{sh(k_1y)}{k_1} \right]\, dt
  \\
  &= -\frac{e^{-k_1|x-y|} -e^{-k_1(x+y)} }{2k_1/r_1} -
  \frac{sh(k_1x)}{k_1} \frac{sh(k_1y)}{k_1}.
  \notag \end{align}

This equation can be solved analytically
\cite{Ra}, but the solution is long. By this reason
we do not give the theory developed in \cite{Ra},
but give the second approach to a study of the properties
of $q(x)$ given $I(k)$ of the form \eqref{e3.4.28}.
This approach is based on the theory of the Riemann problem
\cite{G}.

Equations \eqref{e3.4.26} and \eqref{e3.4.28} imply
\begin{equation}  \label{e3.4.35}
  f(k)f(-k)=\frac{k^2+k^2_1}{k^2+\nu^2_1},
  \qquad \nu^2_1:=k^2_1+r_1.
  \end{equation}
The function
\begin{equation}  \label{e3.4.36}
  f_0(k):= f(k)\, \frac{k+ik_1}{k-ik_1} \not=0
  \quad\hbox{in}\quad \C_+.
  \end{equation}

Write \eqref{e3.4.35} as
\begin{equation}
  f_0(k) \frac{k-ik_1}{k+ik_1} f_0(-k) \frac{k+ik_1}{k-ik_1}
  =\frac{k^2+k^2_1}{k^2+\nu^2_1}.
  \notag
  \end{equation}
Thus
\begin{equation}  \label{e3.4.37}
  f_0(k)=\frac{k^2 +k^2_1}{k^2+\nu^2_1}\,h, 
  \qquad h(k):=\frac{1}{f_0(-k)}.
  \end{equation}
The function $f_0(-k)\not= 0$ in $\C_-$, $f_0(\infty)=1$ in $\C_-$,
so $h:=\frac{1}{f_0(-k)}$ is analytic in $\C_-$.

Consider \eqref{e3.4.37} as a Riemann problem.
One has
\begin{equation}  \label{e3.4.38}
  \ind_{\R} \frac{k^2+k^2_1}{k^2+\nu^2_1}:= \frac{1}{2\pi i}
  \int^\infty_{-\infty} d\ln \frac{k^2+k^2_1}{k^2+\nu^2_1}=0.
  \end{equation}
Therefore (see
\cite{G}) problem \eqref{e3.4.37} is uniquely solvable.
Its solution is:
\begin{equation}  \label{e3.4.39}
  f_0(k)=\frac{k+ik_1}{k+i\nu_1},\qquad
  h(k)=\frac{k-i\nu_1}{k-ik_1},
  \end{equation}
as one can check.

Thus, by \eqref{e3.4.36},
\begin{equation}  \label{e3.4.40}
  f(k)=\frac{k-ik_1}{k+i\nu_1}.
  \end{equation}

The corresponding $S$-matrix is:
\begin{equation}  \label{e3.4.41}
  S(k)=\frac{f(-k)}{f(k)}=
  \frac{(k+ik_1)(k+i\nu_1)}{(k-ik_1)(k-i\nu_1)} 
  \end{equation}

Thus
\begin{equation}\label{e3.4.42}
  F_s(x):=\frac{1}{2\pi} \int^\infty_{-\infty} [1-S(k)]
   e^{ikx} dk=O\left(e^{-k_1x}\right) \quad\hbox{for}\quad x>0,
  \end{equation}
\begin{equation}
  F_d(x)=s_1\,e^{-k_1x},
  \notag
  \end{equation}
and
\begin{equation}  \label{e3.4.43}
  F(x)=F_s(x)+F_d(x)=O\left(e^{-k_1x}\right).
  \end{equation}
Equation \eqref{e1.4.13} implies  $A(x,x)=O\left(e^{-2k_1x}\right)$, so
\begin{equation}  \label{e3.4.44}
  q(x)=O\left(e^{-2k_1x}\right),
  \qquad x\to +\infty.
  \end{equation}

Thus, if $f(0)\not=0$ and $a(t)=0$ then $q(x)$ decays exponentially
at the rate determined by the number $k_1$,
$k_1=\ds\operatorname*{min}_{1\leq j\leq J} k_j$.

If $f(0)=0$, $J=0$, and $a(t)=0$, then
\begin{equation}  \label{e3.4.45}
  I(k)=ik+\frac{ir_0}{k},
  \end{equation}
\begin{equation}  \label{e3.4.46}
  f(k)f(-k) =\frac{k^2}{k^2+r_0}, \qquad r_0>0.
  \end{equation}

Let $f_0(k)=\frac{(k+i)f(k)}{k}$. Then equation \eqref{e3.4.46} implies:
\begin{equation} \label{e3.4.47}
  f_0(k)f_0(-k)= \frac{k^2+1}{k^2+\nu^2_0} ,
  \qquad \nu^2_0:=r_0,
  \end{equation}
and  $f_0(k)\not= 0\quad\hbox{in}\quad \C_+$.

Thus, since $\ind_\R\frac{k^2+1}{k^2+\nu^2_0}=0$, $f_0(k)$ is uniquely
determined by the Riemann problem \eqref{e3.4.47}.

One has:
\begin{equation}
  f_0(k)=\frac{k+i}{k+i\nu_0},\qquad f_0(-k)=\frac{k-i}{k-i\nu_0},
  \notag \end{equation}
and
\begin{equation} \label{e3.4.48}
  f(k) =\frac{k}{k+i\nu_0},
  \quad S(k)=\frac{f(-k)}{f(k)}
       =\frac{k+i\nu_0}{k-i\nu_0},
 \end{equation}

\begin{equation}
  \begin{aligned}
  F_s(x)
   & =\frac{1}{2\pi} \int^\infty_{-\infty}
  \left( 1-\frac{k+i\nu_0}{k-i\nu_0}\right)
  e^{ikx}dk
  \\
  &  =\frac{-2i\nu_0}{2\pi} \int^\infty_{-\infty}
  \frac{e^{ikx}dk}{k-i\nu_0}
  =2\nu_0 e^{-\nu_0x},
  \quad x>0,
  \end{aligned}
  \notag
  \end{equation}
and $F_d(x)=0$.

So one gets:
\begin{equation}  \label{e3.4.49}
  F(x)=F_s(x)=2\nu_0 e^{-\nu_0x},\qquad x>0.
  \end{equation}
Equation \eqref{e1.4.13} yields:
\begin{equation}  \label{e3.4.50}
  A(x,y)+2\nu_0 \int^\infty_x A(x,t) e^{-\nu_0(t+y)} dt
  = -2\nu_0 e^{-\nu_0(x+y)}, \qquad y\geq x\geq 0.
  \end{equation}
Solving \eqref{e3.4.50} yields:
\begin{equation}  \label{e3.4.51}
  A(x,y)=- 2\nu_0 e^{-\nu_0(x+y)} \frac{1}{1+ e^{-2\nu_0x}} .
  \end{equation}
The corresponding potential \eqref{e1.4.12} is
\begin{equation}  \label{e3.4.52}
  q(x)= O\left(e^{-2\nu_0x}\right),\qquad x\to\infty.
  \end{equation}
If $q(x)=O\left(e^{-kx}\right)$, $k>0$, then
$a(t)$ in \eqref{e3.4.1} decays exponentially. Indeed, in this case
$b^\prime(t)$, $A_1(y)$, $b(t)$, $A_1\ast b$ decay exponentially,
so $g(t)$ decays exponentially, and, by \eqref{e3.4.15}, the function
$\frac{\widetilde g(k)-\widetilde g(ik_j)}{k-ik_j}:=\widetilde h$
with $h(t)$ decaying exponentially. We leave the details to the reader.


\chapter{Inverse spectral problem}\label{C:4}

\section{Auxiliary results}\label{S:4.0}
\subsection{Transformation operators} \label{U:4.0.1}
If $A_1$ and $A_2$ are linear operators in a Banach space $X$,
and $T$ is a boundedly invertible linear operator such that $A_1T=TA_2$,
then $T$ is called a transformation (transmutation) operator.
If $A_2f=\lambda f$ then $A_1Tf=\lambda Tf$, so that $T$ sends
eigenfunctions of $A_2$ into eigenfunctions of $A_1$ with the
same eigenvalue. Let $\l_j=-\frac{d^2}{dx^2}+q_j(x)$,
$j=1,2$, be selfadjoint in $H:=L^2(0,\infty)$ operators generated by the
Dirichlet boundary condition at $x=0$. Other selfadjoint boundary
conditions can be considered also, for example,
$u'(0)-hu(0)=0$, $h=\const\geq 0$.

\begin{theorem}\label{T:4.0.1}
Transformation operator for a pair $\{\l_1,\l_2\}$ exists and is of the
form $Tf=(I+K)f$, where the operator $I+K$ is defined in \eqref{e1.1.22}
and the kernel $K(x,y)$ is the unique solution to the problem:
\begin{equation}\label{e4.0.1}
 K_{xx}(x,y)-q_1(x) K(x,y) =K_{yy}-q_2(y)K,
 \end{equation}
\end{theorem}
\begin{equation}\label{e4.0.2}
 K(x,0)=0
 \end{equation}
\begin{equation}\label{e4.0.3}
 K(x,x)=\frac{1}{2} \int^x_0 (q_1-q_2)dy.
 \end{equation}

\begin{proof}
Consider for simplicity the case $q_2=0$, $q_1=q$.
The proof is similar in the case $q_2\not= 0$. If $q_2=0$,
then \eqref{e4.0.3} can be written as
\begin{equation}\label{e4.0.4}
 q(x)=2\frac{dK(x,x)}{dx}, \quad K(0,0)=0.
 \end{equation}

If $\l_1Tf=T\l_2f$ and $Tf=f+\int^x_0 K(x,y)f\,dy$, then
  \begin{align}
  \label{e4.0.5}
  -f''+& q(x)f+qTf-[K(x,x)f]' -\frac{\partial K(x,x)}{\partial x} f
     -\int^x_0 K_{xx}f\,dy
     \\
  &=-f''-\int^x_0 K(x,y) f_{yy} dy
  =-\int^x_0 K_{yy}f\,dy -K(x,y)f'\Big|^x_0 + K_yf\Big|^x_{0.}
  \notag
  \end{align}

Since $f\in D(\l_1)$, $f(0)=0$, and $f$ is arbitrary otherwise,
\eqref{e4.0.5} implies \eqref{e4.0.1},\eqref{e4.0.2} and \eqref{e4.0.4}. 
Conversely, if $K(x,y)$ solves \eqref{e4.0.1}, \eqref{e4.0.2}
and \eqref{e4.0.4}, then $I+K$ is the transformation operator.
To finish the proof of \refT{4.0.1} we need to prove existence of
the solution to \eqref{e4.0.1}, \eqref{e4.0.2} and \eqref{e4.0.4}.
Let $\xi=x+y$, $\eta=x-y$, $K(x,y):=B(\xi,\eta)$. Then
\eqref{e4.0.1}, \eqref{e4.0.2} and \eqref{e4.0.4} can be written as
\begin{equation}\label{e4.0.6}
 B_{\xi\eta}=\frac{1}{4}q \left( \frac{\xi+\eta}{2} \right) B,\,\,\,
 B(\xi,0)=\frac{1}{2} \int^{\xi/2}_0 q(s)ds,\,\,\,
 B(\xi,\xi)=0.
 \end{equation}
Integrate \eqref{e4.0.6} to get
\begin{equation}\label{e4.0.7}
 B_\xi(\xi,\eta)=\frac{1}{4} q \left( \frac{\xi}{2} \right)
 +\frac{1}{4} \int^\eta_0 q  \left( \frac{\xi+\eta}{x} \right)
 B(\xi,\tau)d\tau.
 \end{equation}
Integrate \eqref{e4.0.7} with respect to $\xi$ over $(\eta,\xi)$
and get
\begin{equation}\label{e4.0.8}
 B(\xi,\eta)
 =\frac{1}{4} \int^\xi_\eta q  \left( \frac{s}{2} \right)ds
 +\frac{1}{4} \int^\xi_\eta \int^\eta_0 q \left( \frac{s+\tau}{2} \right)
 B(s,\tau) d\tau ds.
 \end{equation}
This is a Volterra integral equation which has a solution,
this solution is unique, and it can be obtained by iterations.

\refT{4.0.1} is proved.
\end{proof}

\subsection{Spectral function}\label{U:4.0.2}
Consider the problem \eqref{e1.1.0}. The classical result,
going back to Weyl, is:

\begin{theorem}\label{T:4.0.2}
There exists a monotone increasing function $\rho(\lambda)$, 
possibly nonunique,
such that for every $h\in L^2(0,\infty)$, there exists
$\widetilde{h}(\lambda) \in L^2(\R; d\rho)$ such that
\begin{equation}\label{e4.0.9}
 \int^\infty_0 |h|^2dx
 =\int^\infty_{-\infty}|\widetilde{h}|^2 d\rho(\lambda),
   \quad \widetilde{h}(\lambda)
 :=\lim_{n\to\infty} \int^n_0 f(x)\varphi(x,\sqrt{\lambda})dx,
 \end{equation}
where the limit is understood in $L^2(\R,d\rho)$ sense.
If the potential $q$ in \eqref{e1.1.0} generates the
Dirichlet operator $\l$ in the limit point at infinity case, then
$\rho(\lambda)$ is uniquely defined by $q$,
otherwise $\rho(\lambda)$ is defined by $q$ nonuniquely.
The spectral function of $\l$ has the following properties:
\begin{equation}\label{e4.0.10}
 \int^0_{-\infty} e^{|\lambda|^{1/2}} d\rho(\lambda)<\infty,\,\,\,
 \rho(\lambda)=\frac{2\lambda^{3/2}}{3\pi} +o(\lambda^{1/2}),
 \quad \lambda\to +\infty.
 \end{equation}
\end{theorem}

\begin{theorem}\label{T:4.0.3} {\em (Weyl)}.
For any $\lambda$, $\Im\lambda\not=0$, there exists $m(\lambda)$ such that
\begin{equation}\label{e4.0.11}
 W(x,\lambda):=\varphi(x,\lambda) +m(\lambda)\varphi(x,\lambda)
 \in L^2\+R.
 \end{equation}
The function $m(\lambda)$ is analytic in $\C_+$ and in $\C_-$.
\end{theorem}

The function $m(\lambda)$ is called Weyl's function, or 
$m\hbox{-function}$,
and $W$ is Weyl's solution. \refT{4.0.2} and \refT{4.0.3} are proved in
\cite{M}.

\section{Uniqueness theorem} \label{S:4.1}
Let $\rho(\lambda)$ be a non-decreasing function of bounded variation
on every compact subset of the real axis.
Let $h\in L^2_0\+R$, where $L^2_0\+R$ is a subset of $L^2\+R$ functions
which vanish near infinity.
Let $\varphi_0:=\frac{\sin(x\sqrt{\lambda})}{\sqrt{\lambda}}$ and
\begin{equation}\label{e4.1.1}
 H(\lambda)=\int^\infty_0 h(x)\varphi_0(x,\lambda)dx.
 \end{equation}
Our {\bf first assumption} $A_1)$ on $\rho(\lambda)$  is:
\begin{equation}\label{e4.1.2}
 \int^\infty_{-\infty} H^2(\lambda)d\rho(\lambda)=0,
 \quad \Rightarrow h(x)=0.
 \end{equation}
This implication should hold for any $h\in L^2_0\+R$.
It holds, for example, if $d\rho(\lambda)\not=0$ on a set which
has a finite limit point: in this case the entire function of $\lambda$,
$H(\lambda)$, vanishes identically, and thus $h=0$.

Denote by $\calP$ a subset of $\rho(\lambda)$ with the following property:
if $\rho_1, \rho_2\in\calP$,
$\nu:=\rho_1-\rho_2$, and $\calH:=\{H(\lambda):h\in C^\infty_0\+R\}$,
where $H(\lambda)$ is defined in \eqref{e4.1.1}, then
\begin{equation}\label{e4.1.3}
 \left\{\int^\infty_{-\infty} H^2(\lambda) d\nu(\lambda)=0
 \quad \forall H\in \calH\right\}
 \Rightarrow \nu(\lambda)=0.
 \end{equation}
Our {\bf second assumption} $A_2$) on $\rho(\lambda)$ is:
\begin{equation}\label{e4.1.4}
  \rho\in\calP.
  \end{equation}

Let us start with two lemmas.

\begin{lemma}\label{L:4.1.1}
Spectral functions $\rho(\lambda)$ of an operator
$\l_q=-\frac{d^2}{dx^2}+q(x)$ in the limit-point at infinity case
belong to $\calP$.
\end{lemma}

\begin{proof}
Let $\rho_1,\rho_2$ be two spectral functions corresponding to
$\l_1$ and $\l_2$, $\l_j=\l_{q_j}$, $j=1,2$, $\nu=\rho_1-\rho_2$ and
  $(\ast)\ \int^\infty_{-\infty} H^2(\lambda) d\nu=0$
$\forall h\in L^2_0 \+R $.
Let $I+V$ and $I+W$
be the transformation operators corresponding to
$\l_1$ and $\l_2$ respectively, such that
\begin{equation}\label{e4.1.5}
 \varphi_0=(I+V) \varphi_1=(I+W) \varphi_2,
 \end{equation}
where $\varphi_j$ is the regular solution \eqref{e1.1.0}
corresponding to $q_j$. Condition $(\ast)$ implies
\begin{equation}\label{e4.1.6}
 \|(I+V^{\ast})h\| = \|(I+W^\ast)h\|
 \quad \forall h\in L^2(0,b),
 \end{equation}
where, for example,
\begin{equation}\label{e4.1.7}
 Vh=\int^x_a V(x,y)h(y)dy,
 \quad V^\ast h=\int^b_x V(y,x)h(y)dy.
 \end{equation}
It follows from \eqref{e4.1.6} that
\begin{equation}\label{e4.1.8}
 I+V^\ast=U(I+W^\ast),
 \end{equation}
where $U$ is a unitary operator in $L^2(0,b)$.
Indeed, $U$ is an isometry and it is surjective because $I+V^\ast$ is.

To finish the proof, one uses \refL{4.1.2} below and concludes
 from \eqref{e4.1.8}
that $V^\ast=W^\ast$, so $V=W$, $\varphi_1=\varphi_2$, and
$q_1=q_2:=q$.
Since, by assumption, $q$ is in the limit-point at infinity case,
there is only one spectral function $\rho$ corresponding to $q$,
so $\rho_1=\rho_2=\rho$.
\end{proof}

\begin{lemma}\label{L:4.1.2}
If $U$ is unitary and $V$ and $W$ are Volterra operators,
then \eqref{e4.1.8} implies $V=W$.
\end{lemma}

\begin{proof}
From \eqref{e4.1.8} one gets $I+V+(I+W)U^\ast$.
Since $U$ is unitary, one has $(I+V)(I+V^\ast)=(I+W)(I+W^\ast)$.
Because $V$ is a Volterra operator, $(I+V)^{-1}=I+V_1$,
where $V_1$ is also a Volterra (of the same type as $V$ in \eqref{e4.1.7}).
Thus,$(I+V_1)(I+W)=(I+V^\ast)(I+W^\ast_1)$, or
\begin{equation}\label{e4.1.9}
 V_1+W+V_1W=V^\ast +W^\ast_1+V^\ast W^\ast_1  
 \end{equation}
The left-hand side in \eqref{e4.1.9} is a Volterra operator
of the type $V$ in \eqref{e4.1.7}, while its right-hand side
is a Volterra operator of the type $V^\ast$.
Since they are equal, each of them must be equal to zero.
Thus, $V_1(I+W)=-W$, or $(I+V)^{-1}(I+W)=I$, or $V=W$.
\end{proof}

\begin{theorem}\label{T:4.1.3} ({\bf Marchenko})
The spectral function determines $\l_q$ uniquely.
\end{theorem}

\begin{proof}
If $\l_{q_1}$ and $\l_{q_2}$ have the same spectral function $\rho(\lambda)$,
then 
\begin{equation}\label{e4.1.10}
 \|h\|^2=\int^\infty_{-\infty} |H_1(\lambda)|^2d\rho
 =\int^\infty_{-\infty} |H_2(\lambda)|^2 d\rho
 \quad \forall h\in L^2_0(0,b),
 \end{equation}
where
$$
 H_j(\lambda):=\int^b_0 h(x) \varphi_j(x,k)dx,
 \quad k=\sqrt{\lambda}, \quad j=1,2.
 $$
Let $I+K$ be the transformation operator $\varphi_2=(I+K)\varphi_1$,
and $g:=(I+K^\ast)h$.
Then $H_2=(h,\varphi_2)=(h,(I+K)\varphi_1)=(g,\varphi_1)$.
From \eqref{e4.1.10} one gets $\|h\|=\|(I+K^\ast)h\|$.
Thus $I+K^\ast$ is isometric, and, because $K^\ast$ is a Volterra operator,
the range of $I+K^\ast$ is the whole space $L^2(0,b)$.
Therefore $I+K^\ast$ is unitary.
This implies $K^\ast=0$.
Indeed, $(I+K^\ast)^{-1}=I+K$ (unitarity) and $(I+K^\ast)^{-1}=I+V^\ast$
(Volterra property of $K^\ast$). Thus $K=V^\ast$, so $K=V^\ast=0$.
Therefore $\varphi_2=\varphi_1$ and $q_1=q_2$, so $\l_{q_1}=\l_{q_2}$.
\end{proof}

\begin{remark}\label{4.1.4}
If $\rho_1=c\rho_2$, $c=\const>0$, then the above argument is applicable
and shows that $c$ must be equal to $1$, $c=1$ and $q_1=q_2$.
Indeed, the above argument yields the unitarity of the operator  
$\sqrt{c}(I+K^\ast)$,
which implies $c=1$ and $K^\ast=0$.
\end{remark}

Here the following lemma is useful:

\begin{lemma}\label{4.1.5}
If $bI+Q=0$, where $b=\const$ and $Q$ is a compact linear operator,
then $b=0$ and $Q=0$.
\end{lemma}

A simple proof is left to the reader.

\section{Reconstruction procedure}\label{S:4.2}

Assume that $\rho(\lambda)$, the spectral function corresponding to
$\l_q$, is given. How can one reconstruct $\l_q$, that is, to find
$q(x)$? We assume for simplicity the Dirichlet boundary
condition at $x=0$, but the method allows one to reconstruct the boundary
condition without knowing it a priori.

The reconstruction procedure (the Gel'fand-Levitan or GL  procedure)
is given in \eqref{e1.4.1}--\eqref{e1.4.4}.
Its basic step consists of the derivation of equation \eqref{e1.4.4}
and of a study of this equation.

{\it  Let us derive \eqref{e1.4.4} }
We start with the formula
\begin{equation}\label{e4.2.1}
 \int^\infty_{-\infty} \varphi(x,\sqrt{\lambda}) \varphi(y,\sqrt{\lambda})
 d\rho(\lambda)=\delta(x-y),
 \end{equation}
and assume that $L(x,y)$ is a continuous function of $x,y$ in
$[0,b)\times[0,b)$ for any $b\in(0,\infty)$.

If $0\leq y<x$, one gets from \eqref{e4.2.1} the relation:
\begin{equation}\label{e4.2.2}
 \int^\infty_{-\infty} \varphi(x,\sqrt{\lambda}) \varphi(y,\sqrt{\lambda})
 d\rho(\lambda)=0, \quad 0\leq y<x.
 \end{equation}
Using \eqref{e1.1.22}, one gets $\varphi_0=(I+K)^{-1}\varphi$.
Applying $(I+K)^{-1}$ to $\varphi(y,\sqrt{\lambda})$ in \eqref{e4.2.2},
one gets
\begin{equation}\label{e4.2.3}
 0=\int^\infty_{-\infty} \varphi(x,\sqrt{\lambda})
 \varphi_0(y,\sqrt{\lambda})
 d\rho:=I(x,y), \quad 0\leq y<x.
 \end{equation}
The right-hand side can be rewritten as:
\begin{align}\label{e4.2.4}
 I(x,y)
 &=\int^\infty_{-\infty}(\varphi_0+K\varphi_0(x)
   \varphi_0(y,\sqrt{\lambda})d(\rho-\rho_0)
   \notag\\
  &\rule{2in}{0in}+ \int^\infty_{-\infty} (\varphi_0+K\varphi_0) (x)
   \varphi_0(y,\sqrt{\lambda})d\rho_0
   \notag
   \\
  &=L(x,y)+\int^x_0K(x,s) L(s,y)ds +\delta(x-y)
   +\int^x_0K(x,s)\delta(s-y)ds
   \notag
   \\
  &= L(x,y) +\int^x_0 K(x,s) L(s,y)ds + K(x,y), \quad 0\leq y<x.
 \end{align}
From \eqref{e4.2.3} and \eqref{e4.2.4} one gets, using continuity at $y=x$,
equation \eqref{e1.4.4}.

In the above proof the integrals \eqref{e4.2.2}--\eqref{e4.2.4}
are understood in the distributional sense. If the first inequality
\eqref{e4.0.10} holds, then the above integrals over $(-\infty,n)$
are well defined in the classical sense.
If one assumes that the integral in \eqref{e4.2.5} converges to
a function $L(x)$ which is twice differentiable in the classical sense:
\begin{equation} 
 L(x):=\lim_{n\to\infty}L_n(x):=\lim_{n\to\infty}\int^n_{-\infty}
 \frac{1-\cos(x\sqrt{\lambda})} {2\lambda} d\sigma (\lambda),
 \notag
 \end{equation}
then the above proof can be understood in the classical sense,
provided that $(\ast) \sup_{n,x\in(a,b)}|L_n(x)|\leq c(a,b)$ for any
$-\infty<a<b<\infty$.
If $\rho(\lambda)$ is a spectral function corresponding to $\l$,
then the sequence $L_n(x)$ satisfies $(\ast)$. It is known
(see \cite{L}) that the sequence
\begin{equation} 
 \Phi_n(x,y)=\int^n_{-\infty}
 \varphi(x,\sqrt{\lambda}) \varphi(y,\sqrt{\lambda}) d\rho(\lambda)
 -\int^n_{-\infty}
 \frac{\sin(x\sqrt{\lambda})\sin(y\sqrt{\lambda})}{\lambda}
 d\rho_0(\lambda)
 \notag
 \end{equation}
satisfies $(\ast)$ and converges to zero.

\begin{lemma}\label{L:4.2.1}
Assume \eqref{e4.1.2} and suppose that the function
$L(x)\in H^1_{loc} \+R$,
\begin{equation}\label{e4.2.5}
 L(x):=\int^\infty_{-\infty}
 \frac{1-\cos(x\sqrt{\lambda})}{2\lambda} d\sigma(\lambda).
 \end{equation}
Then equation \eqref{e1.4.4} has a solution in $L^2(0,b)$ for
any $b>0$, and this solution is unique.
\end{lemma}

\begin{proof}
Equation \eqref{e1.4.4} is of Fredholm-type: its kernel
\begin{equation}\label{e4.2.6}
 L(x,y)=L(x+y)-L(x-y), \quad L(x,x)=L(2x),\quad L(0)=0,
 \end{equation}
is in $H^1(0,b)\times H^1(0,b)$ for any $b\in (0,\infty)$.
Therefore \refL{4.2.1} is proved if it is proved that the
homogeneous version of \eqref{e1.4.4} has only the trivial solution.
Let
\begin{equation}\label{e4.2.7}
 h(y)+\int^x_0 L(s,y) h(s)ds=0,
 \quad 0\leq y\leq x, \quad h\in L^2(0,x).
 \end{equation}
Because $L(x,y)$ is a real-valued function, one may assume that
$h(y)$ is real-valued. Multiply \eqref{e4.2.7} by $h(y)$,
integrate over $(0,x)$, and use \eqref{e4.1.1}, \eqref{e1.4.3} and
Parseval's equation to get
\begin{equation}\label{e4.2.8}
 0=\|h\|^2 +\int^\infty_{-\infty} H^2(\lambda)d\sigma
  =\|h\|^2 +\int^\infty_{-\infty} H^2(\lambda)d\rho-\|h\|^2
  =\int^\infty_{-\infty} H^2(\lambda)d\rho.
  \end{equation}
From \eqref{e4.1.2} and \eqref{e4.2.8} it follows that $h=0$.
\end{proof}

If the kernel $K(x,y)$ is found from equation \eqref{e1.4.4}, then
$q(x)$ is found by formula \eqref{e4.0.4}.

\section{Invertibility of the reconstruction steps}\label{S:4.3}

Our basic result is:

\begin{theorem}\label{T:4.3.1}
Assume \eqref{e4.1.2}, \eqref{e4.1.3},
and suppose $L(x)\in H^1_{loc}\+R$.
Then each of the steps in \eqref{e1.4.1} is invertible,
so that \eqref{e1.4.5} holds.
\end{theorem}

\begin{proof}
\begin{proof}[1. Step]
$\rho\Rightarrow L$ is done by formula \eqref{e1.4.3}.
Let us prove $L\Rightarrow \rho$. If there are $\rho_1$ and $\rho_2$
corresponding to the same $L(x,y)$, and $\nu:=\rho_1-\rho_2$, then
\begin{equation}\label{e4.3.1}
 0=\int^\infty_{-\infty} \varphi_0(x,\sqrt{\lambda})
 \varphi_0(y,\sqrt{\lambda}) d\nu.
 \end{equation}
Multiply \eqref{e4.3.1} by $h(x)h(y)$, $h\in C^\infty_0\+R$,
use \eqref{e4.1.1} and get
\begin{equation}\label{e4.3.2}
 0=\int^\infty_{-\infty} H^2(\lambda) d\nu(\lambda)
 \quad \forall H\in \calH.
 \end{equation}
By \eqref{e4.1.3} it follows that $\nu=0$, so $\rho_1=\rho_2$.
Thus $L\Rightarrow\rho$.
\end{proof}

\begin{proof}[2. Step]
$L\Rightarrow K$ is done by solving \eqref{e1.4.4}.
\refL{4.2.1} says that $K$ is uniquely determined by $L$.
Let us do the step $K\Rightarrow L$. Put $y=x$ in \eqref{e1.4.4},
use \eqref{e4.2.5} and \eqref{e4.2.6} and get:
\begin{equation}\label{e4.3.3}
 L(2x)+\int^x_0 K(x,s)
 [L(x+s)-L(x-s)]ds =-K(x,x),
 \end{equation}
or
\begin{equation}\label{e4.3.4}
 L(2x)+\int^{2x}_x K(x,y-x) L(y)dy
 -\int^x_0 K(x,x-y) L(y)dy=-K(x,x).
 \end{equation}
This is a Volterra integral equation for $L(x)$ which has a solution
and the solution is unique. Thus the step $K\Rightarrow L$ is done.
The functions $L(x)$ and $K(x,x)$ are of the same smoothness.
\end{proof}

\begin{proof}[3. Step]
 $K\Rightarrow q$ is done by formula \eqref{e4.0.4}, $q(x)$
is one derivative less smooth than $K(x,x)$ and therefore
one derivative less smooth than $L(x)$. Thus $q\in L^2_{loc}\+R$.
The step $q\Rightarrow K$ is done by solving the Goursat problem
\eqref{e4.0.1}, \eqref{e4.0.2}, \eqref{e4.0.4} (with $q_2=0$),
or, equivalently, by solving Volterra equation \eqref{e4.0.8},
which is solvable and has a unique solution.
The corresponding $K(x,y)$ is in $H^1_{loc}(\R_+\times\R_+)$
if $q\in L^2_{loc}\+R$.
\end{proof}

\refT{4.3.1} is proved. 
\end{proof}

Let us prove that the $q$ obtained by formula \eqref{e4.0.4} generates the
function $K_1(x,y)$ identical to the function $K$ obtained in Step 2.
The idea of the proof is to show that both $K$ and $K_1$ solve the problem
\eqref{e4.0.1}, \eqref{e4.0.2}, \eqref{e4.0.4} with the same $q_1=q$
and $q_2=0$. This is clear for $K_1$. In order to prove it for $K$, it is
sufficient to derive from equation \eqref{e1.4.4} equations \eqref{e4.0.1}
and \eqref{e4.0.2} with $q$ given by \eqref{e4.0.4}.
Let us do this.
Equation \eqref{e4.0.2} follows from \eqref{e1.1.4} because
$L(x,0)=0$.
Define $D:=\frac{\partial^2}{\partial x^2}-\frac{\partial^2}{\partial y^2}
:=\partial^2_x-\partial^2_y$.
Apply $D$ to \eqref{e1.4.4} assuming $L(x,y)$ twice differentiable with
respect to $x$ and $y$, in which case $K(x,y)$ is also twice differentiable.
(See \refR{4.3.3}).
By \eqref{e4.2.6}, $DL=0$, so
\begin{equation}
 \begin{aligned}
  DK & +\frac{d}{dx} [K(x,x)L(x,y)] + K_x(x,x)L(x,y) \\
     &+\int^x_0 K_{xx}(x,s) L(s,y)ds
      -\int^x_0 K(x,s)L_{yy} (s,y)ds=0.
 \end{aligned}
 \notag
 \end{equation}
Integrate by parts the last integral, (use \eqref{e4.0.2}), and get
\begin{equation}\label{e4.3.5}
 \begin{aligned}
 (DK)(x,y) & +\int^x_0(DK)(x,s) L(s,y)ds  \\
           & +\dot{K}L+(K_x+K_y) L(x,y)+K(L_x(x,y)-L_s(s,y)|_{s=x})=0,
              \quad 0\leq y\leq x,
 \end{aligned}
 \end{equation}
where $K=K(x,x)$, $L=L(x,y)$, $\dot{K}=\frac{dK(x,x)}{dx}$,
$K_x+K_y=\dot K$, and $L_x(x,y)-L_s(s,y)|_{s=x}=0$.
Subtract from \eqref{e4.3.5} equation \eqref{e1.4.4} multiplied by
$q(x)$, denote $DK(x,y)-q(x) K(x,y):=v(x,y)$, and get:
\begin{equation}\label{e4.3.6}
 v(x,y)+\int^x_0 L(s,y) v(x,s) ds=0,
 \quad 0\leq y\leq x,
 \end{equation}
provided that $-q(x)L(x,y)+2\dot{K} L(x,y)=0$, which is true
because of \eqref{e4.0.4}. Equation \eqref{e4.3.6} has only the
trivial solution by \refL{4.2.1}. Thus $v=0$, and equation \eqref{e4.0.1}
is derived.

We have proved
\begin{lemma}\label{L:4.3.2}
If $L(x,y)$ is twice differentiable continuously or in $L^2\hbox{-sense}$
then the solution $K(x,y)$ of \eqref{e1.4.4} solves \eqref{e4.0.1},
\eqref{e4.0.2} with $q$ given by \eqref{e4.0.4}.
\end{lemma}

\begin{remark}\label{R:4.3.3}
If a Fredholm equation
\begin{equation}\label{e4.3.7}
 (I+A(x)) u=f(x)
 \end{equation}
in a Banach space $X$ depends on
a parameter $x$ continuously in the sense
$\lim_{h\to 0}\|A(x+h)-A(x)\|=0$,
$\lim_{h\to 0}\|f(x+h)-f(x)\|=0$, and at $x=x_0$
equation \eqref{e4.3.7} has $N(I+A(x_0))=\{0\}$,
where $N(A)=\{u:Au=0\}$, then the solution $u(x)$ exists, is unique,
and depends continuously on $x$ in some neighborhood of $x_0$,
$|x-x_0|<r$. If the data, that is, $A(x)$ and $f(x)$, have $m$ derivatives
with respect to $x$, then the solution has the same number of derivatives.
\end{remark}

Derivatives are understood in the strong sense for the elements
of $X$ and in the operator norm for the operator $A(x)$.

\section{Characterization of the class of spectral functions of the
Sturm-Liouville operators}\label{S:4.4}

From \refT{4.3.1} it follows that if \eqref{e4.1.2} holds and
$L(x)\in H^1_{loc}\+R$, then $q\in L^2_{loc}\+R$.
Condition \eqref{e4.1.3} was used only to prove $L\Rightarrow \rho$,
so if one starts with a $q\in L^2_{loc}\+R$, then by diagram
\eqref{e1.4.5} one gets $L(x,y)$ by formula \eqref{e4.2.6}, where
$L(x)\in H^1_{loc}\+R$. If \eqref{e4.1.3} holds, then one gets
from $L(x)$ a unique $\rho(\lambda)$.

Recall that assumption $A_1)$ is (4.2.2). Let $A_3)$ be the assumption
 $L(x)\in H^{m+1}_{loc}(\R_+)$.

\begin{theorem}\label{L:4.5.1}
If $A_1)$ holds, and $\rho$ is a spectral function of $\l_q$, $q\in 
H^m_{loc}(\R_+)$,
then assumption $A_3)$ holds. 
Conversely, if assumptions $A_1)$ and $A_3)$ hold,
then $\rho$ is a spectral function of
$\l_q$, $q\in H^m_{loc}(\R_+)$.
\end{theorem}

\begin{proof}
If $A_1)$ holds and $q\in H^m_{loc}(\R_+)$, then  $L(x)\in 
H^{m+1}_{loc}(\R_+)$ by (4.1.4). If $A_1)$ and $A_3)$ hold, then
$q\in H^m_{loc}(\R_+)$ by (1.5.2), because equation (1.5.4) is uniquely
solvable, and (1.5.5) holds by Theorem 4.4.1. $\Box$
\end{proof}

\section{Relation to the inverse scattering problem}\label{S:4.5}

Assume in the section that $q\in L_{1,1}$. Then the scattering data
$\calS$ are \eqref{e1.1.15} and the spectral function is \eqref{e1.1.19}.

{\textit  Let us show how to get $d\rho$, given $\calS$.}
If $\calS$ is given then $s_j$, $k_j$ and $J$ are known. If one finds
$f(k)$ then $d\rho$ is recovered because
\begin{equation}\label{e4.5.1}
 c_j=-\frac{4k^2_j}{[\dot{f}(ik_j)]^2} \ \frac{1}{s_j},
 \end{equation}
as follows from \eqref{e1.1.18} and \eqref{e1.1.14}. To find $f(k)$,
consider the Riemann problem
\begin{equation}\label{e4.5.2}
 f(k)=S(-k)f(-k), \quad k\in\R, \quad f(\infty)=1,
 \end{equation}
which can be written as (see \eqref{e3.4.3}):
\begin{equation}\label{e4.5.3}
 f_0(k)=S(-k) \frac{w(-k)}{w(k)} f_0(-k)
 \quad\hbox{if\ }\ind S(k)=-2J,
 \end{equation}
\begin{equation}\label{e4.5.4}
 f_0(k)=S(-k) \frac{w(-k)}{w(k)}\ \frac{k+i\kappa}{k-i\kappa} f_0(-k)
 \quad\hbox{if\ }\ind S(k)=-2J-1.
 \end{equation}

Note that $w(-k)=\frac{1}{w(k)}$ if $k\in\R$.
The function $f_0(k)$ is analytic in $\C_+$ and has no zeros in $\C_+$,
and $f_0(-k)$ has similar properties in $\C_-$. Therefore problems
\eqref{e4.5.3} and \eqref{e4.5.4} have unique solutions:
\begin{equation}\label{e4.5.5}
 f_0(k)=\exp
 \left\{ \frac{1}{2\pi i} \int^\infty_{-\infty}
 \frac{\log[S(-t)w^{-2}(t)]dt}{t-k}\right\}
 \quad\hbox{if\ } \ind S(k)=-2J,\quad
 \Im k>0,
 \end{equation}
\begin{equation}\label{e4.5.6}
 \begin{aligned}
 f_0(k)=\exp \left\{\frac{1}{2\pi i} \int^\infty_{-\infty} \right.
    &\left. \frac{\log[S(-t)w^{-2}(t)\frac{t+i\kappa}{t-i\kappa}]}{t-k}
      \right\} dt\\
    &\quad\hbox{if\ } \ind S(k)=-2J-1, \quad\Im k>0,
 \end{aligned}
 \end{equation}
and
\begin{equation}\label{e4.5.7}
 f(k)=f_0(k)w(k)
 \quad\hbox{if\ } \ind S(k)=-2J, \quad \Im k>0,
 \end{equation}
\begin{equation}\label{e4.5.8}
 f(k)=f_0(k) w(k) \frac{k}{k+i\kappa}
 \quad\hbox{if\ } \ind S(k)=-2J-1,\quad \Im k>0.
 \end{equation}
One can calculate $f(x)$  for $k>0$ by taking $k=k+i0$ in
\eqref{e4.5.7} or \eqref{e4.5.8}. Thus, to find $d\rho$, given
$\calS$, one goes through the following steps:
1) one finds $J$, $s_j$, $k_j$, $1\leq j\leq J$;
2) one calculates $\ind S(k):=\mathcal J$.
If $\mathcal J=-2J$, then one calculates $f(k)$ by formulas 
\eqref{e4.5.5},
\eqref{e4.5.7}, where $w(k)$ is defined in \eqref{e3.4.3},
and $c_j$ by formula \eqref{e4.5.1}, and, finally,
$d\rho$ by formula \eqref{e1.1.19}.

If $\mathcal J=-2J-1$, then one calculates $f(k)$ by formulas 
\eqref{e4.5.6}
and \eqref{e4.5.8}, where $\kappa>0$ is an arbitrary number such that
$\kappa \not=k_j$, $1\leq j\leq J$.
If $f(k)$ is found, one calculates $c_j$ by formula \eqref{e4.5.1},
and then $d\rho$ by formula \eqref{e1.1.19}. Note that $f_0(k)$ in
\eqref{e4.5.6} depends on $\kappa$, but $f(k)$ in \eqref{e4.5.8} does not.

This completes the description of the step $\calS\Rightarrow \rho$.

{\textit Let us show how to get $\calS$ given $d\rho(\lambda)$.}

From formula \eqref{e1.1.19} one finds $J$, $k_j$, $c_j$ and $|f(k)|$.
If $|f(0)|\not=0$, then $|f_0(k)|=|f(k)|$ if $k\in\R$.
Thus, if $|f(0)|\not=0$, then $\log f_0(k)$
is analytic in $\C_+$ and vanishes at infinity. It can
be found in $\C_+$ from the values of its real part $\log|f_0(k)|$
by Schwarz's formula for the half-plane:
\begin{equation}\label{e4.5.9}
 \log f_0(k)=\frac{1}{i\pi} \int^\infty_{-\infty}
 \frac{\log|f_0(t)|}{t-k} \ dt,
 \quad \Im k>0.
 \end{equation}
If $f(0)\neq 0$, then $f=f_0 w,$ so
\begin{equation}\label{e4.5.10}
 f(k)=\exp
 \left\{ \frac{1}{i\pi} \int^\infty_{-\infty}
 \frac{\log|f_0(t)|dt}{t-k} \right\} w(k),
 \quad \Im k>0.
 \end{equation}
If $|f(0)|=0$, then the same formula \eqref{e4.5.10} remains valid.
One can see this because $\frac{f(k)}{w(k)}$ is analytic
in $\C_+$, has no zeroes in $\C_+$, tends to $1$ at infinity, and
$|\frac{f(k)}{w(k)}| =|f(k)|$ if $k\in\R$.

Let us {\it summarize the step} $d\rho\Rightarrow\calS$:
one finds $J$, $k_j$, $c_j$, calculates $f(k)$ by formula \eqref{e4.5.10},
and then $S(k)=\frac{f(-k)}{f(k)},$ and $s_j$ are calculated by formula
\eqref{e4.5.1}. To calculate $f(k)$ for $k>0$ one takes $k=k+i0$
in \eqref{e4.5.10} and gets:
\begin{equation}\label{e4.5.11}
 \begin{aligned}
 f(k)
 & =\exp \left\{ \frac{1}{i\pi} \int^\infty_{-\infty}
   \frac{\log|f(t)|dt}{t-k} +\log |f(k)| \right\} w(k)  \\
 & =|f(k)|w(k)\exp \left\{ \frac{1}{i\pi} P\int^\infty_{-\infty}
   \frac{\log|f(t)|dt}{t-k} \right\}, \quad k>0.
 \end{aligned}
 \end{equation}


\chapter{Inverse scattering on half-line} \label{C:5}

\section{Auxiliary material}\label{S:5.0}
\subsection{Transformation operators}\label{U:5.0.1}

\begin{theorem}\label{T:5.0.1}
If $q\in L_{1,1}$, then there exists a unique operator
$I+A$ such that \eqref{e1.1.23} -- \eqref{e1.1.26} hold,
and $A(x,y)$ solves the following Goursat problem:

\begin{equation}\label{e5.0.1}
 A_{xx}-q(x) A=A_{yy}, 0\leq x\leq y\leq \infty,
 \end{equation}
\begin{equation}\label{e5.0.2}
 A(x,x)=\frac{1}{2} \ \int^\infty_x q(s)ds,
 \end{equation}
\begin{equation}\label{e5.0.3}
 \lim_{x+y\to\infty} A(x,y)=\lim_{x+y\to\infty} A_x(x,y)
 =\lim_{x+y\to\infty} A_y(x,y)=0.
 \end{equation}
\end{theorem}

\begin{proof}
Equations \eqref{e5.0.1} and \eqref{e5.0.2} are derived similarly to the
derivation of the similar equations for $K(x,y)$ in \refT{4.0.1}.
Relations \eqref{e5.0.3} follow from the estimates \eqref{e1.1.24} --
\eqref{e1.1.25}, which give more precise information than \eqref{e5.0.3}.
Estimates \eqref{e1.1.24} -- \eqref{e1.1.26} can be derived from the
Volterra equation \eqref{e1.1.26} which is solvable by iterations.
Equation \eqref{e1.1.26} can be derived, for example, similarly
to the derivation of equation \eqref{e4.0.8}, or by substituting
\eqref{e1.1.23} into \eqref{e1.1.5}.

A detailed derivation of all of the results of \refT{5.0.1}
can be found in \cite{M}.
\end{proof}
 
\subsection{Statement of the direct scattering problem on half-axis.
Existence and uniqueness of its solution.} \label{U:5.0.2}

The direct scattering problem on half-line consists of finding the
solution $\psi=\psi(r,k)$ to the equation:
\begin{equation}\label{e5.0.4}
 \psi''+k^2\psi-q(r)\psi=0, \quad r>0,
 \end{equation}
satisfying the boundary conditions at $r=0$ and at $r=\infty$:
\begin{equation}\label{e5.0.5}
 \psi(0)=0,
 \end{equation}
\be\label{e5.0.6}
 \psi(r)=e^{i\delta}\sin(kr+\delta)+o(1), \quad r\to+\infty,
 \end{equation}
where $\delta=\delta(k)$ is called the phase shift, and it has to be
bound. An equivalent formulation of \eqref{e5.0.6} is:
\begin{equation}\label{e5.0.7}
 \psi=\frac{i}{2}[e^{-ikr}-S(k) e^{ikr}] +o(1), \quad r\to\infty,
 \end{equation}
where $S(k)=\frac{f(-k)}{f(k)}=e^{2i\delta(k)}$.
Clearly
\begin{equation}\label{e5.0.8}
 \psi(r,k)=\frac{i}{2} [f(r,-k)-S(k)f(r,k)]
 =a(k)\varphi(r,k), \quad a(k):=\frac{k}{f(k)},
 \end{equation}
where $\varphi(r,k)$ is defined in \eqref{e1.1.0}, see also \eqref{e1.1.9}.
From \eqref{e5.0.8}, \eqref{e1.1.6} and \eqref{e5.0.6} one gets
\begin{equation}\label{e5.0.9}
 \varphi(r,k)=\frac{|f(k)|}{k} \sin (kr+\delta(k)) +o(1),
 \quad r\to\infty.
 \end{equation}
Existence and uniqueness of the scattering solution $\psi(r,k)$
follows from \eqref{e5.0.8} because existence and uniqueness of the
regular solution $\varphi(r,k)$ follows from \eqref{e1.1.0} or from
\eqref{e1.1.8}.

\subsection{Higher angular momenta.}\label{U:5.0.3}
If one studies the three-dimensional scattering problem with a
spherically-symmetric potential $q(x)=q(r)$, $x\in\R^3$, $|x|=r$,
then the scattering solution $\psi(r,\alpha,k)$ solves the problem:
\begin{equation}\label{e5.0.10}
 [\nabla^2+k^2-q(r)]\psi=0 \hbox{\ in\ } R^3
 \end{equation}
\begin{equation}\label{e5.0.11}
 \psi=e^{ik\alpha\cdot x}+A(\alpha',\alpha,k) \frac{e^{ikr}}{r} +o
 \left( \frac{1}{r} \right), r:=|x|\to\infty,
 \alpha':=\frac{x}{r}, \alpha\in S^2.
 \end{equation}
Here $S^2$ is the unit sphere in $\R^3$ $\alpha\in S^2$ is given,
$A(\alpha',\alpha,k)$ is called the scattering amplitude.
If $q=q(r)$, then
$A(\alpha',\alpha,k)=A(\alpha'\cdot\alpha,k)$.
The converse is a theorem of Ramm \cite{R}, p.130.
The scattering solution solves the integral equation:
\begin{equation}\label{e5.0.12}
 \psi=e^{ik\alpha\cdot x}-\int_{\R^3} g(x,y,k) q(y) \psi(y,\alpha,k)dy,
 \quad g:=\frac{e^{ik|x-y|}}{4\pi|x-y|}.
 \end{equation}
It is known that
\begin{equation}\label{e5.0.13}
  e^{ik\alpha\cdot x} =\sum^\infty_{\l=0}
  \frac{4\pi}{k} i^\l \frac{u_\l(kr)}{r} Y_\l (\alpha') 
\overline{Y_\l(\alpha)},
  \quad \alpha':=\frac{x}{r},
  \quad u_\l:=\sqrt{\frac{\pi r}{2}} J_{\l+\frac{1}{2}},
  \end{equation}
$Y_\l(\alpha)$ are orthonormal in $L^2(S^2)$ spherical harmonics,
$Y_l=Y_{\l m}$, $-\l\leq m\leq \l$, and summation over $m$ 
in \eqref{e5.0.13} is understood but not shown, and $J_\l(r)$ is the
Bessel function.

If $q=q(r)$, then
\begin{equation}\label{e5.0.14}
  \psi=\sum^\infty_{\l=0} \frac{4\pi}{k} i^\l \frac{\psi_\l(r,k)}{r}
  Y_\l(\alpha') \overline{Y_\l(\alpha)},
  \end{equation}
where
\begin{equation}\label{e5.0.15}
  \psi''_\l +k^2\psi_\l-q(r)\psi_\l -\frac{\l(\l+1)}{r^2} \psi_\l=0,
  \end{equation}
\begin{equation}\label{e5.0.16}
  \psi_\l=e^{i\delta_\l} \sin
   \left(kr-\frac{\l\pi}{2} +\delta_\l \right) +o(1),
   \quad r\rightarrow\infty,
  \end{equation}
\begin{equation}\label{e5.0.17}
  \psi_\l=O(r^{\l+1}), \quad r\rightarrow 0.
  \end{equation}
Relation \eqref{e5.0.16} is equivalent to
\begin{equation}\label{e5.0.18}
  \psi_\l = \frac{e^{i\frac{\pi}{2}(\l+1)} }{2}
  \left[ e^{-ikr} - e^{i\pi\l} S_\l e^{ikr} \right]
  +o(1), \quad r\rightarrow\infty,
  \end{equation}
similar to \eqref{e5.0.8}, which is \eqref{e5.0.18} with $\l=0$.
If $q=q(r)$, then the scattering amplitude
$A(\alpha',\alpha,)=A(\alpha'\cdot \alpha, k)$ can be
written as
\begin{equation}\label{e5.0.19}
  A(\alpha'\cdot\alpha,k)=\sum^\infty_{\l=0} A_\l (k) Y_\l(\alpha')
  \overline{Y_\l(\alpha)},
  \end{equation}
while in the general case $=q(x)$,  one has
\begin{equation}\label{e5.0.20}
  A(\alpha',\alpha,k)=\sum^\infty_{\l=0} A_l(\alpha,k) Y_\l(\alpha').
  \end{equation}
If $q=q(r)$ then $S_\l$ in \eqref{e5.0.18} are related to $A_\l$
in \eqref{e5.0.19} by the formula
\begin{equation}\label{e5.0.21}
  S_\l=1-\frac{k}{2\pi i}A_\l.
  \end{equation}
In the general case $q=q(x)$, one has a relation between
$S\hbox{-matrix}$ and the scattering amplitude:
\begin{equation}\label{e5.0.22}
  S=I-\frac{k}{2\pi i}A,
  \end{equation}
so that \eqref{e5.0.21} is a consequence of \eqref{e5.0.22} in the case
$q=q(r):S_\l$ are the eigenvalues of $S$ in the eigenbasis
of spherical harmonics. Since $S$ is unitary, one has
$|S_\l|=1$, so $S_\l=e^{2i\delta_\l}$ for some real numbers $\delta_\l$,
which are called phase-shifts. These numbers are the same as in
\eqref{e5.0.17} (cf. \eqref{e5.0.18}). From \eqref{e5.0.21}
one gets
\begin{equation}\label{e5.0.23}
  A_\l(k)=\frac{4\pi}{k} e^{i\delta_\l} \sin (\delta_\l).
  \end{equation}
The Green function $g_\l(r,\rho)$, which solves the equation
\begin{equation}\label{e5.0.24}
  \left(\frac{d^2}{dr^2} +k^2 -\frac{\l(\l+1)}{r^2} \right)
  g_\l= -\delta(r-\rho),
  \quad \frac{\partial g_\l}{\partial r}
  -ik g_\l \underset{r\to+\infty}{\rightarrow} 0,
  \end{equation}
can be written explicitly:
\begin{equation}\label{e5.0.25}
  g_\l (r,\rho) =\left\{
  \begin{array}{llll}
  F^{-1}_{o\l}
    & \varphi_{o\l}(k\rho)f_{o\l}(kr)
      & r\geq \rho,
        & F_{o\l} :=\frac{e^{ \frac{i\l\pi}{2} }}{k^\l},
    \\
  F^{-1}_{o\l}
    &\varphi_{o\l}(kr) f_{o\l}(k\rho),
      & r<\rho,
        & \varphi_{o\l} (kr) =\frac{u_\l(kr)}{k^{\l+1}},
  \end{array}
  \right.
  \end{equation}
and the function $\psi_\l(r,k)$ solves the equation:
\begin{equation}\label{e5.0.26}
  \psi_\l(r,k) =u_\l(kr)
  - \int^\infty_0 g_\l(r,\rho) q(\rho)\psi_\l(\rho,k)d\rho.
  \end{equation}
The function $F_{o\l}(k)$ is the Wronskian $W[f_{o\l}, \varphi_{o\l}]$,
$\varphi_{o\l}(kr)$ is defined in \eqref{e5.0.25} and $f_{o\l}$
is the solution to \eqref{e5.0.15} (with $q=0$) with the asymptotics
\begin{equation}\label{e5.0.27}
  \begin{aligned}
  f_{o\l}=
    & e^{ikr}+o(1),  \quad r\to+\infty, \quad
      f_{o\l}(kr)= e^{i\frac{(\l+1)\pi}{2}} (u_\l(kr)+iv_\l(kr)),
  \\
    & v_\l:=\sqrt{\frac{\pi r}{2}} N_{\l+\frac{1}{2}}(kr).
  \end{aligned}
  \end{equation}
Let $\varphi_\l(r,k)$ be the regular solution to \eqref{e5.0.15} which is
defined by the asymptotics as $r\to 0$:
\begin{equation}\label{e5.0.28}
  \varphi_\l(r,k)=\frac{r^{\l+1}}{(2\l+1)!!} +o(r^{\l+1}),
  \quad r\to 0.
  \end{equation}
Then
\begin{equation}\label{e5.0.29}
 \begin{aligned}
  \psi_\l(r,k)
  & =a_\l(k)\varphi_\l(r,k),\\
  & \varphi_\l(r,k)
  =\frac {|f_\l(0,k)|}{k^{\l+1}}
     \sin\left( kr-\frac{\l\pi}{2}+\delta_\l\right)+o(1),
     r\to\infty.
  \end{aligned}
  \end{equation}

\begin{lemma}\label{L:5.0.2}
One has:
\begin{equation}\label{e5.0.30}
  \sup_{\l=0,1,2,\dots}|a_\l(k)|< \infty, 
  \end{equation}
where $k>0$ is a fixed number.
\end{lemma}

We omit the proof of this lemma.

\subsection{Eigenfunction expansion}\label{U:5.0.4}
We assume that $q\in L_{1,1}$ and $h\in C^\infty_0 \+R $,
$\l h=-h''+q(x)h$, $\lambda=k^2$, let
$g=
\frac{\varphi(x,\sqrt{\lambda})f(y,\sqrt{\lambda}) }{f(\sqrt{\lambda}) }$,
$y\geq x\geq 0$, be the resolvent kernel of $\l:(\l-\lambda) 
g=\delta(x-y)$,
$gh:=g(\lambda)h:=\int^\infty_0 g(x,y,\lambda) h dy$,
and $f_j=f(y,ik_j)$.
Then $\frac{h}{\lambda}=-gh+\frac{1}{\lambda}\l gh$.
Integrate this with respect to $\lambda\in\C$ over
$|\lambda|=N$ and divide by $2\pi i$ to get
\begin{equation} 
  h=-\frac{1}{2\pi i} \int_{|\lambda|=N}
  gh d\lambda +\frac{1}{2\pi i} \int_{|\lambda|=N} \frac{\l gh}{\lambda} 
d\lambda
  := I_1+I_2.
  \notag
  \end{equation}
The function $gh$ is analytic with respect to $\lambda$ on the
complex plane with the cut $(0,\infty)$ except for the points
$\lambda=-k^2_j$, $1\leq j\leq J$, which are simple poles of $gh$,
and $\lim_{N\to\infty} I_2=0$, because $|\l gh|=o(1)$
as $N\to\infty$.
Therefore:
\begin{equation}\label{e5.0.31}
  h=\frac{1}{2\pi i} \int^\infty_0 [g (\lambda+i0) h-g(\lambda-i0)h]
  d\lambda + \sum^J_{j=1} \frac{-1}{2\pi i} \oint_{|\lambda +k^2_j|=\delta}
  ghd\lambda.
  \end{equation}
One has (cf. \eqref{e1.1.9}):
\begin{equation} 
  \begin{aligned}
  ~& \frac{g(\lambda+i0)-g(\lambda-i0)}{2i}
  =\varphi(x,k) \frac{f(-k)f(y,k)-f(y,-k) f(k)}{2i|f(k)|^2}
  \\
  & =\frac{k}{|f(k)|^2} \varphi(x,k) \varphi(y,k),
    \quad k=\sqrt{\lambda}>0.
  \end{aligned}
  \notag
  \end{equation}
Also
\begin{equation} 
  \begin{aligned}
  -\frac{1}{2\pi i}
  & \oint_{|\lambda+k^2_j|=\delta} ghd\lambda = -\res_{\lambda=-k^2_j} gh
  \\
  & = -\int^\infty_0 f_j(y) h(y)dy
    \cdot \frac{\varphi(x,ik_j)}{\dotf(ik_j)} 2ik_j
  =s_jf_j(x) h_j,
  \  h_j:=\int^\infty_0 f_j hdy,
  \end{aligned}
  \notag
  \end{equation}
$s_j$ are defined in \eqref{e1.1.14}, and
\begin{equation}\label{e5.0.32}
  \varphi(x,ik_j)=\frac{f(x,ik_j)}{f'(0,ik_j)}
 :=\frac{f_j(x)}{f'(0,ik_j)}.
  \end{equation}
Therefore
\begin{equation}\label{e5.0.33}
  h(x)=\int^\infty_0
  \left( \int^\infty_0 \varphi(y,k)h(y)dy \right)
  \varphi(x,k) \frac{2k^2dk}{\pi|f(k)|^2}
  +\sum^J_{j=1} s_j f_j(x) h_j.
  \end{equation}
This implies (cf. \eqref{e1.1.19}, \eqref{e1.1.18}, \eqref{e1.1.14}):
\begin{equation}\label{e5.0.34}
  \begin{aligned}
  \delta(x-y)
  & =\frac{2}{\pi} \int^\infty_0 \varphi(x,k)\varphi(y,k)
  \frac{k^2dk}{|f(k)|^2} +\sum^J_{j=1} s_j f_j(x) f_j(y)
  \\
  & =\int^\infty_{-\infty} \varphi(x,\sqrt{\lambda}) \varphi(y,\sqrt{\lambda})
  d\rho(\lambda).
  \end{aligned}
  \end{equation}
We have proved the eigenfunction expansion theorem for
$h\in C^\infty_0 \+R $.
Since this set is dense in $L^2 \+R $, one gets the
theorem for $h\in L^2 \+R $.

\begin{theorem}\label{T:5.0.3}
If $q\in L_{1,1}$, then \eqref{e5.0.33} holds for any
$h\in L^2 \+R $ and the integrals converge in $L^2(\R_+)$ sense.
Parseval's equality is:
\begin{equation}\label{e5.0.35}
  \|h\|^2_{L^2(\R_+)}
  =\sum^J_{j=1} s_j |h_j|^2 +\frac{2}{\pi}
   \int^\infty_0 |\widetilde{h}(k)|^2 \frac{k^2dk}{|f(k)|^2},
   \quad \widetilde{h}:=\int^\infty_0 h(y) \varphi(y,k)dy.
  \end{equation}

\end{theorem}

\section{Statement of the inverse scattering problem
  on the half-line. Uniqueness theorem}\label{S:5.1}

In \refS{1.2} the statement of the ISP is given.
Let us prove the uniqueness theorem.

\begin{theorem}\label{T:5.1.1}
If $q_1, q_2 \in L_{1,1}$ generate the same data \eqref{e1.1.15},
then $q_1=q_2$.
\end{theorem}

\begin{proof}
We prove that the data \eqref{e1.1.15} determine uniquely
$I(k)$, and this implies $q_1=q_2$ by \refT{3.1.2}.

\begin{proof}[Claim 1]
{\it If \eqref{e1.1.15} is given, then $f(k)$ is uniquely determined.}

Assume there are $f_1(k)$ and $f_2(k)$ corresponding to the data
\eqref{e1.1.15}. Then
\begin{equation}\label{e5.1.1}
  \frac{f_1(k)}{f_2(k)} = \frac{f_1(-k)}{f_2(-k)},
  \quad -\infty< k< \infty.
  \end{equation}
The left-hand side of \eqref{e5.1.1} is analytic in $\C_+$
and tends to $1$ as $|k|\to\infty$, $k\in\C_+$,
and the right-hand side of \eqref{e5.1.1} is analytic in $\C_-$
and tends to $1$ as $|k|\to\infty$, $k\in\C_-$.
By analytic continuation $\frac{f_1(k)}{f_2(k)}$ is an analytic function
in $\C$,
which tends to $1$ as $|k|\to\infty$, $k\in\C$.
Thus, by Liouville theorem, $\frac{f_1(k)}{f_2(k)}=1$,
so $f_1=f_2$.
\end{proof}

\begin{proof}[Claim 2]
{\it If \eqref{e1.1.15} is given, then $f'(0,k)$ is uniquely defined.}

Assume there are $f'_1(0,k)$ and $f'_2(0,k)$ corresponding to \eqref{e1.1.15}.
By the Wronskian relation \eqref{e1.1.10}, taking into account that
$f_1(k)=f_2(k):=f(k)$ by Claim 1, one gets
\begin{equation}\label{e5.1.2}
   [f'_1(0,k)-f'_2(0,k)] f(-k)
  -[f'_1(0,-k)-f'_2(0,-k)]f(k)=0.
  \end{equation}
Denote $w(k):=f'_1(0,k)-f'_2(0,k)$. Then:
\begin{equation}\label{e5.1.3}
  \frac{w(k)}{f(k)} =\frac{w(-k)}{f(-k)},\quad k\in\R.
  \end{equation}
The function $\frac{w(k)}{f(k)}$ is analytic in $\C_+$
and tends to zero as $|k|\to\infty$, $k\in\C_+$,
and $\frac{w(-k)}{f(-k)}$ has similar properties in $\C_-$.
It follows that $\frac{w(k)}{f(k)}=0$, so $f'_1(0,k)=f'_2(0,k)$.
Let us check that $\frac{w(k)}{f(k)}$ is analytic in $\C_+$.
One has to check that $w(ik_j)=0$.
This follows from\eqref{e1.1.14}: if $f(k)$, $s_j$
and $k_j$ are given, then $f'(0,ik_j)$ are uniquely determined.

Let us check that $w(k)\to 0$ as $|k|\to\infty$, $k\in\C_+$.
Using \eqref{e3.1.5} it is sufficient to check that $A(0,0)$
is uniquely determined by $f(k)$, because the integral in
\eqref{e3.1.5} tends to zero as $|k|\to\infty$,
$k\in\C_+$ by the Riemann-Lebesgue lemma. From \eqref{e1.4.16},
integrating by parts one gets:
\begin{equation}\label{e5.1.4}
  f(k)=1-\frac{A(0,0)}{ik} -\frac{1}{ik}
  \int^\infty_0 e^{iky} A_y(0,y)dy.
  \end{equation}
Thus
\begin{equation}\label{e5.1.5}
  A(0,0)=-\lim_{k\to\infty} [ik(f(k)-1)].
  \end{equation}
Claim 2 is proved.
\end{proof}

Thus, \refT{5.1.1} is proved.
\end{proof}

\section{Reconstruction procedure}\label{S:5.2}
This procedure is described in \eqref{e1.4.10}.

\begin{proof}[Let us derive equation \eqref{e1.4.13}.]
Our starting point is formula \eqref{e5.0.34}:
\begin{equation}\label{e5.2.1}
  \int^\infty_0 \varphi(x,k)\varphi(y,k)
  \frac{2k^2dk}{\pi|f(k)|^2} + \sum^J_{j=1} s_j f_j(x) f_j(y)=0
  \quad y>x\geq 0.
  \end{equation}
From \eqref{e1.1.9} and \eqref{e1.1.23} one gets:
\begin{equation}\label{e5.2.2}
  \begin{aligned}
  \frac{k\varphi(x,k)}{|f(k)|}
  & =\sin(kx+\delta)
  +\int^\infty_x A(x,y) \sin(ky+\delta)dy
  \\
  & =(I+A) \sin(kx+\delta),
  \quad \delta=\delta(k).
  \end{aligned}
  \end{equation}
Apply to \eqref{e5.2.1} operator $(I+A)^{-1}$, acting on the functions
of $y$, and get:
\begin{equation}\label{e5.2.3}
  \frac{2}{\pi} \int^\infty_0 \frac{k\varphi(x,k)}{|f(k)|}
  \sin(ky+\delta)dk
  +\sum^J_{j=1} s_j f_j(x) e^{-k_jy}=0,
  \quad y>x\geq 0.
  \end{equation}
From \eqref{e5.2.2}, \eqref{e5.2.3}, and \eqref{e1.1.23} with $k=ik_j$,
one gets:
\begin{equation}\label{e5.2.4}
  \begin{aligned}
  (I+A) \left( \frac{2}{\pi} \right.
  & \left.  \int^\infty_0 \sin(kx+\delta) \sin(ky+\delta)dk\right)
  \\
  & +(I+A)\sum^J_{j=1} s_j e^{-k_j(x+y)}=0, y>x\geq 0.
   \end{aligned}
  \end{equation}
One has
\begin{equation}\label{e5.2.5}
  \begin{aligned}
  \frac{2}{\pi}
  & \int^\infty_0 \sin(kx+\delta) \sin(ky+\delta)dk
    =\frac{1}{\pi} \int^\infty_0 \cos[k(x-y)]dk
  \\
  & -\frac{1}{\pi} \int^\infty_0 \cos[k(x+y)+2\delta(k)]
  dk=\delta(x,y) -\frac{1}{2\pi}\int^\infty_{-\infty} (e^{2i\delta(k)}-1)
     e^{ik(x+y)} dk
   \\
  &=\delta(x-y)+\frac{1}{2\pi} \int^\infty_{-\infty} [1-S(k)]
     e^{ik(x+y)} dk.
  \end{aligned}
  \end{equation}
From \eqref{e1.4.11}, \eqref{e5.2.4} and \eqref{e5.2.5}
one gets \eqref{e1.4.13}. By continuity equation \eqref{e1.4.13},
derived for $y>x\geq 0$, remains valid for $y\geq x\geq 0$.
\end{proof}

\begin{theorem}\label{T:5.2.1}
If $q\in L_{1,1}$ and $F$ is defined by \eqref{e1.4.11} then equation
\eqref{e1.4.13} has a solution in $L^1 \xR \cap L^\infty \xR $,
$\R_x:=[x,\infty)$, for any $x\geq 0$, and this solution is unique.
\end{theorem}

Let us outline the steps of the proof.

{\it Step 1. If $q\in L_{1,1}$, then $F(x)$, defined by \eqref{e1.4.11}
satisfies the following estimates:
}

\begin{equation}\label{e5.2.6}
  |F(2x)|\leq c\sigma(x)|, |F(2x)+A(x,x)| \leq c\sigma(x),
  |F'(2x)-\frac{q(x)}{4}|\leq c\sigma^2(x),
  \end{equation}
where $\sigma(x)$ is defined in \eqref{e1.1.24}, and
\begin{equation}\label{e5.2.7}
  \|F\|_{L^2\+R} + \|F\|_{L^1\+R} +\|F\|_{L^\infty\+R}
  +\|xF'(x)\|_{L^1\+R}<\infty,
  \end{equation}
\begin{equation}\label{e5.2.8}
  \int^\infty_0\int^\infty_0 |F(s+y)|dsdy<\infty.
  \end{equation}

{\it Step 2. Equation
\begin{equation}\label{e5.2.9}
  (I+F_x)h:=h(y) +\int^\infty_x h(s) F(s+y)ds=0, \quad y\geq x\geq 0
  \end{equation}
is of Fredholm type in $L^1 \xR $, $L^2 \xR $ and in $L^\infty \xR $.
It has only the trivial solution $h=0$.
}

Using estimates \eqref{e5.2.6} -- \eqref{e5.2.8} and the criteria of
compactness in $L^p (\R_x) $, $p=1,2,\infty$, one checks that $F_x$ is 
compact
in these spaces for any $x\geq 0$. The space $L^1\cap L^\infty\subset L^2$
because $\|h\|_2 \leq \|h\|_1 \|h\|_\infty$,  where $\|h\|_p:=\|h\|_{L^p 
\xR }$.
We need the following lemma:

\begin{lemma}\label{L:5.2.2}
Let $h$ solve \eqref{e5.2.9}. If $h\in L^1:=L^1 \xR, $
then $h\in L^2:=L^2 \xR $. If $h\in L^1$, then $h\in L^\infty$.
If $h\in L^2$, then $h\in L^\infty$.
\end{lemma}

\begin{proof}
If $h$ solves \eqref{e5.2.9}, then 
$\|h\|_\infty \leq \|h\|_1 \sup_{y\geq 2x}
 |F(y)| \leq c(x) \|h\|_1<\infty,$
where $c(x)\to 0$ as $x\to \infty$.
Also
$\|h\|^2_2\leq \int^\infty_x dy \sigma^2(\frac{x+y}{2})
 \|h\|^2_1\leq c_1(x) \|h\|^2_1<\infty$,
$c_1(x)\to 0$ as $x\to\infty$.
So the first claim is proved.
Also
$\|h\|_1\leq \|h\|_1\sup_{s\geq x} \int^\infty_x |F(s+y)|dy=
 c_2(x)\|h\|_1$, $c_2(x)\to 0$ as $x\to\infty$.
If $h\in L^2$, then $\|h\|_\infty \leq \|h\|_2 \sup_{y\geq x}
   \left( \int^\infty_x |F(s+y)|^2ds\right)^{\frac{1}{2}}
   =c_3(x)\|h\|_2$, $c_3(x)\to 0$ as $x\to\infty$.
\end{proof}

\begin{lemma}\label{L:5.2.3}
If $h\in L^1$ solves \eqref{e5.2.9} and $x\geq 0$, then $h=0$.
\end{lemma}

\begin{proof}
By \refL{5.2.2}, $h\in L^2\cap L^\infty$.
It is sufficient to give a proof assuming $x=0$.
The function $F(x)$ is real-valued, so one can assume that $h$ is
real-valued. Multiply \eqref{e5.2.9} by $h$ and integrate over
$(x,\infty)$ to get
\begin{equation}\label{e5.2.10}
  \begin{aligned}
  \|h\|^2 +\frac{1}{2\pi}
    &\int^\infty_{-\infty} [1-S(k)]\tildeh^2(k) dk +\sum_{j=1}^J s_j
  \left( \int^\infty_x e^{-k_js} h(s) ds \right)^2 =0,
  \\
  &  \tildeh:=\int^\infty_x e^{iks} h(s) ds,
  \end{aligned}
  \end{equation}
where $\|h\|=\|h\|_{L^2\+R}$
one gets $\int^\infty_{-\infty} \tildeh^2 (k)dk=0$.
Also,  
$|\frac{1}{2\pi} \int^\infty_{-\infty} S(k)$
$\tildeh^2(k) dk |\leq \frac{1}{2\pi} \int_{-\infty}^\infty
 |\tildeh(k)|^2 dk =\|h\|^2$.
Therefore \eqref{e5.2.10} implies $s_j=0$, 
 $0=h_j:=\int^\infty_x h e^{-k_js}ds$,
$1\leq j\leq J$, and
\begin{equation}\label{e5.2.11}
  (\tildeh, \tildeh)=(\tildeh, S(-k)\tildeh(-k)),
  \end{equation}
where $(\tildeh,\tildeg):=\int^\infty_{-\infty} \tildeh (k)
\overline{\tildeg(k)} dk$. Since $h$ is real valued, one has
$\overline{\tildeh}(-k)$. The unitarity of $S$ implies
$S^{-1}(k)=S(-k)=\overline{S(k)}$, $k\in\R$, and
$\|S(-k)\tildeh(-k)\|=\|\tildeh(-k)\|$.
Because of \eqref{e5.2.11}, one has equality sign in the Cauchy inequality
$(\tildeh, S(-k)\tildeh(-k)\leq \|\tildeh\|^2$.
This means that $\tildeh(k)=S(-k)\tildeh(-k),$
 and \eqref{e1.1.15} implies
\begin{equation}\label{e5.2.12}
  \frac{\tildeh(k)}{f(k)}=\frac{\tildeh(-k)}{f(-k)},
  \qquad k\in\R.
  \end{equation}
Because $h_j=0$, one has $\tildeh(ik_j)=0$, and if
$f(0)\not=0$, then $\frac{\tildeh(k)}{f(k)}$ is analytic in $\C_+$
and vanishes as $|k|\to\infty$, $k\in\infty$, $k\in\C_+$.
Also $\frac{\tildeh(-k)}{f(-k)}$ is analytic in $\C_-$
and vanishes as $|k|\to\infty$, $k\in\C_-$.
Therefore, by analytic continuation,
$\frac{\tildeh(k)}{f(k)}$ is analytic in $\C$
and vanishes as $|k|\to\infty$.
By Liouville theorem, $\frac{\tildeh(k)}{f(k)}=0$,
so $\tildeh(k)=0$ and $h=0$.
If $f(0)=0$, then, by \refT{3.1.3}, $f(k)=ik \tildeA_1(k)$,
$\tildeA_1(0)\not= 0$,
and the above argument works.
\end{proof}

Because $F_x$ is compact in $L^2 \xR $,  the Fredholm alternative
is applicable to \eqref{e5.2.9}, and \refL{5.2.3} implies that
\eqref{e1.4.13} has a solution  in $L^2\xR$ for any $x\geq 0$, and this
solution is unique. Note that the free term in \eqref{e1.4.13}
is $-F(x+y)$, and this function of $y$ belongs to $L^2\xR$
(cf. \eqref{e5.2.8}).
Because $F_x$ is compact in $L^1\xR$, \refL{5.2.3} and \refL{5.2.2}
imply existence and uniqueness of the solution to \eqref{e1.4.13}
in $L^1\xR$ for any $x\geq 0$, and $F(x+y)\in L^1\xR$ for any
$x\geq 0$.
Note that the solution to \eqref{e1.4.13} in $L^1\xR$ is the same
as its solution in $L^2\xR$.
This is established by the argument used in the proof of \refL{5.2.2}.

We give a method for the derivation of the estimates \eqref{e5.2.6} -- 
\eqref{e5.2.8}.
Estimate \eqref{e5.2.8} is an
immediate consequence of the first estimate \eqref{e5.2.6}.
Indeed,
\begin{equation} 
  \begin{aligned}
  \int^\infty_0
     &\int^\infty_0 |F(s+y)|^2 dsdy
  \leq \left( \int^\infty_0 \max_{s\geq 0} |F(s+y)|dy \right)^2
  \\
  & \leq c\left(\int^\infty_0 \int^\infty_{\frac{y}{2}} |q|dt dy\right)^2
  \leq c\left(\int^\infty_0 t|q(t)|dt \right)^2
  <\infty.
  \end{aligned}
  \notag
  \end{equation}
Let us prove the first estimate \eqref{e5.2.6}.
Put in \eqref{e1.4.13} $x=y$:
\begin{equation}\label{e5.2.13}
  A(x,x)+\int^\infty_x A(x,s) F(s+x)dx +F(2x)=0.
  \end{equation}
Thus
\begin{equation}\label{e5.2.14}
  |F(2x)|\leq |A(x,x) | + \int^\infty_x |A(x,s)F(s+x)|dx.
  \end{equation}
From \eqref{e1.1.24} and \eqref{e5.2.14} one gets
\begin{equation}\label{5.2.15}
  \begin{aligned}
  |F(2x)|
    & \leq c\sigma(x)+c\int^\infty_x \sigma \left( \frac{x+s}{2} \right)
  |F(s+x)| ds  \\
    & \leq c\sigma(x) + c\sigma(x) \int^\infty_x |F(s+x)|ds\leq 
c\sigma(x),
  \end{aligned}
  \end{equation}
where $c=\const>0$ stands for various constants and we have used
the estimate
$$\sup_{x\geq 0}\int^\infty_x |F(s+x)|ds
 \leq  \int^\infty_0 |F(t)|dt =c<\infty.
$$
This estimate can be derived from \eqref{e1.1.24}. Write \eqref{e1.1.24} as
\begin{equation}\label{e5.2.16}
  A(x,z-x)+\int^\infty_z A(x,t+x-z) F(t)dt+ F(z)=0,
  \quad z\geq 2x\geq 0.
  \end{equation}
Let us prove that equation \eqref{e5.2.16} is uniquely
solvable for $F$ in $L^p\NR$, $p=\infty$, $p=1$ for all $x\geq 
\frac{N}{2}$,
where $N$ is a sufficiently large number.
In fact, we prove that the operator in \eqref{e5.2.16} has small norm in
$L^p\NR$ in $N$ is sufficiently large. Its norm in $L^\infty\NR$ is not 
more
than
\begin{equation}
  \begin{aligned}
  \sup_{z\geq N}
   & \int^\infty_N |A(x,t+x-z)|dt
   \leq c \int^\infty_{x+\frac{t-N}{2}} |q(s)|ds
   \\
   & \leq c \int^\infty_N dt \int^\infty_t |q(s)|ds
  =c\int^\infty_N (s-N) |q(s)|ds <1
  \end{aligned}
  \notag
  \end{equation}
because $q\in L_{1,1}$. We have used estimate \eqref{e1.1.24} above.
The function $A(x,-x)\in L^\infty\NR$, so our claim is proved for
$p=\infty$. Consider the case $p=1$. One has the following upper estimate
for the norm of the operator in \eqref{e5.2.16} in $L^1\NR$:
$  \sup_{t\geq N} \int ^t_N |A(x,t+x-z)|dz
  \leq \sup_{t\geq N} \int^{t-N}_0  |A(x,x+v)|dv
  \leq \int^\infty_0 dv \int^\infty_{x+\frac{v}{2}} |q|ds
  =2\int^\infty_x (s-x) |q|ds\to 0
  \hbox{\ as\ } x\to \infty.$
Also $\int^\infty_N |A(x,z-x)|dz<\infty$.
Thus equation \eqref{e5.2.16} is uniquely solvable in $L^1\NR$ for
all $x\geq \frac{N}{2}$ if $N$ is sufficiently large.
In order to finish the proof of the first estimate \eqref{e5.2.6} it is
sufficient to prove that $\|F\|_{L^\infty(0,N)} \leq c<\infty$.
This estimate is obvious for $F_d(x)$ (cf. \eqref{e1.4.11}).
Let us prove it for $F_s(x)$.
Using \eqref{e3.4.3}, \eqref{e3.4.5}, \eqref{e3.4.5'}, \eqref{e3.4.6},
one gets
\begin{equation}\label{e5.2.17}
  1-S(k)= \frac{[f(k)-f(-k)] (k+i\kappa)}{f_0(k)w(k) k}
  =[\tildeA(k)-\tildeA(k)]
  (1+\tildeb(k))(1+\tildeg) \left(1+\frac{i\kappa}{k}\right),
  \end{equation}
where all the Fourier transforms are taken of $W^{1,1}\+R$ functions.
Thus, one can conclude that $F_s(x)\in L^\infty\+R$ if one can prove that
$I:=\frac{\tildeA(k)-\tildeA(k)}{k}$
is the Fourier transform of $L^\infty\+R$ function.
One has $I=\int^\infty_0 dy A(y)$
$\frac{e^{iky}-e^{-iky} }{k}$
and
\begin{equation}\label{e5.2.18}
  \begin{aligned}
  \int^\infty_{-\infty} e^{ikx} I(k)dk
  & =\int^\infty_0 dy A(y) \int^\infty_{-\infty}
    \frac{ e^{ik(x+y)}-e^{ik(x-y)} }{k}
  \\
  & =\int^\infty_ 0 dy A(y) i\pi [1-sgn(x-y)]
  =2i\pi\int^\infty_x dy A(y).
  \end{aligned}
  \end{equation}
From \eqref{e1.1.24} it follows that
$\int^\infty_x A(y) dy\in L^\infty\+R$.
We have proved that $\|F\|_{L^\infty\+R} + \|F\|_{L^1\+R}<\infty$.
Differentiate \eqref{e5.2.13} to get
\begin{equation}\label{e5.2.19}
  \begin{aligned}
  2F'(2x)
  &+\dot A(x,x)-A(x,x) F(2x)
  +\int^\infty_x A_x(x,s)  F(s+x)ds
  \\
  &+\int^\infty_x A(x,s) F'(s+x)ds=0,\quad
  \dot A:=\frac{dA(x,x)}{dx},
  \end{aligned}
  \end{equation}
or
\begin{equation} \label{e5.2.20}
  F'(2x)=\frac{q(x)}{4} +A(x,x) F(2x)
  -\frac{1}{2}\int^\infty_x [A_x(x,s)-A_s(x,s)]  F(s+x)ds.
  \end{equation}
One has $\int^\infty_0 x |q|dx<\infty$,
$\int^\infty_0 x|A(x,x)| |F(2x)|dx\leq
 \sup_{x\geq 0} (x|A(x,x)|)\cdot \int^\infty_0$ $|F(2x)|dx\leq c$.
Let us check that
$ I :=\int^\infty_0 x
 |\int^\infty_x [A_x(x,s)-A_s(x,s)] F(s+x)ds|$
 $dx<\infty$.
Use \eqref{e1.1.25} and get
$I \leq c\int^\infty_0 x \sigma(x) \int^\infty_x\sigma$
$ \left( \frac{x+s}{2} \right) |F(s+x)| ds dx
 \leq c\int^\infty_0\sigma(x) dx \int^\infty_0 |F(y)|dy
 \cdot \sup_{x\geq 0, s\geq x} x\sigma \left( \frac{x+s}{2} \right)
 \leq c<\infty$.
The desired estimate is derived.

The third estimate \eqref{e5.2.6},
$|F'(2x)-\frac{q(x)}{4}|\leq c\sigma^2(x)$
follows from \eqref{e5.2.20} because
$|A(x,x)|\leq c\sigma(x)$,  $|F(2x)|\leq c\sigma(x)$,
and
$\int^\infty_x |A_x(x,s)-A_s(x,s)|
|F(s+x)ds\leq c\sigma(x)\int^\infty_x \sigma\left(\frac{x+s}{2}\right)
|F(s+x)|ds\leq c\sigma^2(x) \int^\infty_0 |F(s+x)|ds\leq c\sigma^2(x)$.
The estimate $|F(2x)+ A(x,x)|\leq c\sigma(x)$
follows similarly from \eqref{e5.2.13} and \eqref{e1.1.24}.
\refT{5.2.1} is proved.

\section{Invertibility of the steps of the reconstruction
procedure}\label{S:5.3}

The reconstruction procedure is \eqref{e1.4.10}.
1. The step $\calS\Rightarrow F$ is done by formula \eqref{e1.4.11}.

To do the step $F\Rightarrow\calS$, one takes $x\to -\infty$ in
\eqref{e1.4.11} and finds $s_j$, $k_j$, and $J$. Thus $F_d(x)$ is found and
$F_s=F-F_d$ is found.
From $F_s(x)$ and finds $1-S(k)$ by the inverse Fourier transform.
So $S(k)$ is found and the data $\calS$ (see \eqref{e1.1.15}) is found

2. The step $F\Rightarrow A$  is done by solving equation \eqref{e1.4.13}.
By \refT{5.2.1} this equation is uniquely solvable in
$L^1(\R_x)\cap L^\infty(\R_x)$ for all $x\geq 0$ if $q\in L_{1,1}$,
that is, if $F$ came from $\calS$ corresponding to $q\in L_{1,1}$.

To do the step $A\Rightarrow F$, one finds
$f(k)=1+\int^\infty_0 A(0,y) e^{iky}dy$,
then the numbers $ik_j$, the zeros of $f(k)$ in $\C_+$, the number $J$,
$1\leq j\leq J$, and $S(k)=\frac{f(-k)}{f(k)}$.
The numbers $s_j$ are found by formula \eqref{e1.1.14}, where
\begin{equation} \label{e5.3.1}
 f'(0,ik_j) =-k_j -A(0,0) +\int^\infty_0 A_x(0,y)e^{-k_jy}dy.
 \end{equation}
Thus $A\Rightarrow \calS$ and $\calS \Rightarrow F$ by formula
\eqref{e1.4.11}. We also give a direct way to do the step $A\Rightarrow F$.

Write equation \eqref{e1.4.13} with $z=x+y$, $v=s+y$, as
\begin{equation}\label{e5.3.2}
  (I+B_x)F:=F(z)+ \int^\infty_z A(x,v+x-z) F(v)dv =-A(x,z-x),
  \quad z\geq 2x\geq 0.
  \end{equation}
The norm of the operator $B_x$ in $L^1_{2x}$ is estimated as follows:
\begin{equation}\label{e5.3.3}
 \|B_x\| \leq \sup_{v>0} \int^v_0 |A(x,v+x-z)| dz
 \leq c\sup_{v>0} \int^v_0\sigma \left(x+\frac{v-z}{2}\right) dz
 \leq c \int^\infty_x \sigma(t) dt,
 \end{equation}
where $\sigma(x)=\int^\infty_x|q(t)|dt$
and the estimate \eqref{e1.1.24} was used.
If $x_0$ is sufficiently large then $\|B_x\|<1$ for $x\geq x_0$
because $\int^\infty_x\sigma(t)dt\to 0$ as $x\to\infty$ if $q\in L_{1,1}$.
Therefore equation \eqref{e5.3.2} is uniquely solvable in $L^1_{2x}$ for all
$x\geq x_0$ (by the contraction mapping principle),
and so $F(z)$ is uniquely determined for all $z\geq 2x_0$.

Now rewrite \eqref{e5.3.2} as
\begin{equation}\label{e5.3.4}
 F(z)+\int^{2x_0}_z A(x,v+x-z) F(v)dv
 =-A(x,z-x) -\int^\infty_{2x_0} A(x,v+x-z) F(v)dv.
 \end{equation}

This is a Volterra equation for $F(z)$ on the finite interval $(0,2x_0)$.
It is uniquely solvable since its kernel is a continuous function.
One can put $x=0$ in \eqref{e5.3.4} and the kernel $A(0,v-z)$
is a continuous function of $v$ and $z$, and the right-hand side
of \eqref{e5.3.4} at $x=0$ is a continuous function of $z$.
Thus $F(z)$ is uniquely recovered for all $z\geq 0$ from $A(x,y)$,
$y\geq x\geq 0$. Step $S\Rightarrow F$ is done.

3. The step $A\Rightarrow q$ is done by formula \eqref{e1.4.12}.
The converse step $q\Rightarrow A$ is done by solving Volterra equation
\eqref{e1.1.26}, or, equivalently, the Goursat problem
\eqref{e5.0.1} -- \eqref{e5.0.3}.

We have proved:
\begin{theorem}\label{T:5.3.1} 
If $q\in L_{1,1}$ and $\calS$ are the corresponding data \eqref{e1.1.15},
then each step in \eqref{e1.4.10} is invertible.
In particular, the potential obtained by the procedure \eqref{e1.4.10}
equals to the original potential $q$.
\end{theorem}

\begin{remark}\label{R:5.3.2}
If $q\in L_{1,1}$ and $A_q:=A_q(x,y)$ is the solution to \eqref{e1.1.26},
then $A_q$ satisfies equation \eqref{e1.4.13} and, by the uniqueness
of its solution, $A_q=A$, where $A$ is the function obtained by the
scheme \eqref{e1.4.10}. Therefore, the $q$  obtained by \eqref{e1.4.10}
equals to the original $q$.
\end{remark}

\begin{remark}\label{R:5.3.3}
One can verify directly that the solution $A(x,y)$ to \eqref{e1.4.13}
solves the Goursat problem \eqref{e5.0.1} -- \eqref{e5.0.3}.
This is done as in \refS{4.3}, Step 3.
Therefore $q(x)$, obtained by the scheme \eqref{e1.4.10}, generates
the same $A(x,y)$ which was obtained at the second step
of this scheme, and therefore this $q$ generates the original
scattering data.
\end{remark}

\begin{remark}\label{R:5.3.4}
The uniqueness \refT{5.1.1} does not imply that if one starts with
a $q_0=L_{1,1}$,  computes the corresponding scattering data \eqref{e1.1.15},
and applies inversion scheme \eqref{e1.4.10},
then the $q$ is obtained by this scheme is equal to $q_0$.
Logically it is possible that this $q$  generates data $\calS_1$
which generate by the scheme\eqref{e1.4.10} potential $q_1$, etc.
To close this loop one has to check that $q=q_0$.
This is done in \refT{5.3.1},
because $q_0=-2\frac{dA(x,x)}{dx}=q(x)$.
\end{remark}

\section{Characterization of the scattering data}\label{S:5.4}
In this Section we give a necessary and sufficient condition for the
data \eqref{e1.1.15} to be the scattering data corresponding to
$q\in L_{1,1}$. In \refS{5.7} we give such conditions on $\calS$
for $q$ to be compactly supported, or $q\in L^2(\R_+)$.

\begin{theorem}\label{T:5.4.1}
If $q\in L_{1,1}$, then the following conditions hold:
1) \eqref{e1.1.21}; 
2) $k_j>0$, $s_j>0$, $1\leq j\leq J$, $S(k)=\overline{S(-k)}=S^{-1}(k)$,
$k\geq 0$, $S(\infty)=1$;
3) \eqref{e5.2.7} 
hold. Conversely, if $\calS$ satisfies conditions
1) -- 3), then $\calS$ corresponds to a unique $q\in L_{1,1}$.
\end{theorem}

\begin{proof}
The necessity of conditions 1) -- 3) has been proved in
\refT{5.2.1}. 
Let us prove the sufficiency.
If conditions 1) -- 3) hold, then the scheme 
\eqref{e1.4.10} 
yields a unique potential, as was proved in \refR{5.3.2}.    
Indeed, equation \eqref{e1.4.13} is of Fredholm type in
$L^1(\R_x)$ for every $x\geq 0$ if $F$ satisfies \eqref{e5.2.7}.
Moreover, equation \eqref{e5.2.9} has only the trivial solution if
conditions 1) -- 3) hold.
Every solution to \eqref{e5.2.9} in $L^1(\R_x)$ is also a solution in
$L^2(\R_x)$ and in $L^\infty(\R_x)$, and the proof of the uniqueness
of the solution to \eqref{e5.2.9} under the conditions 1) -- 3) goes as
in \refT{5.2.1}.
The role of $f(k)$ is played by the unique solution of the Riemann
problem:
\begin{equation}\label{e5.4.1}
 f_+(k)=S(-k)f_-(k)
 \end{equation}
which consists of finding two functions $f_+(k)$ and $f_-(k),$ satisfying
(5.5.1), such that $f_+$ is an analytic function in
$\C_+$, $f_+(ik_j)=0$, $\dot f_+(ik_j)\not= 0$, $1\leq j\leq J$,
$f_+(\infty)=1$, and $f_-(k)$ is an analytic function in $\C_-$
such that $f_-(-ik_j)=0$, $\dot f_-(-ik_j)\not= 0$, $1\leq j\leq J$,
$f_-(\infty)=1$, and $f_+(0)=0$ if $\ind S(k)=-2J-1$,
$f_+(0)\not= 0$ if $\ind S(k)=-2J$.
Existence of a solution to (5.5.1) follows from the
non-negativity of $\ind S(-k)=-\ind S(k)$.
Uniqueness of the solution to the above problem is proved as follows.
Denote $f_+(k):=f(k)$ and $f_-(k)=f(-k)$.
Assume that $f_1$ and $f_2$ solve the above problem.
Then (5.5.1) implies
\begin{equation}\label{e5.4.2} 
 \frac{f_1(k)}{f_2(k)} =\frac{f_1(-k)}{f_2(-k)} \,\,\,, k\in\R,
 \quad f_1(ik_j)=f_2(ik_j)=0,
 \quad \dot f_1(ik_j)\not= 0,
 \quad \dot f_2(ik_j)\not=0,
 \quad f_1(\infty)=f_2(\infty)=1.
 \end{equation}
The function $\frac{f_1(k)}{f_2(k)}$ is analytic in $\C_+$
and tends to $1$ at infinity in $\C_+$, 
The function $\frac{f_1(-k)}{f_2(-k)}$ is analytic in $\C_-$
and tends to $1$ at infinity in $\C_-$.
Both functions agree on $\R$. Thus $\frac{f_1(k)}{f_2(k)}$
is analytic in $\C$ and tends to 1 at infinity.
Therefore $f_1(k)=f_2(k)$. To complete the proof we need to check that
$q$, obtained by \eqref{e1.4.10}, belongs to $L_{1,1}$.
In other words, that $q=-2\frac{dA(x,x)}{dx}\in L_{1,1}$.
To prove this, use \eqref{e5.2.19} and \eqref{e5.2.20}.
It is sufficient to check that $F'(2x)\in L_{1,1}$,
$A(x,x)F(2x)\in L_{1,1}$ and
$\int^\infty_x[A_x(x,s)-A_x(x,s)]F(s+x)ds\in L_{1,1}$.
The first inclusion follows from $\|xF'\|_{L^1(\R_+)}<\infty$.
Let us prove that $\lim_{x\to\infty}[x F(x)]=0$.
One has $\int^x_0s F'ds=xF(x)-\int^x_0F ds$.
Because $xF'\in L^1(\R_+)$ and $F\in L^1(\R_+)$ it follows that
the limit $c_0:=lim_{x\to\infty}xF$ exists. This limit has to be zero:
if $F=\frac{c_0}{x}+o(\frac{1}{x})$ as $x\to\infty,$ and $c_0\not= 0$,
then $F\not\in L^1(\R_+)$.
Now $\int^\infty_0 x|F(2x) A(x,x)|dx\leq c\int^\infty_0 |A(x,x)|dx<\infty$.
The last inequality follows from \eqref{e5.2.13}:
since $F(2x)\in L^1(\R_+)$ it is sufficient to check that
$\int^\infty_x A(x,s)F(s+x)ds\in L^1(\R_+)$.
One has
$\int^\infty_0 dx \int^\infty_x |A(x,s)|
|F(s+x)|ds\leq \int^\infty_0dx \sigma_F(2x) \int^\infty_x |A(x,s)|ds\leq 
c.$
Here
\begin{equation}\label{e5.4.3} 
 \sigma_F(x):=\sup_{y\geq x} |F(y)|, \quad \sigma_F\in L^1(\R_+).
 \end{equation}
Note that $\lim_{x\to\infty} x \sigma_F(x)=0$
because $\sigma_F(x)$ is monotonically decreasing and belongs to 
$L^1(\R_+)$.

\end{proof} 


\section{A new equation of Marchenko-type}\label{5.8}
The basic result of this Section is:
\begin{theorem}\label{T:5.8.1}
Equation 
\begin{equation}\label{e5.8.1}
  F(y)+A(y)+\int^\infty_{-\infty} A(t)F(t+y)dt = A(-y),
  \quad -\infty<y<\infty,
  \end{equation}
holds, where $A(y):=A(0,y)$, $A(y)=0$ for $y<0$, $A(x,y)$ is defined in \eqref{e1.1.23} 
and $F(x)$ is defined in \eqref{e1.4.11}.
\end{theorem}

\begin{proof}
Take the Fourier transform of \eqref{e5.8.1} in the sense of distributions 
and get:
\begin{equation}\label{e5.8.2}
  \tildeF(\xi)+\tildeA(\xi)+\tildeA(-\xi)\tildeF(\xi)=\tildeA(-\xi),
\end{equation}
where, by \eqref{e1.4.11},
\begin{equation}\label{e5.8.3}
 \tildeF(\xi)=1-S(-\xi)+2\pi \sum^J_{j=1} s_j \delta(\xi_i+ik_j).
\end{equation}
Use \eqref{e1.4.16}, the equation  $S(\xi)f(\xi)=f(-\xi)$,
add 1 to both sides of \eqref{e5.2.8}, and get:
\begin{equation}\label{e5.8.4}
 f(\xi)+f(-\xi)\tildeF(\xi)=f(-\xi).
\end{equation}
From \eqref{e5.8.3} and \eqref{e5.8.4} one gets:
\begin{equation}\label{e5.8.5}
 f(\xi)=f(-\xi)
 [ S(-\xi)-2\pi \sum^J_{j=1} s_j \delta(\xi+ik_j) ]
=f(\xi)-2\pi \sum^J_{j=1} s_j \delta(\xi+ik_j) f(-\xi)=f(\xi),
\end{equation}
where the equation $\delta(\xi+ik_j)f(-\xi)=0$ was used.
This equation  holds because $f(ik_j)=0$, and the product 
$\delta(\xi+ik_j)f(-\xi)$ makes sense because $f(\xi)$ is analytic in 
$\C_+$.
Equation \eqref{e5.8.5} holds obviously, and since each of our steps was 
invertible, equation \eqref{e5.8.1} holds.
\end{proof}

\begin{remark}\label{R:5.8.2}
Equation \eqref{e5.8.1} has a unique solution $A(y)$, such that 
$A(y)\in L^1(\R_+)$ and $A(y)=0$ for $y<0$.
\end{remark}

\begin{proof}
Equation \eqref{e5.8.1} for $y>0$ is identical with \eqref{e1.4.13} because
$A(-y)=0$ for $y>0$.
Equation \eqref{e1.4.13} has a solution in $L^1(\R_+)$ and this solution is unique, 
see \refT{5.2.1}.
Thus, equation \eqref{e5.8.1} cannot have more than one solution, because
every solution $A(y)\in L^1(\R_+)$, $A(y)=0$ for $y<0$, of \eqref{e5.8.1}
solves \eqref{e1.4.13}, and \eqref{e1.4.13} has no more than one solution.
On the other hand, the solution $A(y)\in L^1(\R_+)$ of \eqref{e1.4.13} does exist, is unique, 
and solves \eqref{e5.8.1}, as was shown in the proof of \refT{5.8.1}. 
This proves \refR{5.8.2}.
\end{proof}


\def\s{{\sigma}}
\def\ra{\rightarrow}
\def\lra{\longrightarrow}
\def\R{{\mathbb R}}
\def\N{{\mathbb N}}
\def\Z{{\mathbb Z}}
\def\C{{\mathbb C}}
\def\oH{\buildrel\circ\over H}
\def\oH1{\buildrel\circ\over H\kern-.02in{}^1}
\def\qed{{\hfill $\Box$}}
\def\l{\ell}
\def\dotf{\dot{f}}
\def\tildeR{\widetilde R}
\def\tildeV{\widetilde v}
\def\tildeW{\widetilde w}
\def\ind{\hbox{ind}}
\def\cop{\bot\hskip-.075in\bot}
\def\bysame{\rule{.5in}{.005in},\ }
\def\Im{\hbox{\,Im\,}}
\def\supp{\hbox{\,supp\,}}
\def\const{\hbox{\,const\,}}

\section{ Inequalities for the transformation operators and applications}\label{S:5.7}

\subsection{Inequalities for $A$ and $F$}\label{U:5.7.1}
The scattering data 
(1.2.17) 
satisfy the following conditions:

A) $k_j, s_j>0, \, S(-k)=\overline {S(k)}=S^{-1}(k), \, k\in \R,
\,S(\infty)=1,$

B) ${\mathcal J}: = ind S(k):=\frac 1 {2\pi}\int_{-\infty}^\infty dlog S(k)$ 
is a nonpositive integer,

C) $F\in L^p$, $p=1$ and $p=\infty$, $xF'\in L^1$, $L^p:=L^p(0,\infty)$.

If one wants to study the characteristic properties of the scattering 
data,  that is, a necessary and sufficient condition on these data 
to guarantee that the corresponding potential belongs to a prescribed 
functional class, then conditions A) and B) are always necessary for
a real-valued $q$ to be in $L_{1,1}$, the usual class in the scattering 
theory, or in some other class for which the scattering theory is 
constructed, and 
a condition of the type C) determines actually the class of potentials 
$q$.
Conditions A) and B) are consequences of the 
selfadjointness of the Hamiltonian, finiteness of its negative spectrum,
and of the unitarity of the $S-\hbox{matrix}$.
Our aim is to derive some inequalities for $F$ and $A$ from equation
\eqref{e1.4.13}.
This allows one to describe the set of $q$, defined by 
(1.5.13). 

Let us assume:
\begin{equation}\label{e5.7.1}  
\sup_{y\geq x}|F(y)|:=\s_F(x)\in L^1, \quad F'\in L_{1,1}.
\end{equation}
 The function $\s_F$ is monotone decreasing, $|F(x)|\leq \s_F(x)$.
Equation \eqref{e1.4.13} 
is of Fredholm type in $L^p_x:=L^p(x,\infty)$ 
$\forall x\geq 0$ and $p=1$.
The norm of the operator $F:=F_x$ in 
\eqref{e1.4.13} 
can be estimated :
\begin{equation}\label{e5.7.2}  
||F_x||\leq \int_x^\infty \s_F(x+y)dy\leq \s_{1F}(2x), \quad 
\s_{1F}(x):=\int_x^\infty \s_F(y)dy.
\end{equation} 
Therefore 
\eqref{e1.4.13} 
is uniquely solvable in $L^1_x$ for any $x\geq x_0$ if
\begin{equation}\label{e5.7.3}
\s_{1F}(2x_0)<1.
\end{equation}
This conclusion is valid for any $F$ satisfying \eqref{e5.7.3}, and conditions 
A), B), and C) are not used.
Assuming \eqref{e5.7.3} and \eqref{e5.7.1} and taking $x\geq x_0$, let us derive 
inequalities  for $A=A(x,y)$. Define 
$$\s_A(x):=\sup_{y\geq x}|A(x,y)|:=||A||.
$$ 
From 
\eqref{e1.4.13} 
one gets:
$$\s_A(x)\leq \s_{F}(2x)+\s_A(x)\sup_{y\geq x}\int_x^\infty \s_F(s+y)ds 
\leq \s_{F}(2x)+\s_A(x)\s_{1F}(2x).
$$
Thus, if \eqref{e5.7.3} holds, then
\begin{equation}\label{e5.7.4}
\s_A(x)\leq c\s_{F}(2x), \quad x\geq x_0.
\end{equation}
 By $c>0$  different constants depending on $x_0$ are denoted.
Let 
$$\s_{1A}(x):=||A||_1:=\int_x^\infty |A(x,s)|ds.$$
Then 
\eqref{e1.4.13} 
yields
$\s_{1A}(x)\leq \s_{1F}(2x)+\s_{1A}(x) \s_{1F}(2x)$. So
\begin{equation}\label{e5.7.5}
\s_{1A}(x)\leq c\s_{1F}(2x), \quad x\geq x_0.
\end{equation}
Differentiate 
\eqref{e1.4.13} 
with respect to $x$ and $y$ and get:
\begin{equation}\label{e5.7.6}
  (I+F_x)A_x(x,y)=A(x,x)F(x+y)-F'(x+y),  
  \quad y \geq x \geq 0,
  \end{equation}
and
\begin{equation}\label{e5.7.7}
  A_y(x,y)+\int_x^\infty A(x,s)F'(s+y)ds=-F'(x+y),
  \quad y \geq x \geq 0.
  \end{equation}
Denote 
\begin{equation}\label{e5.7.8}
\s_{2F}(x):=\int_x^\infty |F'(y)|dy, \quad \s_{2F}(x)\in L^1.
\end{equation}
Then, using \eqref{e5.7.7} and \eqref{e5.7.4}, one gets
\begin{equation}\label{e5.7.9}
||A_y||_1\leq  \int_x^\infty |F'(x+y)|dy+\s_{1A}(x) \sup_{s\geq 
x}\int_x^\infty |F'(s+y)|dy
\leq \s_{2F}(2x) [1+c\s_{1F}(2x)]\leq c\s_{2F}(2x),
  \end{equation}
and using \eqref{e5.7.6} one gets:
$$||A_x||_1\leq A(x,x)\s_{1F}(2x)+\s_{2F}(2x)+||A_x||_1\s_{1F}(2x),$$ 
so
\begin{equation}\label{e5.7.10}
||A_x||_1\leq c[\s_{2F}(2x)+\s_{1F}(2x)\s_{F}(2x)].
 \end{equation}

Let $y=x$ in 
\eqref{e1.4.13},
then differentiate
\eqref{e1.4.13} 
with respect to $x$ and get:
\begin{equation}\label{e5.7.11}
\dot A(x,x)= -2F'(2x)+A(x,x)F(2x)- \int_x^\infty 
A_x(x,s)F(x+s)ds -\int_x^\infty A(x,s)F'(s+x)ds.
  \end{equation}
From \eqref{e5.7.4}, \eqref{e5.7.5},
\eqref{e5.7.10} and \eqref{e5.7.11} one gets:
\begin{equation}\label{e5.7.12}
|\dot A(x,x)|\leq 
2|F'(2x)|+c\s^2_F(2x)+c\s_F(2x)[\s_{2F}(2x)+\s_{1F}(2x)\s_{F}(2x)]
+c\s_F(2x)\s_{2F}(2x).
  \end{equation}
Thus,
\begin{equation}\label{e5.7.13}
x|\dot A(x,x)|\in L^1, 
  \end{equation}
provided that $xF'(2x)\in L^1, \, x\s^2_F(2x)\in L^1,$ and
 $ x\s_F(2x)\s_{2F}(2x)\in L^1.$
Assumption \eqref{e5.7.1} implies $xF'(2x)\in L^1$. If $\s_{F}(2x)\in L^1$,
and $\s_{F}(2x)>0$ decreases monotonically, then $x\s_{F}(x)\to 0$
as $x\to \infty$. Thus $x\s^2_F(2x)\in L^1,$ and
$ \s_{2F}(2x)\in L^1$  because $\int_0^\infty dx\int_x^\infty |F'(y)|dy=
\int_0^\infty |F'(y)|ydy<\infty$, due to \eqref{e5.7.1}.
Thus, \eqref{e5.7.1} implies \eqref{e5.7.4},
\eqref{e5.7.5}, \eqref{e5.7.8}, \eqref{e5.7.9}, and
\eqref{e5.7.12}, while \eqref{e5.7.12} and 
(1.5.13) 
imply $q\in \tildeL_{1,1}$ where $\tildeL_{1,1}=\{q: q={\overline q},\, 
\int^\infty _{x_0}x|q(x)|dx<\infty\}$, and $x_0\geq 0$ satisfies
\eqref{e5.7.3}.

Let us assume now that \eqref{e5.7.4}, \eqref{e5.7.5}, \eqref{e5.7.9},
and \eqref{e5.7.10} hold, where
$\s _F\in L^1$ and $\s_{2F}\in L^1$ are some positive monotone decaying 
functions (which have nothing to do now with the function $F$,
solving equation 
\eqref{e1.4.13}, 
and derive 
estimates for this function $F$. Let us rewrite 
\eqref{e1.4.13} 
as:
\begin{equation}\label{e5.7.14}
F(x+y)+\int^\infty_x A(x,s)F(s+y)ds=-A(x,y), \qquad y\geq x\geq 0.
\end{equation}
Let $x+y=z, s+y=v$. Then,
\begin{equation}\label{e5.7.15}
F(z)+\int^\infty_z A(x,v+x-z)F(v)dv=-A(x,z-x),\qquad z\geq 2x.
\end{equation}
From \eqref{e5.7.15} one gets:
$$\s_F(2x)\leq\s_A(x)+
\s_F(2x)\sup_{z\geq 2x}\int^\infty_z |A(x,v+x-z)|dv\leq\s_A(x)+\s_F(2x)\, 
||A||_1.
$$

Thus, using \eqref{e5.7.5} and \eqref{e5.7.3}, one obtains:
\begin{equation}\label{e5.7.16}
\s_F (2x)\leq c\s_A(x).
\end{equation}
Also from \eqref{e5.7.15} it follows that:
\begin{equation}\label{e5.7.17}
\begin{array}{ll}
\s_{1F}(2x)&:=||F||_1:=\int^\infty_{2x}|F(v)|dv \\
&\leq\int^\infty_{2x}
|A(x,z-x)|dz+\int^\infty_{2x}\int^\infty_z |A(x,v+x-z)| |F(v)|dvdz \\
&\leq
||A||_1 +||F||_1 ||A||_1, \\
&\text {so} \\
& \s_{1F}(2x)\leq c\s_{1A}(x).
\end{array}
\end{equation}
From \eqref{e5.7.6} one gets:
\begin{equation}\label{e5.7.18}
\int^\infty_x |F'(x+y)|dy=\s_{2F}(2x)\leq c\s_A (x) \s_{1A}(x)+||A_x ||+c 
||A_x||_1 \s_{1A}(x).
\end{equation}
Let us summarize the results:

\begin{theorem}\label{T:5.7.1}  
If $x\geq x_0$ and \eqref{e5.7.1} hold, then one has:
\begin{equation}\label{e5.7.19}
\begin{array}{ll}
\s_A(x)\leq c\s_F (2x),\quad \s_{1A}(x)\leq c\s_{1F}(2x),\quad 
||A_y||_1\leq\s_{2F}(2x)(1+c\s_{1F}(2x)),\\
||A_x||_1\leq 
c[\s_{2F}(2x)+\s_{1F}(2x)\s_F (2x)].
\end{array}
\end{equation}
Conversely, if $x\geq x_0$ and
\begin{equation}\label{e5.7.20}
\s_{A}(x)+\s_{1A}(x) +||A_x ||_1 +||A_y||_1 <\infty,
\end{equation}
then
\begin{equation}\label{e5.7.21}
\begin{array}{ll}
\s_F (2x)\leq c\s_{A}(x),\quad \s_{1F}(2x)\leq c\s_{1A}(x),&\\ 
\s_{2F}(x)\leq 
c[\s_{A}(x)\s_{1A}(x) +||A_x||_1 (1+\s_{1A}(x))].
\end{array}
\end{equation}

\end{theorem}

In the next section we replace the assumption $x\geq x_0>0$ by $x\geq 0$.
The argument in this case is based on the Fredholm alternative.

\subsection{Characterization of the scattering data 
revisited}\label{U:5.7.2}

First, let us give necessary and sufficient conditions on 
${\mathcal S}$ for $q$ to 
be in $L_{1,1}$. These conditions are known (\cite{M}, \cite{R},
\cite{R9}, Section 5.5) but we give a short new argument.
We assume throughout that conditions A), B), and C) hold. 
These conditions are known to be necessary for $q\in 
L_{1,1}$. Indeed, conditions A) and B) are obvious, and C) is proved in 
\refT{5.7.1} and \refT{5.7.4}. 
Conditions A), B), and C) are also sufficient for $q\in  L_{1,1}$.
Indeed if they hold, then we prove that equation 
\eqref{e1.4.13} 
has a unique solution in $L^1_x$ 
for all $x\geq 0$. This was proved in 
Theorem 5.3.1,
but we give another proof.

\begin{theorem}\label{T:5.7.2}  
If A), B), and C) hold, then
\eqref{e1.4.13} 
has a solution in
$L^1_x$ for any $x\geq 0$ and this solution is unique.
\end{theorem}

\begin{proof}
Since $F_x$ is compact in $L^1_x,\, \forall x\geq 0$, by
the Fredholm alternative it is sufficient to prove that 
\begin{equation}\label{e5.7.22}
(I+F_x)h=0,\quad h\in L^1_x, 
\end{equation}
implies $h=0$. Let us prove it for $x=0$. The proof is similar for 
$x>0$. If $h\in L^1$, then $h\in L^\infty$ because $||h||_\infty \leq 
||h||_{L^1}\s_F(0)$. If $h\in L^1\cap L^\infty$, then $h\in 
L^2$ because$ ||h||_{L^2}^2\leq ||h||_{L^\infty}||h||_{L^1}$. Thus,
if $h\in L^1$ and solves \eqref{e5.7.22}, then $h\in 
L^2\cap L^1\cap L^\infty$.

Denote $\tildeh=\int^\infty_0 h(x)e^{ikx}dx,\, h\in L^2$. Then, 
\begin{equation}\label{e5.7.23}
\int^\infty_{-\infty}\tildeh^2 dk=0.
\end{equation}
Since $F(x)$ is real-valued, one can assume $h$  real-valued. 
One has, using Parseval's equation:
$$
0=((I+F_0)h,h)=\frac 1{2\pi} ||\tildeh||^2+\frac 
1{2\pi}\int^\infty_{-\infty}[1-S(k)]\tildeh^2(k)dk+\sum^J_{j=1}s_jh^2_j, \quad h_j:=\int^\infty_0e^{-k_jx}h(x)dx.
$$
Thus, using \eqref{e5.7.23}, one gets 
$$h_j=0,\, 1\leq j\leq J,\quad (\tildeh,\tildeh)= 
(S(k)\tildeh,\, \tildeh(-k)), $$ 
where we have used  real-valuedness of $h$, i.e. 
$ \tildeh(-k)=\tildeh(k),\forall k\in R$.

Thus, $(\tildeh, \tildeh)=(\tildeh,S(-k)\tildeh(-k))$, where A) was 
used. Since $||S(-k)||=1$, one has $||\tildeh||^2=\vert (\tildeh,$ $S(-k)\tildeh(-k))\vert\leq ||\tildeh||^2$, so the equality sign is 
attained in the Cauchy inequality. Therefore, $\tildeh(k)=S(-k)
\tildeh(-k)$.

By condition B), the theory of Riemann problem guarantees 
existence and uniqueness of an analytic in $\C_+:=\{k: \Im k>0\}$ function 
$f(k):=f_+(k),\, f(ik_j)=0,\, \dot f(ik_j)\neq 0,\, 1\leq j\leq J, 
\, f(\infty)=1$, such that 
\begin{equation}\label{e5.7.24}
f_+(k)=S(-k)f_-(k), \quad k\in\R,
\end{equation}
and $f_-(k)=f(-k)$ is analytic in $\C_- :=\{k:Imk<0\},\, f_-(\infty)=1$ 
in $\C_-,\, f_-(-ik_j)=0,\, \dot f_-(-ik_j)\neq 0$. Here the property 
$S(-k)=S^{-1}(k),\, \forall k\in \R$ is used.

One has
$$
\psi (k):=\frac{\tildeh(k)}{f(k)}=\frac{\tildeh(-k)}{f(-k},\quad k\in 
\R,\quad h_j:=\tildeh(ik_j)=0,\quad 1\leq j\leq J.
$$
The function $\psi (k)$ is analytic in $\C _+$ and $\psi (-k)$ is analytic 
in $\C _-$, they agree on $\R$, so $\psi(k)$ is analytic in $\C$. Since 
$f(\infty )=1$ and $\tildeh(\infty )=0$, it follows that $\psi\equiv 0$.

Thus, $\tildeh=0$ and, consequently, $h(x)=0$, as claimed.
\refT{5.7.2} is proved. 
\end{proof}

The unique solution to equation 
\eqref{e1.4.14} satisfies the estimates given in 
\refT{5.7.1}.  In the proof of \refT{5.7.1}  
the estimate $x|\dot A(x,x)|\in L^1(x_0,\infty )$ 
was established. So, by 
(1.5.13), 
$xq\in L^1(x_0,\infty)$.

The method developed in Section 5.7.1 gives accurate information 
about the behavior of $q$ near infinity. An immediate consequence of 
\refT{5.7.1}  and \refT{5.7.2}   is:

\begin{theorem}\label{T:5.7.3}  
If A), B), and C) hold, then $q,$ obtained by the 
scheme \eqref{e1.4.10}  belongs to $L_{1,1}(x_0,\infty)$.
\end{theorem}

Investigation of the behavior of $q(x)$ on $(0,x_0)$ requires additional 
argument. Instead of using the contraction mapping principle and 
inequalities, one has to use the Fredholm theorem, which 
says that $||(I+F_x)^{-1}||\leq c$ for any $x\geq 0,$ where the operator 
norm is taken for $F_x$ acting in $L^p_x$, $ p=1$ and $p=\infty$, and the 
constant $c$ does not depend on $x\geq 0$.

Such an analysis yields:

\begin{theorem}\label{T:5.7.4} 
If and only if A), B), and C) hold, then $q\in L_{1,1}$.
\end{theorem}

\begin{proof} It is sufficient to check that
\refT{5.7.1} holds with $x\geq 0$ replacing $x\geq x_0$.
To get \eqref{e5.7.4} with $x_0=0$, one uses
\eqref{e1.4.14} and the  estimate:
\begin{equation}\label{e5.7.25}
||A(x,y)||\leq ||(I+F_x)^{-1}||||F(x+y)||\leq c\s_F(2x), \quad ||\cdot||=
\sup_{y\geq x}|\cdot|,\, x\geq 0,
\end{equation}
where the constant $c>0$ does not depend on $x$. Similarly:
\begin{equation}\label{e5.7.26}
||A(x,y)||_1\leq c\sup_{s\geq x} \int_x^\infty |F(s+y)|dy\leq 
c\s_{1F} (2x), \quad x\geq 0.
\end{equation}
From \eqref{e5.7.6} one gets:
\begin{equation}\label{e5.7.27}
\begin{array} {ll}
||A_x(x,y)||_1\leq c[||F'(x+y)||_1 +A(x,x)||F(x+y)||_1]&\\\leq 
c\s_{2F}(2x)
+c\s_{F}(2x)\s_{1F}(2x),
\quad x\geq 0. 
\end{array}
\end{equation}
From \eqref{e5.7.7} one gets:
\begin{equation}\label{e5.7.28}
||A_y(x,y)||_1\leq c[\s_{2F}(2x) 
+\s_{1F}(2x)\s_{2F}(2x)]\leq \s_{2F}(2x).
\end{equation}
Similarly, from \eqref{e5.7.11} and \eqref{e5.7.24} -- \eqref{e5.7.27}
one gets \eqref{e5.7.12}. Then one checks \eqref{e5.7.13} as in
the proof of \refT{5.7.1}. Consequently \refT{5.7.1} holds with $x_0=0$.
\refT{5.7.4} is proved.
\end{proof}

\subsection{Compactly supported potentials}\label{U:5.7.3}

In this Section necessary and sufficient conditions are
given for $q\in L^a_{1,1}:=\{ q: q={\overline q}, q=0  \hbox { if }
x>a, \int_0^a x|q|dx<\infty\}$.
Recall that the Jost solution is:
\begin{equation}\label{e5.7.29}
f(x,k)=e^{ikx}+\int_x^\infty A(x,y)e^{iky}dy,\quad f(0,k):=f(k).
\end{equation}

\begin{lemma}\label{L:5.7.5} 
If $q\in L^a_{1,1}$, then $f(x,k)=e^{ikx}$ for $x>a$,
$A(x,y)=0$ for $y\geq x\geq a$, $F(x+y)=0$ for $y\geq x\geq a$
(cf \eqref{e1.4.13}), and $F(x)=0$ for $x\geq 2a$.
\end{lemma}

Thus, \eqref{e1.4.13}
with $x=0$ yields $A(0,y):=A(y)=0$ for $x\geq 2a$.
The Jost function 
\begin{equation}\label{e5.7.30}
f(k)=1+\int_0^{2a} A(y)e^{iky}dy,\quad A(y)\in W^{1,1}(0,a),
\end{equation}
is an entire function of exponential type $\leq 2a$, that is,
$|f(k)|\leq ce^{2a|k|}$, $k\in \C$, and $S(k)=f(-k)/f(k)$ is a 
meromorphic function in $\C$. In \eqref{e5.7.30} $W^{l,p}$ is the 
Sobolev space, and the inclusion \eqref{e5.7.30} follows from \refT{5.7.1}. 

Let us formulate the assumption D):

{\it D) the Jost function $f(k)$ is an entire function of exponential type 
$\leq 2a$.}

\begin{theorem}\label{T:5.7.6} 
Assume A), B), C) and D). Then $q\in L^a_{1,1}$.
Conversely, if  $q\in L^a_{1,1}$, then A),B), C) and D) hold.
\end{theorem}

\begin{proof}[Necessity]
If $q\in L_{1,1}$, then A), B) and C) hold by \refT{5.7.4},  and D) is 
proved in \refL{5.7.5}. 
The necessity is proved.

\nd {\it Sufficiency.} If  A), B) and C) hold, then $q\in L_{1,1}$. One has 
to prove that $q=0$ for $x>a$. If D) holds, then from the proof of
\refL{5.7.5}  it follows that $A(y)=0$  for $y\geq 2a$. 

\nd {\it We claim that $F(x)=0$ for $x\geq 2a$.}

If this is proved, then \eqref{e1.4.13} yields $A(x,y)=0$ for $y\geq x\geq 
a$,
and so $q=0$ for $x>a$ by
(1.5.13).

Let us prove the claim. 

Take $x>2a$ in
(1.5.12).
The function $1-S(k)$ is analytic in $\C_+$ except for $J$ 
simple poles at the points
$ik_j$. If $x>2a$ then one can use the Jordan lemma and the residue 
theorem and get:
\begin{equation}\label{e5.7.31}
F_s(x)=\frac 1 {2\pi} \int_{-\infty}^\infty 
[1-S(k)]e^{ikx}dk=-i\sum^J_{j=1}\frac{f(-ik_j)}{\dot 
f(ik_j)}e^{-k_jx},\quad  x>2a.
\end{equation}
Since $f(k)$ is entire, the Wronskian formula 
$$f^\prime (0,k)f(-k)-f^\prime (0,-k)f(k)=2ik$$ 
is valid on $\C$, and at $k=ik_j$ it yields: 
$$f^\prime(0,ik_j)f(-ik_j)=-2k_j,$$
because $f(ik_j)=0$.
This and
\eqref{e5.7.31}
yield
$$
F_s(x)=\sum^J_{j=1}\frac{2ik_j}{f^\prime(0,ik_j)\dot 
f(ik_j)}e^{-k_jx}=-\sum ^J_{j=1}s_je^{-k_jx}=-F_d(x),\quad x>2a.
$$
Thus, $F(x)=F_s(x)+F_d(x)=0$ for $x>2a$. The sufficiency is proved.

\refT{5.7.6} is proved.
\end{proof}

In \cite{M} a condition on ${\mathcal S}$, which guarantees that $q=0$ for 
$x>a$, is given under the assumption that there is no discrete spectrum, 
that is $F=F_s$.

\subsection{Square integrable potentials}\label{U:5.7.4}

Let us introduce conditions
\eqref{e5.7.32} -- \eqref{e5.7.34} 
\begin{equation}\label{e5.7.32}
2ik[f(k)-1+\frac Q{2ik}]\in L^2(\R):=L^2,\quad Q:=\int^\infty _0qds,
\end{equation}
\begin{equation}\label{e5.7.33}
k[1-S(k)+\frac Q{ik}]\in L^2,
\end{equation}
\begin{equation}\label{e5.7.34}
k[|f(k)|^2-1]\in L^2.
\end{equation}

\begin{theorem}\label{T:5.7.7} 
If A), B), C), and any one of the conditions
\eqref{e5.7.32} -- \eqref{e5.7.34} 
hold, then $q\in L^2(\R)$.
\end{theorem}

\begin{proof}
We refer to \cite{R} for the proof.
\end{proof}


\chapter{Inverse scattering problem with fixed-energy phase shifts as the data}\label{C:6}
\section{Introduction} \label{S:6.0}

In \refU{5.0.3} the scattering problem for spherically symmetric $q$ was
formulated, see \eqref{e5.0.15} -- \eqref{e5.0.17}.
The $\delta_\l$ are the fixed-energy $(k=\const>0)$ phase shifts.
Define
\begin{equation}\label{e6.0.1}
 L_r\varphi:=
 \left[ r^2\frac{\partial^2}{\partial r^2} +r^2-r^2 q(r) \right]
 \varphi:= L_{0r} \varphi-r^2 q(r)\varphi,
 \end{equation}
where $\varphi=\varphi_\l(r)$ is a regular solution to
\begin{equation}\label{e6.0.2}
 L_r \varphi_\l =\l(\l+1) \varphi_\l,
 \end{equation}
such that
\begin{equation}\label{e6.0.3}
 \varphi_\l=u_\l + \int^r_0 K(r,\rho)u_\l(\rho)\rho^{-2}d\rho,
 \quad K(r,0)=0,
 \end{equation}
and $u_l=\sqrt{\frac{\pi r}{2}} J_{\l+\frac{1}{2}}(r)$, $J_\l(r)$ is the
Bessel function. In \eqref{e6.0.3} $K(r,\rho)$ is the transformation kernel,
$I+K$ is the transformation operator.
In \eqref{e6.0.2} we assume that $k=1$ without loss of generality.
The $\varphi_\l$ is uniquely defined by its behavior near the origin:
\begin{equation}\label{e6.0.4}
 \varphi_\l(r)=\frac{r^{\l+1}}{(2\l+1)!!} + o(r^{\l+1}), \quad r\to 0.
 \end{equation}
For $u_\l$ we will use the known formula ([GR, 8.411.8]):
\begin{equation}\label{e6.0.5}
  \gamma_\l u_\l:=2^\l \Gamma(\l+1) u_\l(r)=r^{\l+1}
  \int^1_{-1}(1-t^2)^\l e^{irt}dt,
  \end{equation}
where $\Gamma(z)$ is the gamma-function.

The inverse scattering problem with fixed-energy phase shifts
$\{\delta_\l\}_{\l=0,1,2\dots}$ as the data consists of
finding $q(r)$ from these data. We assume throughout
this chapter that $q(r)$ is a real-valued function,
$q(r)=0$ for $r>a$,
\begin{equation}\label{e6.0.6}
 \int^a_0 r^2|q(r)|^2 dr<\infty.
 \end{equation}
Conditions \eqref{e6.0.6} imply that $q\in L^2(B_a)$,
$B_a:=\{x:x\in\R^3,\ |x|\leq a\}$.

In the literature there are books \cite{CS} and \cite{N}
where the so called Newton-Sabatier (NS)
theory  is presented, and many papers were published on this theory,
which attempts to solve the above inverse scattering problem with
fixed-energy phase shifts as the data.
In Section 6.4 it is proved that the NS theory is fundamentally wrong
and is not an inversion method. The main results of this Chapter are
Theorems 6.2.2, 6.3.1, 6.5.1, and {\it the proof of the fact that the 
Newton-Sabatier theory is fundamentally wrong in the sense that its 
foundations are wrong.}

\section{Existence and uniqueness of the transformation
  operators independent of angular momentum} \label{S:6.1}

The existence and uniqueness of $K(r,\rho)$ in \eqref{e6.0.3}
we prove by deriving a Goursat problem for it,
and investigating this problem.
Substitute \eqref{e6.0.3} into \eqref{e6.0.3}, drop index $\l$
for notational simplicity and get


\begin{multline}\label{e6.1.1}
0=-r^2q(r)u+(r^2-r^2q(r))\iro\k u\rho^{-2}d\rho\\
-\iro\k \rho^{-2} L_{0\rho}u d\rho +
r^2\partial^2_r\iro\k u\rho^{-2}d\rho.
\end{multline}

We assume first that $\k$ is twice continuously differentiable with respect
to its variables in the region $0<r<\infty,\quad 0<\rho\leq r$. This 
assumption requires
extra smoothness of $q(r),\quad q(r)\in C^1(0,a)$. If $q(r)$ satisfies condition
(\eqref{e6.0.6}), then equation \eqref{e6.1.7} below has to be understood in the sense of
distributions. Eventually we will work with an integral equation \eqref{e6.1.34}
(see below) for which assumption (\eqref{e6.0.6}) suffices.

Note that
\begin{equation}\label{e6.1.2}
\iro\k\rho^{-2}L_{0\rho}ud\rho=
\iro L_{0\rho}\k u\rho^{-2}d\rho+K(r,r)u_r-K_\rho (r,r)u,
\end{equation}
provided that
\begin{equation}\label{e6.1.3}
K(r,0)=0. 
\end{equation}
We assume \eqref{e6.1.3} to be valid.
Denote
\begin{equation}\label{e6.1.4}
\dot K:=\frac {dK(r,r)}{dr}.
\end{equation}
Then
\begin{multline}
\label{e6.1.5}
r^2\partial^2_r\iro \k u\rho^{-2}d\rho=\dot Ku+K(r,r)u_r-\frac 2r K(r,r)u+\\
K_r(r,r)u+r^2\iro K_{rr}(r,\rho)u\rho^{-2}d\rho.
\end{multline}

Combining \eqref{e6.1.1} -- \eqref{e6.1.5} and writing again $\ul$ in place of $u$, one gets
\begin{multline}
\label{e6.1.6}
0=\iro [L_r\k -L_{0\rho}\k]\ul(\rho)\rho^{-2}d\rho+\ul(r)[-r^2q(r)+
\dot K-\\
\frac{2K_r(r,r)}r+K_r(r,r)+K_\rho(r,r)],\quad \forall r>0,
\quad \ell=0,1,2,....
\end{multline}
Let us prove that \eqref{e6.1.6}  implies:
\begin{equation}\label{e6.1.7}
L_r\k =L_{0\rho}\k,\quad 0<\rho\leq r,
\end{equation}
\begin{equation}\label{e6.1.8}
q(r)=\frac{2\dot K}{r^2}-
\frac{2K(r,r)}r=\frac 2r\frac d{dr}\frac{K(r,r)}r.
\end{equation}
This proof requires a lemma.

\begin{lemma}\label{L:6.1.1}
Assume that $\rho f(\rho)\in L^1(0,r)$
and $\rho A(\rho)\in L^1(0,r)$.  If
\begin{equation}\label{e6.1.9}
0=\iro f(\rho)\ul (\rho)d\rho+\ul (r)A(r) \quad \forall 
\ell=0,1,2,...,
\end{equation}
then
\begin{equation}\label{e6.1.10}
f(\rho)\equiv 0\text{ and } A(r)=0.
\end{equation}
\end{lemma}

\begin{proof}
Equations \eqref{e6.1.9} and \eqref{e6.0.5} imply:
\begin{gather}
0=\int^1_{-1}dt(1-t^2)^\ell
\left(\frac d{idt}\right)^\ell\iro d\rho\rho f(\rho)e^
{i\rho t}+\notag \\
rA(r)\int^1_{-1}(1-t^2)^\ell\left(\frac d{idt}\right)^\ell e^{irt}dt.
\notag
\end{gather}
Therefore
\begin{equation}\label{e6.1.11}
0=\int^1_{-1}dt\frac{d^\ell (t^2-1)^\ell}{dt^\ell}
[\iro d\rho\rho f(\rho)e^
{i\rho t}+rA(r)e^{irt}],\quad l=0,1,2,\dots
\end{equation}
Recall that the Legendre polynomials are defined by the formula
\begin{equation}\label{e6.1.12}
P_\ell (t)=
\frac 1{2^\ell \ell !}\frac{d^\ell}{dt^\ell}(t^2-1)^\ell,
\end{equation}
and they form a complete system in $L^2(-1,1)$.

Therefore \eqref{e6.1.11} implies
\begin{equation}\label{e6.1.13}
\iro d\rho\rho f(\rho)e^{i\rho t}+rA(r)e^{irt}=0\quad \forall t\in [-1,1].
\end{equation}
Equation \eqref{e6.1.13} implies
\begin{equation}\label{e6.1.14}
\iro d\rho\rho f(\rho)e^{i\rho t}=0,\quad \forall t\in [-1,1],
\end{equation}
and
\begin{equation}\label{e6.1.15}
rA(r)=0.
\end{equation}
Therefore $A(r)=0$. Also $f(\rho)=0$ because the left-hand side of \eqref{e6.1.14} is an
entire function of $t$, which vanishes on the interval $[-1,1]$ and,
consequently, it vanishes identically, so that $\rho f(\rho) =0$ and therefore
$f(\rho)\equiv 0$.

\refL{6.1.1} is proved.
\end{proof}

{\it We prove that the problem \eqref{e6.1.7}, \eqref{e6.1.8}, 
\eqref{e6.1.3}, which is a Goursat-type
problem, has a solution and this solution is unique in the class of functions
$\k$, which are twice continuously differentiable with respect to $\rho$ and
$r,\quad 0<r<\infty,\quad 0<\rho\leq r$.}

In this section we assume that $q(r)\in C^1(0,a)$. This assumption implies that
$\k$ is twice continuously differentiable.
If \eqref{e6.0.6} holds, then the
arguments in this section which deal with integral 
equation \eqref{e6.1.34} remain valid.
Specifically, existence and uniqueness of the solution to 
equation \eqref{e6.1.34} is proved under the only 
assumption $\int_0^a r|q(r)|dr<\infty$
as far as the smoothness of $q(r)$ is concerned.

By a limiting argument one can reduce the smoothness requirements on $q$ to the
condition \eqref{e6.0.6}, but in this case equation 
\eqref{e6.1.7} has to be understood in
distributional sense.

Let us rewrite the problem we want to study:
\begin{equation}\label{e6.1.16}
r^2K_{rr}-\rho^2 K_{\rho\rho}+
[r^2-r^2q(r)-\rho^2]\k=0,\quad 0<\rho\leq r,
\end{equation}
\begin{equation}\label{e6.1.17}
K(r,r)=\frac r2\iro sq(s)ds:=g(r),
\end{equation}
\begin{equation}\label{e6.1.18}
K(r,0)=0.
\end{equation}
The difficulty in the study of this Goursat-type problem comes from the fact
that the coefficients in front of the second derivatives of the kernel
$\k$
are variable.          

Let us reduce problem \eqref{e6.1.16} -- \eqref{e6.1.18} to the one with constant coefficients. To
do this, introduce the new variables:
\begin{equation}\label{e6.1.19}
\xi=\ln r+\ln\rho,\quad \eta=\ln r-\ln\rho.
\end{equation}
Note that
\begin{equation}\label{e6.1.20}
r=e^{\frac{\xi +\eta}2},\quad \rho =e^{\frac{\xi-\eta}2},
\end{equation}
\begin{equation}\label{e6.1.21}
\eta\geq0,\quad -\infty<\xi<\infty,
\end{equation}
and
\begin{equation}\label{e6.1.22}
\partial_r=\frac 1r(\partial_\xi +\partial_\eta),\quad \partial_\rho =
\frac 1\rho (\partial_\xi -\partial_\eta).
\end{equation}
Let 
\begin{equation}
\k :=B(\xi,\eta). \notag
\end{equation}
A routine calculation transforms equations \eqref{e6.1.16} -- \eqref{e6.1.18}
to the following ones:
\begin{equation}\label{e6.1.23}
B_{\xi \eta}(\xi,\eta)-\frac 12 B_\eta (\xi,\eta) +Q(\xi,\eta)B=0,
\quad \eta\geq0,\quad
-\infty<\xi<\infty,
\end{equation}
\begin{equation}\label{e6.1.24}
B(\xi,0)=g\left(e^{\frac \xi 2}\right):=G(\xi),
\quad -\infty<\xi<\infty
\end{equation}
\begin{equation}\label{e6.1.25}
B(-\infty,\eta)=0,\quad \eta\geq0,
\end{equation}
where $g(r)$ is defined in \eqref{e6.1.17}.

Here we have defined
\begin{equation}\label{e6.1.26}
Q(\xi,\eta):=\frac 14
\left[e^{\xi +\eta} -e^{\xi +\eta}q\left(e^{\frac{\xi+\eta}2}\right)
-e^{\xi-\eta}\right],
\end{equation}
and took into account that $\rho=r$ implies $\eta =0$,
while $\rho =0$ implies,
for any fixed $\eta\geq 0$, that $\xi=-\infty$.

Note that
\begin{equation}\label{e6.1.27}
\sup_{-\infty<\xi<\infty}
e^{-\frac \xi 2}G(\xi)<c,
\end{equation}
\begin{equation}\label{e6.1.28}
\sup_{0\leq\eta\leq B}
\int^A_{-\infty}|Q(s,\eta)|ds\leq c(A,B),
\end{equation}
for any $A\in\mathbb R$ and $B>0$, where $c(A,B)>0$ is a constant.

To get rid of the second term on the left-hand side of \eqref{e6.1.23},
let us introduce
the new kernel $L(\xi,\eta)$ by the formula:
\begin{equation}\label{e6.1.29}
L(\xi,\eta):=B(\xi,\eta)e^{-\frac \xi 2}.
\end{equation}
Then \eqref{e6.1.23}-- \eqref{e6.1.25} can be written as:
\begin{equation}\label{e6.1.30}
 L_{\eta \xi}(\xi,\eta )+Q(\xi,\eta)L(\xi,\eta)=0,\quad \eta\geq 0,\quad
-\infty<\xi<\infty,
\end{equation}
\begin{equation}\label{e6.1.31}
L(\xi,0)=e^{-\frac\xi 2}G(\xi):=b(\xi):=
\frac 12\int^{e^{\frac \xi 2}}_0 s q(s) ds,
\quad -\infty<\xi<\infty,
\end{equation}
\begin{equation}\label{e6.1.32}
L(-\infty,\eta)=0,\quad \eta\geq 0.
\end{equation}
We want to prove existence and uniqueness of the solution
to \eqref{e6.1.30} -- \eqref{e6.1.32}.
In order to choose a convenient Banach space in which to work, let us transform
problem \eqref{e6.1.30} -- \eqref{e6.1.32} to an equivalent Volterra-type integral equation.
 
Integrate \eqref{e6.1.30} with respect to $\eta$ from $0$ to $\eta$ and use \eqref{e6.1.31} to get
\begin{equation}\label{e6.1.33}
L_\xi (\xi,\eta)-b^\prime (\xi)+\int^\eta_0 Q(\xi,t)L(\xi, t)dt=0.
\end{equation}
Integrate \eqref{e6.1.33} with respect to $\xi$ from $-\infty$ to $\xi$ and use \eqref{e6.1.33} to
get
\begin{equation}\label{e6.1.34}
L(\xi,\eta)=-\int^\xi_{-\infty}ds\int^\eta_0dtQ(s,t)L(s,t)+b(\xi):=VL+b,
\end{equation}
where
\begin{equation}\label{e6.1.35}
VL:=-\int^\xi _{-\infty}ds\int^\eta_0dtQ(s,t)L(s,t).
\end{equation}
Consider the space $X$ of continuous functions $L(\xi,\eta)$, defined in the
half-plane $\eta\geq 0,\quad -\infty<\xi<\infty$, such that for any $B>0$ and
any $-\infty<A<\infty$ one has
\begin{equation}\label{e6.1.36}
\Vert L\Vert :=\Vert L\Vert_{AB}:=
\sup_{\substack{-\infty<s\leq A \\ 0\leq t\leq B}}
\left(e^{-\gamma t}|L(s,t)|\right)
< \infty,
\end{equation}
where $\gamma>0$ is a number which will be chosen later so that the
operator $V$ in \eqref{e6.1.34} will be a contraction mapping on the Banach space of
functions with norm \eqref{e6.1.36} for a fixed pair $A,B$. To choose $\gamma>0$, let us
estimate the norm of $V$. One has:
\label{e6.1.37}
\begin{multline}
\Vert VL\Vert\leq
\sup_{ -\infty<\xi\leq A, 0\leq\eta\leq B}
\left(\int^\xi_{-\infty}ds\int^\eta_0dt|Q(s,t)|e^{-\gamma(\eta -t)}e^{-\gamma t}
|L(s,t)|\right)\\
\leq\Vert L\Vert
\sup_{-\infty <\xi\leq A,0\leq\eta\leq B}
\int^{\xi}_{-\infty}ds\int^\eta_0dt\left(2e^{s+t}+e^{s+t}
|q\left( e^{\frac{s+t}2}\right)|\right)
e^{-\gamma(\eta -t)}\leq\frac c\gamma\Vert L \Vert,
\end{multline}
where $c>0$ is a constant depending on $A,B$ and $\int^a_0r|q(r)|dr$.
Indeed, one has:
\begin{equation}\label{e6.1.38}
2\int^A_{-\infty}ds\int^\eta_0dte^{s+t-\gamma(\eta-t)}=2e^A\int^\eta_0dte^{t-
\gamma (\eta -t)}dt\leq 2e^{A+B}\frac{1-e^{-\gamma B}}\gamma=\frac{c_1}\gamma,
\end{equation}
and, using the substitution $\sigma=e^{\frac{s+t}2}$, one gets:
\begin{multline}
\int^A_{-\infty}ds\int^\eta_0dte^{s+t}|q(e^{\frac{s+t}2})|e^
{-\gamma(\eta -t)}=\\
=\int^\eta_0dte^{-\gamma (\eta -t)}\int^A_{-\infty}dse^{s+t}
\vert q\left(e^{\frac{s+t}2}\right)\vert=\\
=2\int^\eta_0dte^{-\gamma(\eta -t)}\int^{e^\frac{A+t}2}_0d\sigma\sigma|
q(\sigma )|=\\
=\frac{2(1-e^{-\gamma B})}\gamma\int^a_0d\sigma\sigma |q(\sigma)|:=
\frac{c_2}\gamma.
\end{multline}
From these estimates inequality (6.2.38) follows.

It follows from (6.2.38) that $V$ is a contraction mapping in the space
$X_{AB}$ of continuous functions in the region $-\infty<\xi\leq A,\quad
0\leq\eta\leq B$, with the norm \eqref{e6.1.36} provided that
\begin{equation}\label{e6.1.40}
  \gamma>c.
  \end{equation}
Therefore equation \eqref{e6.1.34} has a unique solution $L(\xi,\eta)$ in the region
\begin{equation}\label{e6.1.41}
  -\infty<\xi<A,\qquad 0\leq\eta\leq B
\end{equation}
for any real $A$ and $B>0$
if \eqref{e6.1.40} holds. This means that the above solution is defined for any
$\xi\in \mathbb R$ and any $\eta\geq 0$.

Equation \eqref{e6.1.34} is equivalent to problem
\eqref{e6.1.30} -- \eqref{e6.1.32} and,
by \eqref{e6.1.29}, one has:
\begin{equation}\label{e6.1.42}
B(\xi,\eta)=L(\xi,\eta)e^{\frac \xi 2}.
\end{equation}
Therefore we have proved the
existence and uniqueness of $B(\xi,\eta)$, that is, of
the kernel $\k =B(\xi,\eta )$ of the transformation operator \eqref{e6.0.3}. Recall
that $r$ and $\rho$ are related to $\xi$ and $\eta$ by formulas \eqref{e6.1.20}.

Let us formulate the result:
\begin{theorem}\label{T:6.1.2} 
The kernel $K(r,\rho)$
of the transformation operator \eqref{e6.0.3} solves
problem \eqref{e6.1.16} -- \eqref{e6.1.18}. The solution to this 
problem does exist and is unique
in the class of twice continuously
differentiable functions for any potential $q(r)\in C^1(0,a)$. If $q(r)\in 
L^\infty (0,a)$, then $\k$ has first
derivatives which are bounded and equation \eqref{e6.1.16} 
has to be understood in the
sense of distributions. The following estimate holds for any $r>0$:
\begin{equation}\label{e6.1.43}
\iro|\k |\rho^{-1}d\rho<\infty.
\end{equation}
\end{theorem}


\begin{proof}[Proof of \refT{6.1.2}] 
We have already proved all the assertions of \refT{6.1.2}
except for the estimate \eqref{e6.1.43}. Let us prove this estimate.

Note that
\begin{equation}\label{e6.1.54}
\iro |\k |\rho^{-1}d\rho=
r\int^\infty_0|L(2\ln r-\eta,\eta)|e^{-\frac \eta 2}d\eta<\infty
\end{equation}
Indeed, if $r>0$ is fixed, then, by \eqref{e6.1.20}, $\xi +\eta =2\ln r=const$. 
Therefore
$d\xi =-d\eta$, and $\rho^{-1}d\rho=
\frac 12(d\xi-d\eta)=-d\eta,\quad \xi =2\ln r-\eta$. Thus:
\begin{equation}\label{e6.1.55}
\iro |\k|\rho^{-1}d\rho=
\int^\infty_0|L(2\ln r-\eta,\eta)|e^{\frac{2\ln r-\eta}2}d\eta =r
\int^\infty_0|L(2\ln r-\eta,\eta)|e^{-\frac \eta 2}d\eta. 
\end{equation}
The following estimate holds:
\begin{equation}\label{e6.1.56}
|L(\xi,\eta)|\leq c e^{(2+\epsilon_1) [\eta \mu_1 (\xi +
\eta)]^{\frac 12 +\epsilon_2}}, 
\end{equation}
where $\epsilon_j>0, \, j=1,2,$
are arbitrarily small numbers and $\mu_1$ is defined in
formula \eqref{e6.1.60} below, see also formula \eqref{e6.1.58} for the
definition of $\mu$.

Estimate \eqref{e6.1.56} is proved  below, in \refT{6.1.2}.

From \eqref{e6.1.55} and estimate \eqref{e6.1.64} (see below) estimate \eqref{e6.1.43} follows.
Indeed, denote by $I$ the integral on the right-hand side of
\eqref{e6.1.55}. Then, by \eqref{e6.1.64} one gets:
\begin{equation} 
I\leq 2+2\sum_{1}^{\infty}\frac {[2\mu_1(2\log r)]^n}{n!}=2 \exp
[2\mu_1(2\log r)]< \infty.
\end{equation} 
\refT{6.1.2} is proved.
\end{proof}

\begin{theorem}\label{T:6.1.3} 
Estimate \eqref{e6.1.56} holds.
\end{theorem}
\begin{proof}[Proof of \refT{6.1.3}] 
From \eqref{e6.1.34} one gets:
\begin{equation}\label{e6.1.57}
m(\xi,\eta)\leq c_0+(Wm)(\xi,\eta),\qquad
m(\xi,\eta):=|L(\xi,\eta)|,  
\end{equation}
where $c_0=
\sup_{-\infty<\xi<\infty} 
|b(\xi)|\leq\frac
12\int^a_0s
|q(s)|ds$ (see \eqref{e6.1.31}), and
\begin{equation}\label{e6.1.58}
Wm:=\int^\xi_{-\infty}ds\int^\eta_0dt\mu (s+t)m(s,t),
\quad \mu(s):=\frac 12 e^s\left(1+|q(e^{\frac s2})|\right). 
\end{equation}
It is sufficient to consider inequality \eqref{e6.1.57} with $c_0=1$: 
if $c_0=1$ and the
solution $m_0(\xi,\eta)$  to \eqref{e6.1.57} satisfies \eqref{e6.1.56}
with $c=c_1$, then the solution
$m(\xi,\eta)$ of \eqref{e6.1.57} with any
$c_0>0$ satisfies \eqref{e6.1.56} with $c=c_0c_1$.

Therefore, assume that $c_0=1$, then \eqref{e6.1.57} reduces to:
\begin{equation}\label{e6.1.59}
m(\xi,\eta)\leq1+(Wm)(\xi,\eta).
\end{equation}
Inequality \eqref{e6.1.56} follows from \eqref{e6.1.59} by iterations. Let us give the details.

Note that
\begin{equation}
W1=\int^\xi_{-\infty}ds\int^\eta_0dt\mu (s+t)=\int^\eta_0dt\int^\xi_{-\infty}
ds\mu (s+t)=\int^\eta_0dt\mu_1(\xi +t)\leq\eta\mu_1(\xi+\eta).
\end{equation}
Here we have used the notation
\begin{equation}\label{e6.1.60}
\mu_1(\xi)=\int^\xi_{-\infty}\mu (s) ds,
\end{equation}
and the fact that $\mu_1(s)$ is a monotonically increasing function, since
$\mu
(s)>0$. Note also that $\mu_1(s)<\infty$ for any $s, \,-\infty<s<\infty$.

Furthermore,
\begin{equation}\label{e6.1.61}
 W^21\leq\int^\xi_{-\infty}ds\int^\eta_0dt\mu (s+t)t\mu_1(s+t)
 \leq\int^\eta_0dtt
 \int^\xi_{-\infty}ds\mu (s+t)\mu_1(s+t)
 =\frac{\eta^2}{2!}\frac{\mu^2_1(\xi+\eta)}{2!}.
 \end{equation}
Let us prove by induction that
\begin{equation}\label{e6.1.62}
W^n1\leq\frac{\eta^n}{n!}\frac{\mu^n_1(\xi +\eta)}{n!}.
\end{equation}
For $n=1$ and $n=2$ we have checked \eqref{e6.1.62}. Suppose \eqref{e6.1.62} holds for some $n$,
then
\begin{equation}\label{e6.1.63}
W^{n+1}1\leq W\left(\frac{\eta^n}{n!}\frac{\mu^n_1(\xi+\eta)}{n!}\right)
=\int^\eta_0dt\frac{t^n}{n!}\int^\xi_{-\infty}ds
\mu (s+t)\frac{\mu^n_1(s+t)}{n!}
\leq\frac{\eta^{n+1}}{(n+1)!}\frac{\mu^{n+1}_1(\xi+\eta)}{(n+1)!}.
\end{equation}
By induction, estimate \eqref{e6.1.61} is proved for all $n=1,2,3,...$. Therefore \eqref{e6.1.59}
implies
\begin{equation}\label{e6.1.64}
m(\xi,\eta)\leq 1+\sum^\infty_{n=1}\frac{\eta^n}{n!}
\frac{\mu^n_1(\xi +\eta)}
{n!}\leq c e^{(2+\epsilon_1)
[\eta \mu_1 (\eta+\xi)]^{\frac 12+\epsilon_2}},
\end{equation}
where we have used Theorem 2 from [Lev, section 1.2], namely
the order of the entire function $F(z):=1+\sum_{n=1}^\infty \frac
{z^n}{(n!)^2}$ is $\frac 12$ and its type is 2. The constant
$c>0$ in \eqref{e6.1.56} depends on $\epsilon_j, \, j=1,2.$

Recall that the order of an entire function $F(z)$ is the number
$\rho:=\limsup_{r\to \infty}\frac {ln\,ln\,M_F(r)}{ln\,r}$,
where $M_F(r):=max_{|z|=r}|F(z)|$. The type of $F(z)$ is the number
$\sigma:=\limsup_{r\to \infty}\frac {ln\, M_F(r)}{r^\rho}$.
It is known [Lev], that if $F(z)=\sum_{n=0}^\infty c_nz^n$ is
an entire function, then its order $\rho$ and type $\sigma$
can be calculated by the formulas:
\begin{equation}
\rho=\limsup _{n\to \infty} \frac {n\, ln\,n}{ln\,\frac 1{|c_n|}},
\quad \sigma=
\frac { \limsup_{n\to \infty}( n|c_n|^{\frac {\rho} {n}})}{e\rho}.
\end{equation}
If $c_n=\frac 1 {(n!)^2}$, then the above formulas yield $\rho=\frac 12$
and $\sigma=2$.
\refT{6.1.3} is proved. 
\end{proof}


\section{Uniqueness theorem.} \label{S:6.2}

Denote by $\calL$  any fixed subset of the set $\N$ of integers
$\{0,1,2,\dots\}$ with the property:
\begin{equation}\label{e6.2.1}
 \sum_{\substack{\l\in\calL \\ \l\not= 0}}
 \frac{1}{\l}=\infty
 \end{equation}

\begin{theorem}\label{T:6.2.1} ({\bf [R10]})
Assume that $q$ satisfies \eqref{e6.0.6} and \eqref{e6.2.1}
holds. Then the data
$\{\delta_\l\}_{\forall \l\in\calL}$ determine $q$
uniquely.
\end{theorem}

The idea of the proof is based on property C-type argument.

\underbar{Step 1}:
If $q_1$ and $q_2$ generate the same data
$\{\delta_\l\}_{\forall \l\in\calL}$, then the following orthogonality
relation holds for $p:=q_1-q_2$:
\begin{equation}\label{e6.2.2}
 h(\l):=\int^a_0 p(r) \phi_{1\l}(r) \phi_{2\l}(r) dr=0
 \quad \forall \l\in\calL,
 \end{equation}
where $\phi_{j\l}$ is the scattering solution corresponding to
$q_j$, $j=1,2$.

\underbar{Step 2}:
Define
$ h_1(\l):=2^{2\l}[\Gamma(\l+1)]^2 h(\l)$, where $\Gamma$ is the 
Gamma-function. Check that $h_1(\l)$ is holomorphic in 
$\Pi_+:=\{\ell: Re\ell>0\},\quad
\ell=\sigma+i\tau,\quad \sigma \geq 0,$
and $\tau$ are real numbers, $h_1(\ell)\in N$
(where $N$ is the Nevanlinna class in $\Pi_+$), that is
$$
\sup_{0<r<1}\int^\pi_{-\pi}\log^+| h_1(\frac{1-re^{i \varphi}}
{1+re^{i\varphi}}) |d\varphi<\infty,
$$
where $\log^+x= \left\{
\aligned \log x&\text { if } \log x>0,\\
              0&\text{ if }\log x\leq 0.\endaligned
\right.$
If $h_1\in N$ vanishes $\forall \l\in \calL$, then $h_1=0$ in $\Pi_+$,
and, by property $C_\varphi$, $p(r)=0$. Theorem 6.3.1 is proved. $\Box$

\section{Why is the Newton-Sabatier (NS) procedure fundamentally  
wrong?}\label{S:6.3}



The NS procedure is described in \cite{N} and \cite{CS}.
A vast bibliography of this topic  
is given in \cite{CS} and \cite{N}.
  
Below two cases are discussed.  
The first case deals with the inverse scattering problem with   
fixed-energy phase shifts as the data. This problem is understood  
as follows: an unknown spherically symmetric potential $q$ from an a  
priori fixed class, say $L_{1,1}$, a standard scattering class,  
generates fixed-energy phase shifts  $\delta_l, l=0,1,2, \dots,$.  
The inverse scattering problem consists of recovery of  $q$  
from these data.  
  
The second case deals with a different problem: given some  
numbers  $\delta_l, l=0,1,2, \dots,$, which are assumed to  
be fixed-energy phase shifts of some potential  $q$, from a class not  
specified, find some potential  $q_1$, which generates   
fixed-energy phase shifts equal to  $\delta_l, l=0,1,2, \dots,$.  
This potential  $q_1$ may have no physical interest because  
of its non-physical" behavior at infinity or other  undesirable  
properties.                     
   
We first discuss NS procedure assuming that it is intended to solve  
the inverse scattering problem in case 1.   
Then we discuss NS procedure assuming that it is intended  
to solve the problem in case 2.

\smallskip

\noindent{\bf Discussion of case 1:}

\smallskip
In \cite{N2} and \cite{N} a procedure was proposed by R. Newton for
inverting fixed-energy phase shifts $\delta_l, l=0,1,2, \dots,$  
corresponding to an unknown spherically symmetric potential $q(r)$.  
R. Newton did not specify the class of potentials for which  
he tried to develop an inversion theory and did not formulate   
and proved any results which would justify the inversion procedure  
he proposed (NS procedure).   
His arguments are based on the following claim, which  
is implicit in his works, but crucial for the validity of  
NS procedure:   
  
\begin{claimN1}\label{CL:N1}
\begin{equation}\label{e6.3.1}
 K(r,s) = f(r,s) - \int^r_0 K(r,t) f(t,s) \frac{dt}{t^2}, \quad   
 0\leq s  \leq r<\infty, 
 \end{equation}
is uniquely solvable for all $r>0$.
\end{claimN1}

Here
\begin{equation}\label{e6.3.2}
 f(r,s) := \sum^\infty_{l=0} c_l u_l (r) u_l (s), \quad  
  u_l := \sqrt{\frac{\pi r}{2}} J_{l + \frac{1}{2}}(r),  
  \end{equation}  
$ c_l$ are real numbers,  
the energy $k^2$ is fixed: $k=1$ is taken without loss of generality,  
$J_{l + \frac{1}{2}}(r)$ are the Bessel functions. If  equation  
 \eqref{e6.3.1} is uniquely solvable for  all $r>0$, then the potential $q_1,$  
that NS procedure yields, is defined by the formula:  
\begin{equation}\label{e6.3.3} 
 q_1(r) = -\frac{2}{r} \frac{d}{dr} \frac{K(r,r)}{r}. 
 \end{equation} 
The R. Newton's ansatz \eqref{e6.3.1} - \eqref{e6.3.2}
for the transformation  kernel $K(r,s)$ of the   
Schroedinger operator, corresponding to  
some $q(r)$, namely, that $K(r,s)$ is the unique solution to  
\eqref{e6.3.1} - \eqref{e6.3.2},
{\it is  not correct for a generic potential}, as follows from   
our argument below (see the justification of Conclusions).   
  
{\it If for some $r>0$ equation \eqref{e6.3.1} 
is not uniquely solvable, then NS  
procedure breaks down: it leads to locally non-integrable potentials  
for which the scattering theory is, in general, not available  
(see \cite{R9} for a proof of the above statement) .}  
  
In the original paper \cite{N2} and in his book  
\cite{N} R. Newton did not   
study the question,  fundamental for any   
inversion theory: does the reconstructed potential $q_1$ generate the   
data from which it was reconstructed?  
  
In \cite[p.~205]{CS}, there are two claims:  
  
Claim i) that $q_1(r)$ generates the original shifts $\{\delta_l\}$  
"provided that $\{\delta_l\}$ are not "exceptional"",  
and   

Claim ii) that NS procedure "yields one (only one) potential which
decays faster than $r^{-\frac{3}{2}}$" and generates the original  
phase shifts  
$\{\delta_l\}$.     
  
If one considers NS procedure as a solution to inverse scattering  
problem of finding an unknown potential $q$ from a certain class, for  
example $q(r)\in L_{1,1} := \{ q : q= \overline q, \int^\infty_0 r  
|q(r)|dr < \infty \}$, from the  fixed-energy phase shifts, generated  
by this  $q$,  
then   
the proof, given in \cite{CS}, of Claim i) is not convincing:  
it is not clear why the potential $q_1$, obtained by NS procedure,  
has the transformation operator generated by the potential  
corresponding to   
 the original data, that  
is, to the given fixed-energy phase shifts. In fact, as follows from  
\refP{6.3.1} 
below, the potential  $q_1$ cannot generate the  
kernel $K(r,s)$ of the transformation operator corresponding to  
a generic original potential    
$q(r)\in L_{1,1} := \{ q : q= \overline q, \int^\infty_0 r |q(r)|     
dr < \infty \}$.   
  
Claim ii) is incorrect because the original generic  
potential $q(r)\in L_{1,1}$ generates the phase shifts  
$\{\delta_l\},$ and if $q_1(r),$  
the potential obtained by NS procedure and therefore not  
equal to  $q(r)$ by 
\refP{6.3.1} 
generates the same phase shifts  
$\{\delta_l\}$, then one has two different potentials $q(r)$ and  
$q_1(r)$, which  
both decay faster than $r^{-\frac{3}{2}}$ and both generate the  
original phase shifts $\{\delta_l\},$ contrary to Claim ii).  
  
Our aim is to formulate and justify the following  
  
{\bf Conclusions:} {\it  Claim N1 and ansatz \eqref{e6.3.1} - 
\eqref{e6.3.2} 
are not proved by R.~Newton and,  in general,  are wrong.  
Moreover, one cannot approximate with a prescribed accuracy  
in the norm $||q||:$ $=\int_0^\infty r|q(r)|dr$ a  
generic potential   
$q(r) \in L_{1,1}$  
by the potentials which might possibly  
be obtained by the NS procedure. Therefore NS procedure  
cannot be justified even as an approximate inversion procedure.
The NS procedure is fundamentally wrong in the sense that its foundations 
are wrong.}  
  
\medskip  
  
\noindent{\bf Let us justify these conclusions:}  
  
\medskip  
  
\noindent  Claim N1 formulated above and basic for NS procedure,  
is wrong, in general, for the following reason:  
  
Given fixed-energy phase shifts, corresponding to a generic potential  
$q \in L_{1,1}$,  
one either cannot carry through NS procedure because:  
  
a) the system (12.2.5a) in \cite{CS}, which should determine numbers  
$c_l$ in formula \eqref{e6.3.2}, given the phase shifts $\delta_l,$  
may be not solvable, or  
  
b) if the above system is solvable, equation \eqref{e6.3.1}  
 may be not (uniquely) solvable for some $r>0$,   
and in this case NS procedure breaks down since it  
yields a potential which is not locally integrable (see \cite{R9}  
for a proof).  
  
If equation \eqref{e6.3.1} is solvable for all $r>0$ and yields  
a potential $q_1$ by formula \eqref{e6.3.3}, then this potential is not equal  
to the original generic potential $q \in L_{1,1}$, as follows from   
\refP{6.3.1} 
which is proved in \cite{R9} (see also \cite{ARS}):  
  
\begin{proposition}\label{P:6.3.1} 
If equation \eqref{e6.3.1} is solvable for  
all $r>0$  
and yields a potential $q_1$ by formula \eqref{e6.3.3}, then this  $q_1$  
is a restriction to $(0, \infty) $ of a function analytic  
in a neighborhood of $(0, \infty) $.
\end{proposition}
  
Since a generic potential $q\in L_{1,1}$ is not  
 a restriction to $(0, \infty) $ of an analytic function,  
one concludes that even if equation \eqref{e6.3.1} is solvable for  
all $r>0$, the potential $q_1$, defined by formula \eqref{e6.3.3},  
is not equal to the original generic potential  $q\in L_{1,1}$  
and therefore the inverse scattering problem  
of finding an unknown  $q\in L_{1,1}$ from its fixed-energy phase  
shifts is not solved by NS procedure.  
  
The ansatz \eqref{e6.3.1} - \eqref{e6.3.2} for the transformation kernel is, in general,  
incorrect, as follows also from \refP{6.3.1} 
  
Indeed, if the ansatz \eqref{e6.3.1} - \eqref{e6.3.2} would be true and formula \eqref{e6.3.3} 
would yield the original generic $q$,  
that is $q_1=q$, this would contradict 
\refP{6.3.1} 
If formula \eqref{e6.3.3} would yield a $q_1$ which is different from the  
original generic $q$, then NS procedure does not solve the inverse  
scattering problem formulated above. Note also that  
it is proved in \cite{R10} that independent of  
the angular momenta $l$ transformation operator, corresponding to a  
generic $q\in L_{1,1}$ does exist, is unique, and is defined by  
a kernel $K(r,s)$ which   
cannot have representation \eqref{e6.3.2}, since it yields by the formula    
similar to \eqref{e6.3.3} the original generic potential $q$, which is not  
a restriction of an analytic in a neighborhood of $(0,\infty)$  
function to $(0,\infty)$.  
  
The conclusion, concerning impossibility of approximation  
of a generic  $q \in L_{1,1}$ by potentials  $q_1$, which can possibly  
be obtained by NS procedure, 
is proved in \refCL{Claim1} section 2, see proof of \refCL{Claim1}
there.  
  
Thus, our conclusions are justified. \qed  
  
Let us give some additional comments concerning NS procedure.   
  
  Uniqueness of the solution to the inverse problem in case 1  
was first proved by A.~G.~Ramm in 1987 (see \cite{R7})  
for a class of compactly supported potentials, while   
R. Newton's procedure was published in \cite{N2}, when no uniqueness  
results   
for this inverse problem were known.  
It is still an open problem if for the standard in scattering  
theory class of $ L_{1,1}$ potentials the uniqueness theorem   
for the solution of the above inverse scattering problem holds.   
  
We discuss the   
inverse scattering problem with fixed-energy phase shifts (as   
the data) for potentials $q \in L_{1,1} $, because   
 only for this class of potentials a general theorem  
of existence and  
uniqueness of the transformation operators, independent of   
the angular momenta $l$, has been proved,  see \cite{R10}.   
 In \cite{N2}, \cite{N}, and in \cite{CS} this result was not  
formulated and proved, and it was not clear for what class of  
potentials   
the transformation operators, independent of $l$, do exist.  
For slowly decaying potentials the existence of the transformation  
operators, independent of $l$, is not established, in general, and  
the potentials, discussed in \cite{CS} and \cite{N} in connection with NS  
procedure, are slowly decaying.  
  
Starting with \cite{N2}, \cite{N}, and \cite{CS} Claim N1  
was not proved or the proofs given (see \cite{CT})  
were incorrect (see \cite{R11}). This equation is uniquely  
solvable for  
sufficiently small $r>0$, but, in general, {\it it may be not  
solvable  for some $r>0$, and if it is solvable for  
all $r>0$, then it yields by formula \eqref{e6.3.3}  a potential  
 $q_1$, which is not equal  
to the original generic potential $q\in L_{1,1}$, as follows from  
\refP{6.3.1} }
  
Existence of "transparent" potentials is often cited in the literature.  
A  "transparent" potential is a potential which is not equal to zero  
identically, but generates the fixed-energy shifts which are all equal to  
zero.  
  
{\it In \cite[p.~207]{CS}, there is a remark concerning the existence  
of "transparent" potentials. This remark is not justified because it is  
not proved  
that for the values $c_l$, used in  \cite[p.~207]{CS}, 
equation \eqref{e6.3.1} is  
solvable for all $r>0$. If it is not solvable even for one $r>0$,  
then NS procedure breaks down and the existence of transparent  
potentials is not established.}  
  
 In the proof, given for the existence of the "transparent"  
potentials in \cite[p.~197]{CS}, formula (12.3.5), is used.  
This formula involves a certain infinite matrix $M$.   
It is claimed in \cite[p.~197]{CS},  that this  
matrix $M$ has the property $MM=I$, where $I$ is the unit matrix,  
and on \cite[p.~198]{CS}, formula (12.3.10), it is claimed that a vector  
$v \neq 0$ exists such that $Mv=0$. However, then $MMv=0$ and at the  
same time $MMv=v \neq 0$, which is a contradiction. The difficulties  
come from the claims about infinite matrices, which are not  
formulated clearly: it is not clear in what space $M$, as an operator,   
acts, what is the domain of definition of $M$, and on what set of  
vectors formula (12.3.5) in \cite{CS} holds.   
  
The construction of the "transparent"   
potential in \cite{CS} is based on the following logic: take all the fixed-energy  
shifts equal to zero and find the corresponding $c_l$ from  
the infinite linear algebraic system (12.2.7) in \cite{CS}; then  
construct the kernel $f(r,s)$ by formula \eqref{e6.3.2} and solve equation \eqref{e6.3.1}  
for all $r>0$; finally construct the  "transparent" potential by formula  
\eqref{e6.3.3}. As was noted above, it is not proved that equation \eqref{e6.3.1}  
with the constructed above kernel  $f(r,s)$ is solvable for all  
$r>0$.   
Therefore the existence of the  
"transparent" potentials is not established.   
   
The physicists have been using NS procedure without questioning  
its validity for several decades.  
Apparently the physicists still believe that NS procedure is  
``an analog of the Gel'fand-Levitan method" for inverse  
scattering problem with fixed-energy phase shifts as the data.  
In fact, the NS procedure is not a valid inversion method.  
Since modifications of NS procedure are still used by  
some physicists, who believe that this procedure  
is an inversion theory, the author pointed out some  
questions concerning this procedure in \cite{ARS} and \cite{R9}  
and wrote this paper.  
  
This concludes the discussion of case 1. \qed

\smallskip

\noindent{\bf Discussion of case 2:}  

\smallskip
{\it Suppose now that one wants just to construct a   
potential $q_1$, which generates the phase shifts corresponding to  
 some $q$.}  
  
This problem is actually {\it not an inverse scattering problem} because  
one does not recover an original potential from the scattering  
data, but rather wants to construct some potential which  
generates these data and may have no physical meaning.  
Therefore this problem is much less interesting practically than  
the inverse scattering problem.  
   
{\it  However, NS procedure does not solve this  
problem either: there is no guarantee that this procedure is  
applicable, that is, that the steps a) and b), described   
in the justification of the conclusions,  
can be done, in particular, that equation \eqref{e6.3.1} is   
uniquely solvable for all $r>0$.}  
  
If these steps can be done, then one needs to check that  
the potential $q_1$, obtained by formula \eqref{e6.3.3},  generates  
the original phase shifts. This was not done in \cite{N2} and  
\cite{N}.  
  
This concludes the discussion of case 2. \qed

\smallskip

The rest of the paper contains formulation and proof  
of \refR{Remark1} and \refCL{Claim1}.  
    
It was mentioned in \cite{N3} that if $Q:=\int^\infty_0 r q(r) dr\neq  
0,$ then the numbers $c_l$ in formula \eqref{e6.3.2}  cannot   
satisfy the condition  $\sum_0^\infty  
|c_l|<\infty$.     
This observation can be obtained also from the following   
  
\begin{remark}\label{R:Remark1}
For any potential $q(r) \in L_{1,1}$ such that  
$Q:= \int^\infty_0 rq(r) dr \neq 0$ the basic equation  
\eqref{e6.3.1} is not solvable for some $r>0$ and any choice of $c_l$  
such that   
$\sum^\infty_{l=0} |c_l| < \infty.$
\end{remark}
  
\medskip  
  
Since generically, for  $q \in L_{1,1},$ one has  
 $Q\neq 0$, this gives an additional  
illustration to the conclusion that equation \eqref{e6.3.1},  
in general, is not solvable for some $r>0$.   
Conditions $\sum^\infty_{l=0} |c_l| < \infty$ and $Q\neq 0$  
are incompatible.   
  
In \cite[p.~196]{CS},   
a weaker condition $\sum^\infty_{l=0} l^{-2} |c_l| < \infty$  
is used, but in the  
examples (\cite[pp.~189-191]{CS}), $c_l = 0$ for all $l \geq l_0 > 0$,  
so that $\sum^\infty_{l=0} |c_l| < \infty$ in all of these examples.    
  
\smallskip  
  
\begin{claim}\label{CL:Claim1} 
The set of the potentials $v(r)\in L_{1,1},$  
which can possibly be obtained by the NS procedure,  
is not dense (in the norm $\| q \| := \int^\infty_0 r |q(r)| dr$)  
in the set $L_{1,1}$.
\end{claim}

Let us prove \refR{Remark1} and \refCL{Claim1}.

\begin{proof}[Proof of \refR{Remark1}.]
Writing \eqref{e6.3.3} as  
$K(r,r) = -\frac{r}{2} \int^r_0 sq_1(s) ds $  
and assuming $Q \neq 0,$  
one gets the following relation:  
\begin{equation}\label{e6.3.4}
 K(r,r) = -\frac{Qr}{2} \left[ 1+ o(1) \right] \to \infty \hbox{\ as\ }  
 r \to \infty.
 \end{equation}
  
If \eqref{e6.3.1} is solvable for all $r>0$, then from \eqref{e6.3.2} and \eqref{e6.3.1}   
it follows that  
$K(r,s) = \sum^\infty_{l=0} c_l \varphi_l (r)$ $ u_l (s),$  
where  
$\varphi_l (r) := u_l (r) - \int^r_0 K(r,t) u_l (t) \frac{dt}{t^2}$,
so that $I - K$ is a transformation operator,  
where $K$ is the operator with kernel $K(r,s)$,  
$\varphi^{\prime \prime}_l + \varphi_l - \frac{l(l+1)}{r^2}  
  \varphi_l - q_1(r) \varphi_l = 0, $  
$q_1(r)$ is given by \eqref{e6.3.3}, $\varphi_l  = O(r^{l+1})$, as $r \to 
0,$  
\begin{equation}
  u_l (r) \sim \sin \left(r - \frac{l \pi}{2} \right), \quad
  \varphi_l (r) \sim |F_l|\sin \left( r-\frac{l \pi}{2} + \delta_l  
  \right)  \hbox{\ as\ } r \to \infty,
  \notag
 \end{equation}
where $\delta_l$ are the phase shifts at $k=1$  
and $F_l$ is the Jost function at $k=1$.   
One can prove that $\sup_l |F_l|<\infty$. Thus,  
if $\sum^\infty_{l=0} |c_l| <\infty,$ then  
\begin{equation}\label{e6.3.5}
  K(r,r) = O(1) \hbox{\ as\ } r \to \infty.
  \end{equation}
If $Q\neq 0$ then \eqref{e6.3.5} contradicts \eqref{e6.3.4}. It follows 
that if $Q \neq  
0$ then equation  
\eqref{e6.3.1} cannot be uniquely solvable for all $r>0$, so that NS procedure  
cannot be carried through if $Q \neq 0$ and $\sum^\infty_{l=0}  
|c_l|<\infty.$   
This proves \refR{Remark1}.
\end{proof}  
 
\begin{proof}[Proof of \refCL{Claim1}.]
Suppose that $v(r)\in L_{1,1}$ and $Q_v :=  
\int_0^\infty rv(r) dr = 0$,  
because otherwise NS procedure cannot be carried through as was  
proved in \refR{Remark1}.  
  
If $Q_v=0$, then there is also no guarantee that  NS procedure can  
be  carried  through. However, we claim that if one assumes that it   
can be carried through,  
then the set of potentials, which can possibly be   
obtained by NS procedure,  
is not dense in $L_{1,1}$ in the norm  
$\| q \| := \int^\infty_0 r |q(r)| dr$. In fact, any potential $q$ such  
that $Q:= \int^\infty_0 r q(r) dr \neq 0,$  
and the set of such potentials is dense in  $L_{1,1}$, cannot be  
approximated with a prescribed accuracy by  
the potentials which can be possibly obtained by the NS procedure.  
  
Let us prove this. Suppose that $q \in L_{1,1}$,  
\begin{equation}
  Q_q := \int^\infty_0 rq(r) dr \neq 0, \hbox{\ and\ }  
  \|v_n - q \| \to 0 \hbox{\ as\ } n \to \infty,
  \notag
  \end{equation}  
where the potentials $v_n \in L_{1,1}$ are obtained by the NS procedure,  
so that  
 $ Q_n := \int^\infty_0 rv_n (r) dr = 0.$
We assume $v_n \in L_{1,1}$ because otherwise $v_n$ obviously cannot  
converge in the norm $||\cdot||$ to $q\in L_{1,1}$.   
Define a linear bounded on $L_{1,1}$ functional  
\begin{equation}
 f(q) := \int^\infty_0 rq(r) dr, \quad |f(q)| \leq \| q \|,
 \notag
 \end{equation}
 where  
$\| q \| := \int^\infty_0 r|q(r)| dr$. The potentials  
$v \in L_{1,1}$, which can possibly be obtained by the NS procedure, belong  
to the null-space of $f$, that is $f(v) = 0.$   
  
If $\lim_{n \to \infty} \| v_n-q \| = 0$, then  
$\lim_{n \to \infty} |f(q-v_n)| \leq \lim_{n \to \infty}  
  \| q-v_n \| = 0. $  
Since $f$ is a linear bounded functional and $ f(v_n)=0$, one gets:   
$f(q-v_n) = f(q) - f(v_n) = f(q)$. So if $f(q) \neq 0$ then  
 $ \lim_{n \to \infty} |f(q-v_n)| = |f(q)| \neq 0.$
Therefore, no potential $q \in L_{1,1}$ with $Q_q \neq 0$ can be  
approximated  
arbitrarily accurately by a potential $v(r)\in L_{1,1}$ which can possibly  
be  obtained by the NS procedure. \refCL{Claim1} is proved.
\end{proof}


\section{Formula for the radius of the support of the
  potential in terms of scattering data}        \label{S:6.4}



The aim of this section is to prove the formula for the radius of the 
support of the potential in terms of the phase shifts.
Let us make the following assumption.

\noindent {\bf Assumption (A)}:
{\it the potential $q(r)$, $r=|x|$, is  spherically
symmetric, real-valued, $\int^a_0 |q|^2 dr<\infty$,
and $q(r)=0$ for $r>a$, but $q(r)\neq 0$ on 
$(a-\varepsilon, a)$ for all sufficiently small $\varepsilon>0$.}

The number $a>0$ we
call the radius of compactness of the potential, or simply the
radius of the
potential. Let $A(\alpha',\alpha)$ denote the scattering
amplitude corresponding to the potential $q$ at a fixed energy
$k^2>0$. Without loss of generality let us take $k=1$
in what follows. By $\alpha',\alpha\in S^2$ the unit vectors in
the direction of the scattered, respectively, incident wave, are
meant, $S^2$ is the unit sphere in ${\mathbb R}^3$.
Let us use formulas \eqref{e5.0.19} and \eqref{e5.0.20}.

It is of interest to obtain some information about $q$ from the
(fixed-energy) scattering data, that is, from the scattering
amplitude $A(\alpha',\alpha)$, or, equivalently, from the
coefficients $A_\l (\alpha)$. Very few results of such type are
known.

A result of such type is a
necessary and sufficient condition for $q(x)=q(|x|)$: it was
proved \cite[p.131]{R}, that $q(x)=q(|x|)$ if and only
if $A(\alpha',\alpha)=A(\alpha'\cdot\alpha)$. Of course, the
necessity of this condition was a common knowledge, but the
sufficiency, that is, the implication: 
$A(\alpha',\alpha)=A(\alpha'\cdot\alpha)\Rightarrow 
q(x)=q(|x|)$, is a new result \cite{R2}.

A (modified) conjecture from \cite[p.356]{R}
says that if the potential $q(x)$ is
compactly supported, and $a>0$ is its radius (defined for 
non-spherically symmetric potentials in the same way as for the
spherically symmetric), then
\begin{equation}\label{e6.4.1}
 a=\overline{\lim}_{\l \to\infty}
 \left( \frac{2\l }{e}
   \left[ \sup_{\substack {
           \alpha\in S^2 \\ -\l \leq m\leq\l}}
   |A_{\l  m}(\alpha)|\right]^{\frac{1}{2\l}}\right)
   =\overline{\lim}_{\l \to\infty}
   \left( \frac{2\l}{e} |\delta_\l |^{\frac{ 1}{2\l}} \right),
 \end{equation}
where $\delta_\l$ are the fixed-energy $(k=1)$ phase shifts.
We prove \eqref{e6.4.1} for the spherically symmetric
potentials $q=q(r)$. 

If $q=q(r)$ then
$ A_{\l  m}(\alpha)=\tildea_\l  Y_{\l  m}(\alpha) $
where $\tildea_\l $ depends only on $\l $ and $k$, but
not on $\alpha$ or $\alpha'$. Since $k=1$ is fixed, $\tildea_\l $ 
depends only on $\l $ for $q=q(r)$.
Assuming $q=q(r)$, one takes $A(\alpha',\alpha)=A(\alpha'\cdot\alpha)$ and
calculates
$ A_{\l  m}(\alpha)=\int_{S^2}A(\alpha'\cdot\alpha) 
\overline{Y_{\l  m}(\alpha')}\,d\alpha' 
=\tildea_\l 
\overline{Y_{\l  m}(\alpha)}$,
where
$\tildea_\l := 
\frac{2\pi}{C^{(\frac12)}_\l (1)} 
\int^1_{-1} A(t)C^{(\frac12)}_\l (t)\,dt,
\quad \l =0,1,2,\dots $
Here we have used formula (14.4.46) in \cite[p.413]{RK}, 
and $C^{(p)}_\l (t)$ are the Gegenbauer polynomials 
(see \cite[p.408]{RK}).
Since $C^{(\frac12)}_\l =P_\l (t)$,
$P_\l (1)=1$, where $P_\l (t)$ are the Legendre polynomials 
(see, e.g., \cite[p.409]{RK}),
one gets:
$ \tildea_\l =2\pi\int^1_{-1} A(t)P_\l (t)\,dt$.

Formula \eqref{e6.4.1} for $q=q(r)$ can be written as
$a=\overline{\lim}_{\l \to\infty}\left ( \frac{2\l +1}{e} 
|\tildea_\l |^{\frac{1}{2\l }}\right)$.

Indeed, 
$\sup_{\substack{ \alpha\in S^2 \\-\l \leq m\leq\l }}
|Y_{\l  m}|=O$
$\left(\l ^{\frac12}\right)$,
as is well known (see, e.g., [MP, p.261]). Thus
$\overline{\lim}_{\l \to\infty} $
$\left(\sup_{\substack{ \alpha\in S^2 \\ -\l \leq m\leq\l }}
|Y_{\l  m}(\alpha)|\right)^{\frac{1}{\l }}=1$, 
and formula for \eqref{e6.4.1} yields:
\begin{equation}\label{e6.4.2}
  a=\frac{2}{e} \overline{\lim}_{\l \to\infty}
 \left(\l  |\tildea_\l |^{\frac{1}{2\l }} \right).
 \end{equation}
Note that assumption (A) implies the following assumption:

{\bf Assumption (A$^\prime$)}:  {\it the potential $q(r)$
does not change sign in some left
neighborhood of the point $a$.}

This assumption in practice is not restrictive, however, as
shown in \cite[p.282]{R}, the potentials which oscillate infinitely often
in a neighborhood of the right end of their support, may
have some new properties which the potentials without this property
do not have. For example, it is proved in \cite[p.282]{R}, that
such infinitely oscillating potentials may have infinitely many
purely imaginary resonances, while the potentials which
do not change sign in a neighborhood of the right end of their
support cannot have infinitely many purely imaginary resonances.
Therefore it is of interest to find out if assumption $\hbox{A}'$
is necessary for the validity of \eqref{e6.4.2}.

The main result is:

\begin{theorem}\label{T:6.4.1}  
Let assumption (A) hold.
Then formula \eqref{e6.4.2} holds with
$\overline {\lim}$ replaced by $\lim$.
\end{theorem}

This result can be stated equivalently in terms of the fixed-energy
phase shift $\delta_\l$:
\begin{equation}\label{e6.4.3}
 \lim_{\l  \to \infty}
 \left( \frac{2\l +1}{e}|\delta_\l |^{\frac{1}{2\l }}\right) =a.
 \end{equation}

Below, we prove an auxiliary result:

\begin{lemma}\label{L:6.4.1}    
If $q=q(r)\in L^2(0,\infty)$, $q(r)$ is real-valued and
 does not change sign in some
interval $(a_1,a]$ where $a_1<a$, and $a$ is the radius of $q$, then
\begin{equation}\label{e6.4.4}
 a={\lim}_{m\to\infty}\left\vert\int^\infty_0
 q(r)r^m\,dr\right\vert^{\frac{1}{m}}, m=1,2,....
 \end{equation}
\end{lemma}

Below we prove \eqref{e6.4.3} and, therefore, \eqref{e6.4.1} for
spherically symmetric potentials. 

\begin{proof}[Proof of \refL{6.4.1}]
First, we obtain a slightly different result than \eqref{e6.4.4}
as an immediate consequence of the Paley-Wiener theorem.
Namely, we prove \refL{6.4.1} with a continuous parameter $t$ replacing
the integer $m$ and $\overline {\lim}$ replacing $\lim$.
This is done for $q(r)\in L^2(0,a)$ and without additional
assumptions about $q$.
However, we are not able to prove \refL{6.4.1}  assuming only
that $q(r)\in L^2(0,a)$.
 
Since $q(r)$ is compactly supported, one can write
\begin{equation}\label{e6.4.5}
 I(t):=\int^\infty_0 q(r)r^t\,dr=
 \int^a_0 q(r)e^{t\ln r}dr=
 \int^{\ln a}_{-\infty}q(e^u)e^ue^{tu}du. 
 \end{equation}
Let us recall that Paley-Wiener theorem 
implies the following claim (see \cite{Lev}): 

{\it 
If $f(z)=\int^{b_2}_{b_1}g(u)e^{-iuz}du$,  $[b_1,b_2]$ is the
smallest interval
containing the support of $g(u)$, and $g(u)\in L^2(b_1, b_2)$, then
\begin{equation}\label{e6.4.6}
 b_2=\overline{\lim}_{t\to+\infty}
 \left( t^{-1}\ln|f(it)|\right)=
 \overline{\lim}_{t\to+\infty} 
 \frac{\ln|\int^{b_2}_{b_1}g(u)e^{tu}du|}{t}.
 \end{equation}
}
Thus, using \eqref{e6.4.5} and \eqref{e6.4.6}, one gets:
\begin{equation}\label{e6.4.7}
 \ln a=\overline{\lim}_{t\to+\infty}\left(
 t^{-1}\ln\left|\int^{\ln a}_{-\infty}
 q(e^u)e^ue^{tu}du\right|\right).
 \end{equation}

Formula \eqref{e6.4.7} is similar to
\eqref{e6.4.4} with $m$ replaced by $t$ and
$\lim$ replaced by $\overline {\lim}$.

\begin{remark}\label{R:6.4.2}  
We have used formula \eqref{e6.4.6}  with $b_1=-\infty$, while in the 
Paley-Wiener theorem
it is assumed that $b_1>-\infty$. However, for
$b_1<b_2$, $g\not\equiv 0$ on $[b_2-\varepsilon,b_2]$ for any 
$\varepsilon>0$, one has:
\begin{equation} 
  \int^{b_2}_{-\infty} g(u)e^{tu}du=
  \int^{b_1}_{-\infty}g(u)e^{tu}du +
  \int^{b_2}_{b_1}g(u)e^{tu}du:=h_1(t)+h_2(t).
  \notag
  \end{equation}
\end{remark}

Thus  
$\lim_{t\to\infty}\frac{h_1(t)}{h_2(t)}=0$, and 

\begin{equation} 
\begin{aligned}
  \overline{\lim}_{t\to\infty}
  &
  \frac{\ln|h_1(t)+h_2(t)|}{t}
  \\
  =& \overline{\lim}_{t\to\infty}
  \frac{\ln|h_2(t)|}{t}+\lim_{t\to\infty}\frac{\ln|1+o(1)|}{t}
  =
\overline{\lim}_{t\to\infty}
\frac{\ln|h_2(t)|}{t}=\ln a.
\end{aligned}
\notag
\end{equation}

Therefore formula \eqref{e6.4.7} follows.

To prove \eqref{e6.4.4},
we use a different approach independent
of the Paley-Wiener theorem. 
We will use \eqref{e6.4.4}  below, in formula \eqref{e6.4.19}.
In this formula the role of $q(r)$ in \eqref{e6.4.4} 
is played by $rq(r)[1+\epsilon(r,\l )]$, where $\epsilon=O(\frac 1
\l )$. Let us prove \eqref{e6.4.4}.
        
Assume without loss of generality that $q\geq 0$ near $a$. Let
$I:=\int_0^a q(r) r^mdr=\int_0^{a_1} q(r) r^mdr+\int_{a_1}^a q(r) r^mdr:=I_1
+I_2$. We have $|I_1|<ca_1^m, \, c_1(a-\eta)^m<I_2<c_2a^m$, where
$\eta$ is an arbitrary small positive number. Thus, $I>0$ for all
sufficiently large $m$, and $I^{1/m}=I_2^{1/m}(1+\frac {I_1}{I_2})^{1/m}.$
One has $a-\eta\leq I_2^{1/m}\leq a$ and $\frac {I_1}{I_2}\to 0$ as $m\to \infty$.
Since $\eta$ is arbitrary small, it follows that $\lim _{m\to
\infty}I^{1/m}=a$.
This completes the proof of \eqref{e6.4.4}. \refL{6.4.1} is proved.
\end{proof}

\begin{proof}[Proof of formula \eqref{e6.4.3}]
From \eqref{e5.0.19} and \eqref{e5.0.23}
denoting $a_\l:=e^{i\delta_\l} \sin\delta_\l$, one gets
$  A(\alpha'\cdot\alpha)=\sum^\infty_{\ell=0}
 \tildea_\ell \overline{Y_\ell(\alpha)}
 Y_\ell(\alpha'):=4\pi\sum^\infty_{\ell=0}
 a_\ell \overline{Y_\ell(\alpha)} Y_\ell(\alpha')$,
where, $a_{\ell}:=\frac {\tildea_{\ell}}{4\pi}$, $k=1$, and
$ a_\ell=\frac{e^{2i\delta_\ell}-1}{2i} =
e^{i\delta_\ell}\sin\delta_\ell,$

\begin{equation}\label{e6.4.8}
  a_\l =-\int^\infty_0dr u_\l (r)q(r)\psi_\l (r),
  \end{equation}
where $u_\l (r)=rj_\l (r)\sim\sin \left(r-\frac{\l \pi}{2}\right)$
as $r\to\infty$, $j_\l (r)$ are the spherical Bessel functions, 
$j_\l (r):=\sqrt{\frac{\pi}{2r}} J_{\l +\frac12}(r)$, and
$\psi_\l (r)$
solves \eqref{e5.0.15} - \eqref{e5.0.17}, and the integral
\begin{equation}\label{e6.4.9}
 \psi_\l (r)=u_\l (r)+
 \int^\infty_0g_\l (r,s)q(s)\psi_\l (s)ds,\qquad k=1,
 \end{equation}
where
\begin{equation}\label{e6.4.10}
 g_\l (r,s)=-u_\l (r)w_\l (s),\qquad r<s;\qquad
 g_\l (r,s)=g_\l  (s,r),
 \end{equation}
\begin{equation}\label{e6.4.11}
 w_\l (s):=
 i\sqrt{\frac{\pi s}{2}}H^{(1)}_{\l +\frac12}(s),\qquad 
 u_\l (r)=
 \sqrt{\frac{\pi r}{2}}J_{\l +\frac12}(r),
 \end{equation}
and $H^{(1)}_\l $ is the Hankel function.

It is known \cite[p.407]{RK} that
\begin{equation}\label{e6.4.12}
 \begin{aligned}
 J_{\nu}(r) & \sim\left(\frac{er}{2\nu}\right)^{\nu}
 \frac{1}{\sqrt{2\pi \nu}},\quad  H^{(1)}_{\nu}(r)\sim
 -i\sqrt{\frac{2}{\pi \nu}} \left(\frac{er}{2\nu}\right)^{-\nu},
 \\
  J_{\nu}(r)& H^{(1)}_{\nu}(r)\sim-\frac{i}{\pi \nu},
 \nu \to+\infty,
 \end{aligned}
 \end{equation}
and \cite[Appendix 4]{AR}:
\begin{equation}\label{e6.4.13}
  \left|J_{\nu}(r)H^{(1)}_{\nu}(r)\right|<
  \left(\nu^2-\frac{1}{16}\right)^{-\frac14},\qquad 
  \nu >\frac14.
  \end{equation}
It follows from \eqref{e6.4.12} that $u_\l (r)$ does not have zeros on any
fixed interval $(0,a]$ if $\l $ is sufficiently large. Define
$v_\l (r):= \frac{\psi_\l (r)}{u_\l (r)}$.
Then \eqref{e6.4.9} yields
\begin{equation}\label{e6.4.14}
 v_\l (r)=1+\int^{a}_0
 \frac{g_\l (r,s)u_\l (s)}{u_\l (r)} q(s)v_\l (s) ds.
 \end{equation}
From \eqref{e6.4.10} and \eqref{e6.4.12} one gets
\begin{equation}\label{e6.4.15}
  g_\l (r,s)\sim \frac{r}{2\l +1}
  \left(\frac{r}{s}\right)^\l, \qquad r<s, 
  \qquad \l \to+\infty,
  \end{equation}
\begin{equation}\label{e6.4.16}
  \frac{u_\l (s)}{u_\l (r)} \sim
 \left(\frac{s}{r}\right)^{\l +1}, \qquad 
 \l \to+\infty.
 \end{equation}
Thus
\begin{equation}\label{e6.4.17}
  g_\l (r,s) \frac{u_\l (s)}{u_\l (r)} \sim
  \frac{s}{2\l +1}.
  \end{equation}
This implies that for sufficiently large $\l $ equation \eqref{e6.4.14}
has small kernel and therefore is uniquely solvable in $C(0,a)$
and one has 
\begin{equation}\label{e6.4.18}
 \psi_\l (r)=u_\l (r)
 \left[1+O\left(\frac{1}{\l }\right)\right]
 \ \text{as }\l \to+\infty,\quad 0\leq r\leq a,
 \end{equation}
uniformly with respect to $r\in[0,a]$.

{\it In the book \cite{N} formula (12.180), which gives the
asymptotic behavior of $S_{\l }$ for large $\l $,
is misleading: the remainder in this formula is of order
which is much greater, in general, than the order of the
main term in this formula.} That is why we 
had to find a different approach, which yielded 
formula \eqref{e6.4.18}.

From \eqref{e6.4.8}, \eqref{e6.4.11}, \eqref{e6.4.12},
and \eqref{e6.4.18} one has:
\begin{equation}\label{e6.4.19}
 \begin{aligned}
 a_\l
 = & -\int^\infty_0dr\,q(r)u_\l ^2(r)
 \left[1+O\left(\frac{1}{\l }\right)\right]
 \\
 = & -\int^a_0dr\,q(r)r^2r^{2\l }
 \left[1+O\left(\frac{1}{\l }\right)\right]
 \frac{1}{4\l +2}
 \left(\frac{e}{2\l +1}\right)^{2\l +1}.
 \end{aligned}
 \end{equation}
Therefore, using \eqref{e6.4.4}, one gets:
\begin{equation}\label{e6.4.20}
 \lim _{\l \to\infty}
 \left(\frac{2\l +1}{e} |a_\l |^{\frac{1}{2\l }}\right)
 =\lim_{\l \to\infty}
 \left| \int^a_0 dr\,q(r)r^2r^{2\l } \right|^{\frac{1}{2\l }}=a.
 \end{equation}
\refT{6.4.1} is proved.
\end{proof} 

\begin{remark}\label{R:6.4.3} 
Since $\delta_\l \to 0$ as
$\l \to+\infty$, and $\sin\delta_\l \sim\delta_\l $, 
$e^{i\delta_\l }\sim 1$, as $\delta_\l \to 0$, formulas \eqref{e6.4.20}
and $a_\l=e^{i\delta_\l}\sin \delta_\l$ imply
$\lim_{\l  \to \infty} \left(\frac{2\l +1}{e}
|\delta_\l |^{\frac{1}{2\l }} \right)=a$, 
where $\delta_\l $ is the phase shift at a fixed positive
energy. This is formula \eqref{e6.4.3}.
\end{remark}


\chapter{Inverse scattering with ``incomplete data''}\label{C:7}
\section{Uniqueness results}\label{S:7.1}
Consider equation \eqref{e1.1.2} on the interval $[0,1]$ with boundary
conditions $u(0)=u(1)=1$ (or some other selfadjoint homogeneous
separated boundary conditions), and $q=\barq$, $q\in L^1[0,1]$.
Fix $0<b\leq 1$. Assume $q(x)$ on $[b,1]$ is known and a subset
$\{\lambda_{m(n)}\}_{\forall n=1,2,3,\dots}$ of the eigenvalues
$\lambda_n=k^2_n$ of the operator $\l$ corresponding to the
chosen boundary conditions is known.
Here
\begin{equation}\label{e7.1.1}
 \frac{m(n)}{n}=\frac{1}{\sigma}(1+\varepsilon_n),
 \quad \sigma=\const>0,
 \quad |\varepsilon_n|<1,
 \quad \varepsilon_n\to 0.
 \end{equation}
We assume sometimes that
\begin{equation}\label{e7.1.2}
 \sum^\infty_{n=1} |\varepsilon_n|<\infty.
 \end{equation}

\begin{theorem}\label{T:7.1.1}
If \eqref{e7.1.1} holds and $\sigma>2b$, then the data
$\{q(x), b\leq x\leq 1; \{\lambda_{m(n)}\}_{\forall n}\}$ 
determine $q(x)$ on $[0,b]$ uniquely.
If \eqref{e7.1.1} and \eqref{e7.1.2} hold, the same conclusion holds also if
$\sigma=2b$.
\end{theorem}

The number $\sigma$ is ``the percentage" of the spectrum of $\l$ which is
sufficient to determine $q$ on $[0,b]$ if $\sigma\geq 2b$ and \eqref{e7.1.2} holds.
For example, if $\sigma=1$ and $b=\frac{1}{2}$, then ``one spectrum"
determines $q$ on the half-interval $[0,\frac{1}{2}]$.
If $b=\frac{1}{4}$, $\sigma=\frac{1}{2}$, then ``half of the spectrum"
determines $q$ on $[0,\frac{1}{4}]$. Of course, $q$ is assumed known
on $[b,1]$.
If $b=1$, $\sigma=2$, then ``two spectra" determines $q$ on the whole
interval. By ``two spectra" one means the set $\{\lambda_n\}\cup\{\mu_n\}$,
where $\{\mu_n\}$ is the set of eigenvalues of $\l$ corresponding to the same
boundary condition $u(0)=0$ at one end, say at $x=0$, and some other
selfadjoint boundary condition at the other end, say $u'(1)=0$
or $u'(1)+hu(1)=0$, $h=\const>0$. The last result is a well-known theorem of
Borg, which was strengthened in \cite{M},
where it is proved that not only the potential but the boundary conditions 
as
well are uniquely determined by two spectra. A version of
``one spectrum" result was mentioned in \cite[p.81]{L1}.

\begin{proof}[Proof of Theorem 7.1.1]
First, assume $\sigma > 2 b $. If there are $q_1$ and $q_2$
which produce the same data, then as above, one gets
\begin{equation}\label{e7.1.3}
  G(\lambda):=g(k):= \int^ b_0 p(x) \varphi_1 (x,k) \varphi_2 (x,k)\, dx
  = (\varphi_1 w^\prime - \varphi_1^\prime  w)\Big|^ b_0
  = (\varphi_1 w^\prime -\varphi_1^\prime w )\Big|_{x=b },
  \end{equation}
where $w:= \varphi_1-\varphi_2$, $p:=q_1 - q_2$, $k= \sqrt \lambda$.
Thus
\begin{equation}\label{e7.1.4}
  g(k) = 0\ \hbox{at}\ k
  = \pm \sqrt{\lambda_{m(n)}}: = \pm k_n.
  \end{equation}

The function $G(\lambda)$ is an entire function of $\lambda$
of order $\frac{1}{2}$
(see \eqref{e1.1.10} with $k=\sqrt\lambda$),
and is
an entire even function of
$k$ of exponential type $\leq 2  b $. One has
\begin{equation}\label{e7.1.5}
  |g(k)| \leq c \frac{e^{2 b  |Imk|}}{1+|k|^2}.
  \end{equation}
The indicator of $g$ is defined by the formula
\begin{equation}\label{e7.1.6}
  h(\theta):=h_g (\theta):=
  \overline{\lim_{r\to\infty}}
   \frac {\ln|g (r e^{i\theta})| }{r},
  \end{equation}
where $k=r e^{i \theta}$. Since $|Imk| = r|\sin \theta|$, one gets from
\eqref{e7.1.5} and \eqref{e7.1.6} the following estimate
\begin{equation}\label{e7.1.7}
  h(\theta) \leq 2b |\sin \theta|.
  \end{equation}

It is known \cite[formula (4.16)]{Lev} that for any entire function
 $g(k) \not\equiv 0$ of
exponential type one has:
\begin{equation}\label{e7.1.8}
  \lim_{\overline{r\to\infty}}
  \frac{n(r )}{r } \leq \frac{1}{2 \pi} \int^{2 \pi}_0 h_g
  (\theta)\, d \theta,
\end{equation}
where $n(r )$ is the number of zeros of $g(k)$ in the disk
$|k| \leq r $. From \eqref{e7.1.7} one gets
\begin{equation}\label{e7.1.9}
  \frac {1}{2\pi} \int ^{2\pi}_0 h_g(\theta)\, d\theta \leq
  \frac {2 b}{2 \pi} \int^{2 \pi}_0 |\sin \theta|\,d\theta
  = \frac{4  b }{\pi}
  \end{equation}
From \eqref{e7.1.2} and the known asymptotics of the Dirichlet eigenvalues:
\begin{equation}\label{e7.1.10}
  \lambda_n = (\pi n)^2 + c + o(1),
  \quad n \rightarrow\infty, \quad c=const,
  \end{equation}
one gets for the number of zeros the estimate
\begin{equation}\label{e7.1.11}
  n(r) \geq 2 \sum_{ \frac{n\pi}{\sigma}
     \left[ 1+0 \left( \frac{1}{n^2} \right) \right] <r}
  1=2\frac{\sigma r }{\pi}[1+o(1)],
  \quad r  \rightarrow \infty.
  \end{equation}

 From \eqref{e7.1.8}, \eqref{e7.1.9} and \eqref{e7.1.11} it follows that
\begin{equation}\label{e7.1.12}
   \sigma \leq 2 b.
   \end{equation}

Therefore, if $\sigma > 2  b $, then $g(k) \equiv 0$. If $g(k) \equiv 0$
then, by property $C_\varphi$, $p(x)=0$.
\refT{7.1.1} is proved in
the case $\sigma > 2  b $.

Assume now that $\sigma = 2  b $ and
\begin{equation}\label{e7.1.13}
   \sum_{n=1}^\infty |\varepsilon_n| < \infty.
   \end{equation}
We {\it claim} that if an entire function $G(\lambda)$ in \eqref{e7.1.3} of order
$\frac{1}{2}$ vanishes at the points
\begin{equation}\label{e7.1.14}
   \lambda_n = \frac{n^2 \pi^2}{\sigma^2} (1+\varepsilon_n),
    \end{equation}
and \eqref{e7.1.13} holds, then $G(\lambda) \equiv 0$. If this is proved,
then \refT{7.1.1} is proved as above.

Let us prove the claim. Define
\begin{equation}\label{e7.1.15}
  \Phi(\lambda):=\prod^\infty_{n=1} \left( 1-\frac{\lambda}{\lambda_n}\right)
  \end{equation}
and recall that
\begin{equation}\label{e7.1.16}
  \Phi_0(\lambda):=\frac{\sin(\sigma\sqrt{\lambda})}{\sigma\sqrt{\lambda}}
  =\prod^\infty_{n=1} \left( 1-\frac{\lambda}{\mu_n}\right),\quad
  \mu_n:=\frac{n^2\pi^2}{\sigma^2}.
   \end{equation}

Since $ G(\lambda_n) = 0$, the function
\begin{equation}\label{e7.1.17}
  w(\lambda): = \frac {G(\lambda)}{\Phi(\lambda)}
  \end{equation}
is entire, of order $\leq\frac{1}{2}$.
Let us use a Phragmen-Lindel\"of lemma.

\begin{lemma}\label{L:7.1.2}
\cite[Theorem 1.22]{Lev} If an entire function $w(\lambda)$ of order $<1$ 
has the property $sup_{-\infty<y<\infty}$  $|w(iy)|\leq c$,
then $w(\lambda) \equiv c$.
If, in addition $w(iy) \rightarrow 0$
as $y \rightarrow +\infty$, then $w(\lambda) \equiv 0.$
\end{lemma}

We use this lemma to prove that $w(\lambda) \equiv 0$.
If this is proved then $G(\lambda) \equiv 0$ and \refT{7.1.1} is proved.

The function $w(\lambda)$ is entire of order $\frac{1}{2} < 1$.

Let us check that
\begin{equation}\label{e7.1.18}
   \sup_{-\infty<y<\infty} |w(iy)| < \infty,
   \end{equation}
and that
\begin{equation}\label{e7.1.19}
   |w(iy)| \rightarrow 0\ \hbox{as}\ y \rightarrow +\infty.
   \end{equation}

One has, using \eqref{e7.1.5}, \eqref{e7.1.15}, \eqref{e7.1.16} and taking into
account that $\sigma = 2 b $:
\begin{equation}\label{e7.1.20}
 \begin{aligned}
 |w(iy)| = &
   \left| \frac{G(iy)}{\Phi(iy)} \frac{\Phi_0 (iy)}{\Phi_0(iy)} \right|
 \leq \frac{ e^{2b|Im\sqrt{iy}|} }{ (1+|y|)}
   \left(\frac{ e^{\sigma|Im\sqrt{iy}|} }{ 1+|y|^{\frac{1}{2}} }\right)^{-1}
  \left( \prod^\infty_{h=1}
    \frac{ 1+\frac{y^2}{\mu^2_n} }{ 1+\frac{y^2}{\lambda^2_n} }
       \right)^{ \frac{1}{2} }  \notag \\
 \leq &
    \frac{c}{ 1+|y|^{\frac{1}{2}} }
  \left( \prod_{ \{n:\mu_n\leq\lambda_n\} }
    \frac{\lambda_n^2}{\mu^2_n} \right)^{ \frac{1}{2} }
  \leq \frac{c}{ 1+|y|^{\frac{1}{2}} }
    \prod_{ \{n:\mu_n\leq\lambda_n\} }
    \left( 1+|\varepsilon_n|\right)
  \leq \frac{c_1}{ 1+|y|^{\frac{1}{2}} }\ . \notag
  \end{aligned}
   \end{equation}

Here we have used elementary inequalities:
\begin{equation}\label{e7.1.21}
   \frac {1+a}{1+ d} \leq \frac {a}{ d }\quad \hbox{if}\quad
   a \geq  d> 0;
   \quad \frac {1+a}{1+ d} \leq 1\quad \hbox{if}\quad 0\leq a \leq  d ,
   \end{equation}
with
  $a:=\frac {y^2}{\mu_n^2}$, $d:= \frac {y^2}{\lambda_n^2}$,
and the assumption \eqref{e7.1.13}.

We also used the relation:
\begin{equation}
 \left| \frac{\sin(\sigma \sqrt{iy})}{\sigma \sqrt{iy}} \right|
 \sim \frac {e^{\sigma |Im \sqrt{iy}|}}{2 \sigma |\sqrt{iy}|}
 \quad \hbox{as}\quad y \rightarrow +\infty.
 \notag
 \end{equation}
Estimate \eqref{e7.1.20} implies \eqref{e7.1.18} and \eqref{e7.1.19}.
An estimate similar to \eqref{e7.1.20} has been used in
the literature (see \cite {GS}).

\refT{7.1.1}    
is proved.
\end{proof}


\section{Uniqueness results: compactly supported potentials}\label{L:7.2}

Consider the inverse scattering problem of \refS{5.1} and assume
\begin{equation}\label{e7.2.1}
 \hbox{$q=0$ for $x\geq a>0$.}
 \end{equation}

\begin{theorem}\label{T:7.2.1}
If $q\in L_{1,1}$ satisfies \eqref{e7.2.1}, then any one of the data $S(k)$, $\delta(k)$, $f(k)$, $f'(k)$,
determine $q$ uniquely.
\end{theorem}

\begin{proof}
We prove first $S(k)\Rightarrow q$.
Note that without assumption \eqref{e7.2.1}, or an assumption which implies that $f(k)$ 
is an entire function on $\C$, the result does not hold. If \eqref{e7.2.1} holds,
or even a weaker assumption:
\begin{equation}\label{e7.2.2}
 |q(x)|\leq c_1 e^{-c_2|x|^\gamma}, \quad \gamma>1, \quad c_1,c_2>0,
 \end{equation}
then $f(k)$, the Jost function \eqref{e1.1.4}, is an entire function of 
$k$, and $S(k)$ is a meromorphic 
function on $\C$ with the only poles in $\C_+$ at the points $ik_j$, $1\leq j\leq J$.
Thus, $k_j$ and $J$ are determined by $S(k)$.
Using \eqref{e1.1.14} and \eqref{e1.1.10}, which holds for all $k\in\C$, because $f(k)$ and $f(x,k)$ are entire functions of $k$, one finds $s_j=i\res_{k=ik_j} S(k)$. Thus all the data \eqref{e1.1.15} are found from $S(k)$ if \eqref{e7.2.1} (or \eqref{e7.2.2}) holds. If the data \eqref{e1.1.15} are known, then $q$ is uniquely determined, see \refT{5.1.1}.

If $\delta(k)$ is given, then $S(k)=e^{2i\delta(k)}$, so 
$\delta(k)\Rightarrow q$.
If $f(k)$ is given then $S(k)=\frac{f(-k)}{f(k)}$, so $f(k)\Rightarrow q$.
If $f'(0,k)$ is given, then one can uniquely find $f(k)$ from \eqref{e1.1.10}.
Indeed, assume there are two $f(k)$, $f_1$ and $f_2$, corresponding to the given
$f'(0,k)$. Subtract from \eqref{e1.1.10} with $f=f_1$ equation 
\eqref{e1.1.10}
with $f=f_2$, denote $f_1-f_2:=w$, and get
$(\ast)\  f'(0,k)w(-k)=f'(0,-k)w(k)$ 
or $\frac{w(k)}{f'(0,k)} = \frac{w(-k)}{f'(0,-k)}$.
Since $w(\infty)=0$, and $f'(0,k)=ik-A(0,0) +\int^\infty_0 A_x(0,y) 
e^{iky}dy$, 
one can conclude that $w=0$ if one can check that $\frac{w(k)}{f'(0,k)}$ 
is analytic in $\C_+$.
The function $f'(0,k)$ has at most finitely many zeros in $\C_+$, 
and these zeros are simple. From $(\ast)$ one concludes that if 
$f'(0,\kappa)=0$, $\kappa\in\C_+$,  then $w(\kappa)=0$, because if 
$f'(0,\kappa)=0$
then $f'(0,-\kappa)\not= 0$ (see \eqref{e1.1.10}). 
Thus $\frac{w(k)}{f'(0,k)}$ is analytic in $\C_+$.
Similarly $\frac{w(-k)}{f'(0,-k)}$ is analytic in $\C_+$.
These two functions agree on the real axis, so, by analytic continuation, 
the function
$\frac{w(k)}{f'(0,k)}$ is analytic in $\C_+$ and vanishes at infinity.
Thus it vanishes identically. So $w(k)=0$, $f_1=f_2$, and $f(k)$ is uniquely determined
by $f'(0,k)$.
Thus \refT{7.2.1} is proved.
\end{proof}

\section{Inverse scattering on the full line by a potential vanishing on a half-line}\label{S:7.3}

The scattering problem on the full line consists of finding the solution to:
\begin{equation}\label{e7.3.1}
 lu-k^2u=0, \quad x\in\R,
 \end{equation}
\begin{equation}\label{e7.3.2}
 u=e^{ikx}+r(k)e^{-ikx} +o(1), \quad x\to -\infty,
 \end{equation}
\begin{equation}\label{e7.3.3}
 u=t(k) e^{ikx}+o(1), \quad x\to +\infty,
 \end{equation}
where $r(k)$ and $t(k)$ are, respectively, the reflection and transmission coefficients. The above scattering
problem describes plane wave scattering by a potential, the plane wave is incident from $-\infty$ in the 
positive direction of the $x\hbox{-axis}$. The inverse scattering problem consists of finding $q(x)$ given the 
scattering data
\begin{equation}\label{e7.3.4}
 \{r(k), k_j, s_j, 1\leq j\leq J\},
 \end{equation}
where $s_j>0$ are norming constants, $k_j>0$, and $-k^2_j$ are the  negative eigenvalues of the operator $\l_o$.

It is known \cite{M}, that the data \eqref{e7.3.4} determine 
$q\in L_{1,1}(\R):=\{q:q=\barq,\ \int^\infty_{-\infty} (1+|x|)|q|dx<\infty\}$
uniquely. Assume that
\begin{equation}\label{e7.3.5}
 q(x)=0, \quad x<0.
 \end{equation}

\begin{theorem}\label{T:7.3.1}
If $q\in L_{1,1}(\R)$ and \eqref{e7.3.5} holds, then $\{r(k)\}_{\forall k>0}$ determines $q$ uniquely.
\end{theorem}

\begin{proof}
If \eqref{e7.3.5} holds, then $u=e^{ikx} +r(k) e^{-ikx}$ for $x<0$,
and $u=t(k) f(x,k)$ for $x>0$, where $f(k,x)$ is the Jost solution 
\eqref{e1.1.4}. Thus
\begin{equation}
 \frac{ik(1-r(k))}{1+r(k)} = \frac{u'(-0,k)}{u(-0,k)} = 
\frac{u'(+0,k)}{u(+0,k)}
 =\frac{f'(0,k)}{f(k)} :=I(k).
 \end{equation}
Therefore $r(k)$ determines $I(k)$, 
so by \refT{3.1.2} $q$ is uniquely determined.
\end{proof}

\chapter{Recovery of quarkonium systems}\label{C:8}

\section{Statement of the inverse problem}\label{S:8.1}

The problem discussed in this Section is: to what extent does the
spectrum of a quarkonium system together with other
experimental data determines the interquark potential?
This problem was discussed in \cite{TQR}, where one can find further
references. The method given in \cite{TQR} for solving this problem
is this: one has few scattering data $E_j$, $s_j$, which will be defined
precisely later, and one constructs, using the known results of inverse
scattering theory, a Bargmann potential with the same scattering data
and considers this a solution to the problem. {\it This approach is
wrong} because the scattering theory is applicable to the potentials
which tend to zero at infinity, while our confining potentials
 grow to infinity at infinity, and no Bargmann potential can approximate
a confining potential on the whole semiaxis $(0, \infty )$.
Our aim  is to give an algorithm which is
consistent and yields a solution to the above problem.
The algorithm is based on the Gel'fand-Levitan procedure of \refS{4.2}.

Let us formulate the problem precisely. Consider the Schroedinger
equation
\begin{equation}\label{e8.1.1}
- \nabla^2 \psi_j + q(r)\psi_j=E_j \psi_j  \text { in } \mathbb R^3,  
\end{equation}
where $q(r)$ is a real-valued spherically symmetric potential,
$r:=|x|, x\in \mathbb R^3$,
\begin{equation}\label{e8.1.2}
q(r)=r+p(r), \quad p(r)=o(1) \,\,\text {as } r\to \infty. 
\end{equation}
The functions $\psi_j(x), \, ||\psi_j||_{L^2(\mathbb R^3)}=1,$
are the bound states, $E_j$ are the energies of these states.
 We define $u_j(r):=r\psi_j(r)$, which correspond to
$s$-waves, and consider the resulting equation for $u_j$:
\begin{equation}\label{e8.1.3}
\l u_j:=-u_j^{\prime\prime}+q(r)u_j=E_ju_j,\quad
 r>0,\quad u_j(0)=0,\quad ||u_j||_{L^2(0,\infty)}=1. 
\end{equation}
One can measure the energies $E_j$ of the bound states
and the quantities $s_j=u^\prime_j(0)$ experimentally.

Therefore the following inverse problem (IP) is of interest:

(IP): given:
\begin{equation}\label{e8.1.4}
\{E_j,s_j\}_{\forall j=1,2,...}
\end{equation}
can one recover $p(r)$?

In \cite{TQR} this question was considered but the approach in \cite{TQR} is inconsistent
and no exact results are obtained. The inconsistency of the approach in \cite{TQR}
is the following: on the one hand \cite{TQR} uses the inverse scattering theory
which is applicable to the potentials decaying sufficiently rapidly
at infinity, on the other hand, \cite{TQR} is concerned with potentials which grow
to infinity as $r\rightarrow +\infty$. It is nevertheless of some interest
that numerical results in \cite{TQR} seem to give some approximation of the potentials
in a neighborhood of the origin.

Here we present a rigorous approach to 
IP and prove
the following result:
\begin{theorem}\label{T:8.1.1}
IP has at most one solution and the potential $q(r)$
can be reconstructed from data \eqref{e8.1.4} algorithmically.
\end{theorem}

The reconstruction algorithm is based on the Gel'fand-Levitan
procedure for the reconstruction of $q(x)$ from the spectral function.
We show that the data \eqref{e8.1.4} allow one to write the spectral function of the
selfadjoint in $L^2(0,\infty)$ operator $\l,$ defined by the differential
expression \eqref{e8.1.3} and the boundary condition \eqref{e8.1.3} at zero.

In \refS{8.2} proofs are given and the recovery procedure is described.

Since in experiments one has only finitely many data $\{E_j,s_j\}_{1\leq j
\leq J}$, the question arises:

{\it How does one use these data for the recovery
of the potential?}

We give the following recipe: the unknown confining potential is assumed
to be of the form \eqref{e8.1.2}, and it is assumed that for $j> J$ the 
data
$\{E_j,s_j\}_{j> J}$ for this potential are the same as for the
unperturbed potential $q_0(r)=r$. In this case an easy algorithm is given
 for finding $q(r)$.

This algorithm is described in \refS{8.3}.

\section{Proofs}\label{S:8.2}
We prove \refT{8.1.1} by reducing (IP) to problem
of recovery of $q(r)$ from the spectral function.

Let us recall that the selfadjoint operator $L$ has discrete spectrum
since $q(r)\to  +\infty$. The formula for the number of eigenvalues
(energies of the bound states), not exceeding $\lambda$, is known: 
\begin{equation}
 \sum_{E_j<\lambda}1:=N(\lambda)\sim\frac 1\pi \int_{q(r)<\lambda}
 [\lambda -q(r)]^{\frac 12} dr.
 \notag
 \end{equation}
This formula yields, under the assumption $q(r)\sim r$ as $r\rightarrow
\infty$, the following asymptotics of the eigenvalues: 
\begin{equation}
 E_j\sim(\frac{3\pi}2 j)^
 {\frac 23}\quad \text { as } j\rightarrow +\infty.
 \notag
 \end{equation}
The  spectral function $\rho (\lambda)$ of the operator $L$ is defined 
 by the formula
\begin{equation}\label{e8.2.1}
\rho (\lambda)=\sum_{E_j< \lambda}\frac 1{\alpha_j},
\end{equation}
where $\alpha_j$ are the normalizing constants:
\begin{equation}\label{e8.2.2}
\alpha_j:=\int^\infty_0\phi^2_j(r)dr. 
\end{equation}
Here $\phi_j(r):=\phi(r,E_j)$ and $\phi (r,E)$ is the unique solution
of the problem:
\begin{equation}\label{e8.2.3}
L\phi:=-\phi^{\prime \prime}+q(r)\phi =
E \phi, \, r>0,\, \phi(0,E)=0,\ \phi'(0,E)=1. 
\end{equation}
If $E=E_j$, then $\phi_j=\phi(r,E_j)\in L^2(0,\infty)$.
The function $\phi(r,E)$ is the unique solution to
the Volterra integral equation:
\begin{equation}\label{e8.2.4}
\phi(r,E)=\frac {\sin(\sqrt E r)}{\sqrt E} +
\int_0^r \frac {\sin[\sqrt E (r-y)]}
{\sqrt E}q(y)\phi(y,E)dy. 
\end{equation}
For any fixed $r$ the function $\phi$ is an entire function
of $E$ of order $\frac 12$, that is, $|\phi|<c\exp (c|E|^{1/2}),$
where $c$ denotes various positive constants. At $E=E_j$,
where $E_j$ are the eigenvalues of \eqref{e8.1.3}, one has
 $\phi(r,E_j):=\phi_j\in L^2(0,\infty)$.
In fact, if $q(r)\sim cr^a,\,\, a>0,$ then $|\phi_j|<c\exp(-\gamma r)$
for some $\gamma >0$.

Let us relate $\alpha_j $ and $s_j$. From \eqref{e8.2.3} with $E=E_j$ and
from \eqref{e8.1.3}, it follows that
\begin{equation}\label{e8.2.5}
\phi_j=\frac {u_j}{s_j}. 
\end{equation}
Therefore
\begin{equation}\label{e8.2.6}
\alpha_j:=||\phi_j||^2_{L^2(0,\infty)}=\frac 1 {s_j^2}. 
\end{equation}
Thus data \eqref{e8.1.4} define uniquely the spectral function
 of the operator $L$ by the formula:
\begin{equation}\label{e8.2.7}
\rho(\lambda):=\sum_{E_j< \lambda} s_j^2. 
\end{equation}
Given $\rho(\lambda)$, one can use the Gel'fand-Levitan (GL)
method for recovery of $q(r)$.
According to this method, define
\begin{equation}\label{e8.2.8}
\sigma(\lambda):=
         \rho (\lambda)-\rho_0(\lambda), 
\end{equation}
where $\rho_0(\lambda)$ is the spectral function of the unperturbed problem,
which in our case is the problem with $q(r)=r$,
then set
\begin{equation}\label{e8.2.9}
L(x,y):=\int_{-\infty}^\infty \phi_0(x,\lambda) \phi_0 (y, \lambda) 
d\sigma (\lambda), 
\end{equation}
where $\phi_0(x, \lambda)$ are the  eigenfunctions of the 
problem \eqref{e8.2.3} with $q(r)=r$,
and solve the second kind Fredholm integral equation for
the kernel $K(x,y)$:
\begin{equation}\label{e8.2.10}
K(x,y)+\int_0^xK(x,t)L(t,y)dt=-L(x,y), \quad 0\leq y \leq x. 
\end{equation}
The kernel $L(x,y)$ in equation (8.2.10) is given by formula (8.2.9).
If $K(x,y)$ solves (8.2.10),
then
\begin{equation}\label{e8.2.11}
p(r)= 2 \frac {d K(r,r)}{dr},\qquad r>0. 
\end{equation}

\section{Reconstruction method}\label{S:8.3}

Let us describe the algorithm we propose for recovery of the function
$q(x)$ from few experimental data $\{E_j, \, s_j\}_{1\leq j \leq J}$.
Denote by $ \{E^0_j, s_j^0\}_{1\leq j \leq J}$ the data corresponding to
$q_0:=r$. These data are known and the corresponding eigenfunctions
\eqref{e8.1.3} can be expressed in terms of Airy function $Ai(r)$, which solves
the equation $w^{\prime \prime}-rw=0$ and decays at $+\infty$, see \cite{Leb}.
The spectral function of the operator $L_0$ corresponding to $q=q_0:=r$ is
\begin{equation}\label{e8.3.1}
\rho_0(\lambda):=\sum_{E_j^0< \lambda} (s_j^0)^2 . 
\end{equation}
Define
\begin{equation}\label{e8.3.2}
\rho(\lambda):=\rho_0(\lambda) +\sigma (\lambda), 
\end{equation}
\begin{equation}\label{e8.3.3}
\sigma (\lambda):=\sum_{E_j< \lambda}s_j^2-
\sum_{E_j^0< \lambda}(s_j^0)^2, 
\end{equation}
and
\begin{equation}\label{e8.3.4}
L(x,y):=\sum_{j=1}^J s^2_j\phi(x,E_j)\phi(y,E_j)-
\sum_{j=1}^J(s_j^0)^2\phi_j(x)\phi_j(y), 
\end{equation}
where $\phi(x,E) $ can be obtained by solving the Volterra equation
\eqref{e8.2.5} with $q(r)=q_0(r):=r$ and represented in the form:
\begin{equation}\label{e8.3.5}
\phi(x,E)=\frac {\sin (E^{1/2} x)}{E^{1/2}} +\int_0^x
K(x,y) \frac {\sin (E^{1/2}y)}{E^{1/2}}dy, 
\end{equation}
where $K(x,y)$ is the transformation kernel corresponding
to the potential $q(r)=q_0(r):=r$,
and $\phi_j$ are the eigenfunctions of the unperturbed problem:
\begin{equation}\label{e8.3.6}
-\phi_j^{\prime \prime}+r\phi_j=E_j\phi_j \quad r>0, \quad\phi_j(0)=0,\quad
\phi_j^{\prime}(0)=1. 
\end{equation}
Note that for $E\neq E_j^0$ the functions \eqref{e8.3.5} do not belong to
$L^2(0,\infty)$, but $\phi(0,E)=0$.
We denoted in this section the eigenfunctions of the
{\it unperturbed problem}
by $\phi_j$ rather than  $\phi_{0j}$ for simplicity of notations, since
the eigenfunctions of the perturbed problem are not used in this section.
One has: $\phi_j(r)=c_j Ai(r-E_j^0)$, where $c_j=
[Ai^{\prime}(-E_j^0)]^{-1}$,  $E_j^0>0$ is the $j-$th positive root
if the equation $Ai(-E)=0$ and, by formula \eqref{e8.2.6},
one has $s_j^0=[c_j^2 \int_0^\infty
Ai^2(r-E_j^0)dr]^{-1/2}.$
 These
formulas make the calculation of $\phi_j(x), \, E_j^0$ and $s_j^0$ easy
since the tables of Airy functions are available \cite{Leb}.

The equation analogous to \eqref{e8.2.10} is:
\begin{equation}\label{e8.3.7}
K(x,y)+\sum_{j=1}^{2J} c_j \Psi_j(y) \int_0^xK(x,t)\Psi_j(t)dt=
-\sum_{j=1}^{2J} c_j\Psi_j(x)\Psi_j(y), 
\end{equation}
where $\Psi_j(t):=\phi(t,E_j), c_j=s_j^2, 1\leq j \leq J,$
and $\Psi_j(t)=\phi_{j-J}(t), c_j=(s_{j-J}^0)^2, J+1\leq j \leq 2J.$
Equation \eqref{e8.3.7} has degenerate kernel and therefore can be
reduced to a linear algebraic system.

If $K(x,y)$ is found from \eqref{e8.3.7}, then
\begin{equation}\label{e8.3.8}
p(r)=2\frac d{dr} K(r,r), \quad q(r)=r+p(r). 
\end{equation}
Equation \eqref{e8.2.10} and, in particular \eqref{e8.3.7}, is uniquely solvable by the
Fredholm alternative: the homogeneous version of \eqref{e8.2.10}
has only the trivial solution.
Indeed, if $h+\int_0^xL(t,y)h(t)dt=0, 0\leq y \leq x,$
then $||h||^2 +\int_{-\infty}^\infty |\tildeh|^2 [d\rho(\lambda)-
\rho_0(\lambda)]=0,$ so that, by Parseval equality,
 $ \int_{-\infty}^{\infty} |\tildeh|^2 d \rho (\lambda)=0.$
Here $\tildeh:=\int_0^x h(t)\phi(t,\lambda)dt$,
where $\phi(t,\lambda)$ are defined by \eqref{e8.3.5}. 
This implies that $\tildeh(E_j)=0$ for all $j=1,2,....$
Since $\tildeh(\lambda)$ is an entire function of exponential type
$\leq x$, and since the density of the sequence $E_j$ is
infinite, i.e., $\lim_{\lambda \to \infty}\frac 
{N(\lambda)}{\lambda}=\infty$, 
because $E_j=O(j^{2/3})$, as
was shown in the beginning of \refS{8.2}, it follows that 
$\tildeh=0$ and consequently $h(t)=0$, as claimed.

In conclusion consider the case when $E_j=E_j^0, s_j=s_j^0$
for all $j\geq 1$, and $\{E_0, s_0\}$ is the new 
eigenvalue, $E_0<E_1^0,$ 
with the corresponding data $s_0$.
In this case 
$L(t,y)=s_0^2 \phi_0(t,E_0)\phi_0(y,E_0)$, so that
equation \eqref{e8.2.10} takes the form
\begin{equation}
 K(x,y)+s_0^2 \phi_0(y) \int_0^x K(x,t)\phi_0(t,E_0)dt
=-s_0^2 \phi_0(x,E_0)\phi_0(y,E_0).
\notag
\end{equation}

Thus,  one gets:
\begin{equation}
p(r)=-2 \frac {d}{dr} \frac {s_0^2 \phi^2_0(x,E_0)}{1+
s_0^2 \int_0^x \phi_0^2(t,E_0)dt}.
  \notag
\end{equation}


\chapter{Krein's method in inverse scattering} \label{C:9}


\section{Introduction and description of the method}\label{S:9.1}

Consider inverse scattering problem studied in \refC{5}
and for simplicity {\it assume that there are no bound states.}
This assumption is removed in \refS{9.4}.

This chapter is a commentary to Krein's paper \cite{K1}.
It contains not only a detailed proof of the results announced in
\cite{K1} but also a proof of the new results not mentioned in
\cite{K1}.
In particular, it contains an analysis of the invertibility of the
steps in the inversion procedure based on Krein's 
results, and a proof of
the consistency of this procedure, that is, a proof of the fact
that the reconstructed potential generates the scattering data
from which it was reconstructed. A numerical scheme for solving
inverse scattering problem, based on Krein's inversion method, is 
proposed,
and its advantages compared with the Marchenko and Gel'fand-Levitan 
methods are discussed. 
Some of the results are stated in
\refT{9.1.1} -- \refT{9.1.4} below.

Consider the equation for a function $\Gamma_x(t,s)$:
\begin{equation}\label{e9.1.1}
  (I+H_x)\Gamma_x:=\Gamma_x(t,s)
  +\int^x_0H(t-u) \Gamma_x(u,s)du
  =H(t-s), \quad 0\leq t,s\leq x.
  \end{equation}
Equation \eqref{e9.1.1} shows that 
$\Gamma_x=(I+H_x)^{-1}H=I-(I+H_x)^{-1}$, so
\begin{equation}\label{e9.1.2}
  (I+H_x)^{-1}=I-\Gamma_x 
  \end{equation}
in operator form, and
\begin{equation}\label{e9.1.3}
  H_x=(I-\Gamma_x)^{-1}-I. 
  \end{equation} 
Let us assume that $H(t)$ is a real-valued even function
$$
 H(-t)=H(t),\quad H(t)\in L^1(\R)\cap L^2(\R), 
$$
\begin{equation}\label{e9.1.4}
  1+\tildeH(k)>0,\quad
  \tildeH(k):=\int^\infty_{-\infty} H(t)e^{ikt} dt
    =2\int^\infty_0\cos(kt)H(t)dt.
  \end{equation}
Then \eqref{e9.1.1} is uniquely solvable for any $x>0$,
and there exists a limit
\begin{equation}\label{e9.1.5}
  \Gamma(t,s)=\lim_{x\to\infty} \Gamma_x(t,s):=\Gamma_{\infty}(t,s),
\quad t,s\geq 0,
 \end{equation}
where $\Gamma(t,s)$ solves the equation
\begin{equation}\label{e9.1.6}
  \Gamma(t,s)+\int^\infty_0 H(t-u) \Gamma(u,s)du=H(t-s),\quad
  0\leq t, s<\infty.
  \end{equation}
Given $H(t)$, one solves \eqref{e9.1.1}, finds $\Gamma_{2x}(s,0)$, then 
defines
\begin{equation}\label{e9.1.7}
  \psi(x,k):=\frac{E(x,k)-E(x,-k)}{2i},
  \end{equation}
where
\begin{equation}\label{e9.1.8}
  E(x,k):= e^{ikx}
   \left[ 1-\int^{2x}_0 \Gamma_{2x}(s,0) e^{-iks}ds \right].
  \end{equation}
Formula \eqref{e9.1.8} gives a one-to-one correspondence between
$E(x,k)$ and $\Gamma_{2x}(s,0)$.
\begin{remark}\label{R:9.1.1}
In \cite{K1}  $\Gamma_{2x}(0,s)$ is used in place
of $\Gamma_{2x}(s,0)$ in the definition of $E(x,k)$.
By formula \eqref{e9.2.21} (see \refS{9.2} below) one has
$\Gamma_x(0,x)=\Gamma_x(x,0)$,
but $\Gamma_x(0,s)\neq \Gamma_x(s,0)$ in general.
The theory presented below cannot be constructed with 
$\Gamma_{2x}(0,s)$ in place of $ \Gamma_{2x}(s,0)$ in formula (9.1.8).
\end{remark}

Note that
\begin{equation}\label{e9.1.9}
  E(x, k)=e^{ ikx}f(- k)+o(1), \quad x\to+\infty,
  \end{equation}
where
\begin{equation}\label{e9.1.10}
  f(k):=1-\int^\infty_0 \Gamma(s) e^{iks}ds,
  \end{equation}
and
\begin{equation}\label{e9.1.11}
  \Gamma(s):=\lim_{x\to +\infty} \Gamma_x(s,0):=\Gamma_{\infty}(s,0).
  \end{equation}
Furthermore,
\begin{equation}\label{e9.1.12}
  \psi(x,k)=\frac{e^{ikx} f(-k)-e^{-ikx}f(k)}{2i}
   +o(1),\quad x\to+\infty.
  \end{equation}
Note that $\psi(x,k)=|f(k)|\sin (kx+\delta(k)) + o(1), \quad x\to +\infty$,
where $  f(k)=|f(k)|e^{-i\delta(k)}, \quad \delta(k)=-\delta(-k),\quad k\in \R$.

The function $\delta(k)$ is called the phase shift. One has $S(k)=e^{2i\delta(k)}$.

We have changed the notations from \cite{K1} in order to
show the physical meaning of the function \eqref{e9.1.9}:
$f(k)$ is the Jost function of the scattering theory.
The function $\frac {\psi(x,k)}{f(k)}$ is the solution
to the scattering problem: it solves equation (1.2.3), and satisfies the
correct boundary conditions: $\frac {\psi(0,k)}{f(k)}=0$, and
 $\frac {\psi(x,k)}{f(k)}=e^{i\delta(k)}\sin(kx+\delta(k)) +o(1)$
as $x\to \infty$. 

Krein \cite{K1} calls $S(k):=\frac{f(-k)}{f(k)}$  
 the $S$-function, and $S(k)$ is the $S$-matrix used in physics. 

Assuming no bound states, one can solve 
{\it the inverse scattering  problem (ISP):} 
{\it Given $S(k)\,\,\forall  k>0$, find $q(x)$}. 

{\it A solution of the ISP, based on the results 
of \cite{K1}, consists of four steps:}
\begin{description}
\item{1)} Given $S(k)$, find $f(k)$ by solving the 
Riemann problem \eqref{e9.2.37}.
\item{2)} Given $f(k)$, calculate $H(t)$ using the formula
\begin{equation}\label{e9.1.13}
  1+\tildeH=1+\int^\infty_{-\infty} H(t)e^{ikt}dt=\frac{1}{|f(k)|^2}.  
  \end{equation}
\item{3)} Given $H(t)$, solve \eqref{e9.1.6} for
  $\Gamma_x(t,s)$ and then find $\Gamma_{2x}(2x,0)$, $0\leq
x<\infty$.
\item{4)} Define
\begin{equation}\label{e9.1.14}
  a(x)=2\Gamma_{2x}(2x,0),  
  \end{equation}
where
\begin{equation}\label{e9.1.15}
  a(0)=2H(0), 
  \end{equation}
and calculate the potential
\begin{equation}\label{e9.1.16}
  q(x)=a^2(x)+a^\prime(x), \quad a(0)=2H(0).
  \end{equation}
\end{description}

One can also calculate $q(x)$ by the formula:
\begin{equation}\label{e9.1.17}
 q(x)=2\frac d {dx}[\Gamma_{2x}(2x,0)-\Gamma_{2x}(0,0)]. 
 \end{equation}
Indeed, $2\Gamma_{2x}(2x,0)=a(x),$ see \eqref{e9.1.14}, 
$\frac{d}{dx}\Gamma_{2x}(0,0)=-2\Gamma_{2x}(2x,0)\Gamma_{2x}(0,2x)$, 
see \eqref{e9.2.22}, and
$\Gamma_{2x}(2x,0)$ $=\Gamma_{2x}(0,2x)$, see \eqref{e9.2.21}.

There is an alternative way, based
on the Wiener-Levy theorem, to do step 1).
Namely, given $S(k)$, find $\delta(k),$ the phase shift, 
then calculate the function 
$g(t):= -\frac{2}{\pi}\int_0^{\infty}\delta(k)\sin (kt)dk,$
and finally calculate $f(k)=\exp \left(\int_0^{\infty}g(t)e^{ikt}dk\right)$.

The potential $q\in L_{1,1}$ 
generates the $S$-matrix $S(k),$ with which we started, provided that
the following conditions \eqref{e9.1.18} -- \eqref{e9.1.21} hold:
\begin{equation}\label{e9.1.18}
  S(k)=\overline{S(-k)}=S^{-1}(k), \quad k\in\R,
 \end{equation}
the overbar stands for complex conjugation,
and
\begin{equation}\label{e9.1.19}
  \ ind_\R S(k)=0,  
  \end{equation}
\begin{equation}\label{e9.1.20}
  ||F(x)||_{L^\infty(\R_+)} +||F(x)||_{L^1(\R_+)}
   +||xF^\prime(x)||_{L^1(\R_+)}<\infty,  
  \end{equation}
where
\begin{equation}\label{e9.1.21}
  F(x):=\frac{1}{2\pi}\int^\infty_{-\infty} [1-S(k)] e^{ikx} dk. 
  \end{equation}
By the index \eqref{e9.1.19} one means the increment of the
argument of $S(k)$  ( when $k$ runs from $-\infty$ to $+\infty$
along the real axis) divided by $2\pi$.
The function \eqref{e9.1.7} satisfies equation \eqref{e1.1.4}.
Recall that {\it we have assumed that there are no bound states.}

In \refS{9.2} the above method is justified and the following
theorems are proved:

\begin{theorem}\label{T:9.1.1}  
 If \eqref{e9.1.18} -- \eqref{e9.1.20} hold, then $q(x)$ defined by \eqref{e9.1.16} is the unique
 solution to ISP and this $q(x)$ has $S(k)$ as the scattering matrix.
\end{theorem}

\begin{theorem}\label{T:9.1.2} 
 The function $f(k)$, defined by \eqref{e9.1.10}, is the Jost function corresponding
 to potential \eqref{e9.1.16}.
\end{theorem}

\begin{theorem}\label{T:9.1.3} 
 Condition \eqref{e9.1.4} implies that equation \eqref{e9.1.1} is
 solvable for all $x\geq 0$ and its solution is unique.
\end{theorem}

\begin{theorem}\label{T:9.1.4} 
If condition \eqref{e9.1.4} holds, then relation
\eqref{e9.1.11} holds and $\Gamma(s):=\Gamma_{\infty}(s,0)$ 
is the unique solution to the equation
\begin{equation}\label{e9.1.22}
  \Gamma(s)+\int^\infty_0 H(s-u)\Gamma(u)du=H(s), \quad s\geq 0. 
  \end{equation}
\end{theorem}

The diagram explaining the inversion method for solving ISP, based
on Krein's results, can be shown now:
\begin{equation}\label{e9.1.23}
  S(k)\operatorname*\Rightarrow^{ (9.2.34)}_{s_1}
  f(k)\operatorname*\Rightarrow^{(9.1.13)}_{s_2}
  H(t)\operatorname*\Rightarrow^{(9.1.1)}_{s_3}
  \Gamma_x(t,s)\operatorname*\Rightarrow^{\hbox{(trivial)}}_{s_4}
  \Gamma_{2x}(2x,0)\operatorname*\Rightarrow^{(9.1.14)}_{s_5}
  a(x)\operatorname*\Rightarrow^{(9.1.16)}_{s_6} q(x). 
  \end{equation}
In this diagram $s_m$ denotes step number $m$. Steps $s_2$, $s_4$,
$s_5$ and $s_6$ are trivial. Step $s_1$ is almost trivial: it
requires solving a Riemann problem with index zero and can be
done analytically, in closed form. Step $s_3$ is the
basic (non-trivial) step which requires solving a family of
Fredholm-type linear integral equations \eqref{e9.1.1}.
These equations are uniquely solvable if assumption \eqref{e9.1.4} holds, 
or if
assumptions \eqref{e9.1.18} -- \eqref{e9.1.20} hold.

In \refS{9.2} we analyze the invertibility of 
the steps in diagram \eqref{e9.1.23}.
Note also that, if one assumes \eqref{e9.1.18} -- \eqref{e9.1.20}, diagram \eqref{e9.1.23} can be
used for solving the inverse problems of finding $q(x)$ from
the following data:

\begin{description}
\item[a)] from $f(k)$, $\forall k>0$,
\item[b)] from $|f(k)|^2$, $\forall k>0$, 
or
\item[c)] from the spectral function $d\rho(\lambda)$.
\end{description}

Indeed, if \eqref{e9.1.18} -- \eqref{e9.1.20} hold, then a) and b) are 
contained in diagram \eqref{e9.1.23}, and c) follows from the known  formula 
$  d\rho(\lambda)=\left\{ \begin{array}{rl}
  \frac{\sqrt{\lambda}}{\pi} \frac{d\lambda} { |f(\sqrt{\lambda} ) |^2 },
    & \quad \lambda>0, \\  0, & \quad \lambda<0.   \end{array}\right. $ 
Let $\lambda=k^2$.
Then (still assuming \eqref{e9.1.19}) one has:  
$d\rho=\frac{2k^2}{\pi}  \frac{1}{|f(k)|^2} dk, \quad k>0$.

Note that the general case of the inverse scattering problem on the
half-axis, when $ \ind_\R S(k):=\nu\not= 0$,
can be reduced to the case $\nu=0$ by the procedure
described in Section 9.4, provided that $S(k)$ is the $S-$matrix
corresponding to a potential $q\in L_{1,1}(\R_+)$.
Necessary and sufficient conditions for this
are conditions \eqref{e9.1.18} -- \eqref{e9.1.20}.

\refS{9.3} contains a discussion of the numerical 
aspects of the inversion procedure based on Krein's method.
There are advantages in using this procedure (as compared with the
Gel'fand-Levitan procedure): integral equation \eqref{e9.1.1},
solving of which constitutes the basic step in the Krein inversion
method, is a Fredholm convolution-type equation.
Solving such an equation numerically leads to inversion of
Toeplitz matrices, which can be done efficiently and with much
less computer time than solving the Gel'fand-Levitan equation 
\eqref{e1.4.4}.
Combining Krein's and Marchenko's inversion methods yields an
efficient way to solve inverse scattering problems.

Indeed, for small $x$ equation \eqref{e9.1.1} can be solved by iterations since
the norm of the integral operator in \eqref{e9.1.1} is less than 1 for
sufficiently small $x$, say $0<x<x_0$.
Thus $q(x)$ can be calculated for
$0\leq x\leq \frac{x_0}{2}$ by diagram \eqref{e9.1.23}.

For $x\geq x_0>0$ one can solve by iterations Marchenko's 
equation \eqref{e1.4.13} for the kernel $A(x,y)$,
where, if \eqref{e9.1.19} holds, the function $F(x)$ is defined by the \eqref{e1.4.11}
with $F_d=0$.

Indeed, for $x>0$ the norm of the operator in \eqref{e1.4.11} is less than 
$1$ and it tends to $0$ as $x\to +\infty$.

Finally let us discuss the following question: in the justification of
both the Gel'fand-Levitan and Marchenko methods, the eigenfunction
expansion theorem and the Parseval relation play the fundamental
role. In contrast, the Krein method apparently does not use the
eigenfunction expansion theorem and the Parseval relation.
However, implicitly, this method is also based on such relations.
Namely, assumption \eqref{e9.1.4} implies that 
the $S$-matrix corresponding to the potential \eqref{e9.1.16}, has index $0$. 
If, in addition, this potential is in $L_{1,1}(\R_+)$, then conditions \eqref{e9.1.18} 
and \eqref{e9.1.20} are satisfied as well, and the eigenfunction expansion theorem and 
Parseval's equality hold. Necessary and sufficient conditions, imposed directly on the
function $H(t)$, which guarantee that conditions \eqref{e9.1.18} -- \eqref{e9.1.20} hold,
are not known.
However,  it follows that conditions
\eqref{e9.1.18} -- \eqref{e9.1.20} hold if and only if $H(t)$ is such that the diagram
\eqref{e9.1.23} leads to a $q(x)\in L_{1,1}(\R_+)$.
Alternatively, conditions \eqref{e9.1.18} -- \eqref{e9.1.20} hold (and consequently,
$q(x)\in L_{1,1}(\R_+)$) if and only if condition \eqref{e9.1.4} holds and
the function $f(k)$, which is uniquely defined as the solution to the Riemann problem
\begin{equation}\label{e9.1.24}
  \Phi_+(k)=[1+\tildeH(k)]^{-1}\Phi_-(k),\quad k\in\R, 
 \end{equation}
by the formula $ f(k)=\Phi_+(k)$,
generates the $S$-matrix $S(k)$ by formula \eqref{e9.1.15}, and this
$S(k)$ satisfies conditions \eqref{e9.1.18} -- \eqref{e9.1.20}.
Although the above conditions are  verifiable, they are not quite
satisfactory because they are implicit, they are not formulated in terms of
structural properties of the function $H(t)$ (such as smoothness,
rate of decay, etc.).

In \refS{9.2} \refT{9.1.1} -- \refT{9.1.4} are proved.
In \refS{9.3} numerical aspects of the inversion method based on Krein's results are discussed.
In \refS{9.4} the ISP with bound states is discussed.
In \refS{9.5} a relation between Krein's and Gel'fand-Levitan's methods is 
explained.

\section{Proofs}\label{S:9.2}
\begin{proof}[Proof of \refT{9.1.3}]
If $v\in L^2(0,x)$, then
\begin{equation}\label{e9.2.1}
 (v+H_xv,v)=\frac{1}{2\pi} 
 [(\tildev,\tildev)_{L^2(\R)}+(\tildeH\tildev,\tildev)_{L^2(\R)}] 
 \end{equation}
where the Parseval equality was used,
$ \tildev:=\int^x_0 v(s)e^{iks}ds$,   
\begin{equation}\label{e9.2.2}
  (v,v)=\int^x_0 |v|^2ds=(v,v)_{L^2(\R)}.
  \end{equation}
Thus $I+H_x$ is a positive definite selfadjoint operator in the Hilbert
space $L^2(0,x)$ if \eqref{e9.1.4} holds. Note that, since $H(t)\in L^1(\R)$,
one has $\tildeH(k)\to 0$ as $|k|\to\infty$, so \eqref{e9.1.4} implies
\begin{equation}\label{e9.2.3}
  1+\tildeH(k)\geq c>0.  
  \end{equation}
A positive definite selfadjoint operator in a Hilbert space is boundedly
invertible. \refT{9.1.3} is proved.
\end{proof}

Our argument shows that
\begin{equation}\label{e9.2.4}
 \sup_{x\geq 0} ||(I+H_x)^{-1}||_{L^2(\R)}\leq c^{-1}.  
  \end{equation}
Before we prove \refT{9.1.4}, let us prove a simple  lemma.
For  results of this type, see \cite{K2}.

\begin{lemma}\label{L:9.2.1}
If (9.1.4) holds, then the operator
\begin{equation}\label{e9.2.5}
  H\varphi:=\int^\infty_0 H(t-u)\varphi(u) du  
  \end{equation}
is a bounded operator in $L^p(\R_+)$, $p=1,2,\infty$.

For  $\Gamma_x(u,s)\in L^1(\R_+)$ one has
\begin{equation}\label{e9.2.6}
  ||\int^\infty_x du
    H(t-u) \Gamma_x(u,s)||_{L^{2}(0,x)}\leq c_1\int^\infty_x du
    |\Gamma_x(u,s)|.  
   \end{equation}
\end{lemma}

\begin{proof}
Let $||\varphi||_p:=||\varphi||_{L^p(\R_+)}$.
One has
\begin{equation}\label{e9.2.7}
  ||H\varphi||_1\leq \sup_{u\in\R_+} \int^\infty_0 dt
    |H(t-u)| \int^\infty_0 |\varphi(u)| du
    \leq \int^\infty_{-\infty} |H(s)|ds ||\varphi||_1
    =2||H||_1\, ||\varphi||_1,  
  \end{equation}
where we have used the assumption $H(t)=H(-t)$.
Similarly,
\begin{equation}\label{e9.2.8}
  ||H\varphi||_\infty\leq 2||H||_1\, ||\varphi||_\infty. 
  \end{equation}
Finally, using Parseval's equality, one gets:
\begin{equation}\label{e9.2.9}
2\pi  ||H\varphi||^2_2 =||\tildeH\tilde\varphi_+||^2_{L^2(\R)}
  \leq   \sup_{k\in\R} |\tildeH(k)|^2 ||\varphi||^2_2,  
  \end{equation}
where
\begin{equation}\label{e9.2.10}
  \varphi_+(x):=\left\{ 
     \begin{array}{rl}\varphi(x), &\ x\geq 0,\\0,& \ x<0.\end{array}\right.  
  \end{equation}
Since $|\tildeH(k)|\leq 2||H||_1$ one gets from \eqref{e9.2.9} the estimate:
\begin{equation}\label{e9.2.11}
  ||H\varphi||_2 \leq   \sqrt{2/\pi}||H||_1\ ||\varphi||_2.  
 \end{equation}
To prove \eqref{e9.2.6}, one notes that 
\begin{multline}
  \int_0^xdt|\int^\infty_x du    H(t-u) \Gamma_x(u,s)|^2\leq 
  \sup_{u,v\geq x}\int_0^xdt|H(t-u) H(t-v)|
  (\int_x^{\infty}|\Gamma_x(u,s)|du)^2
  \notag \\
\leq c_1(\int^\infty_x du
    |\Gamma_x(u,s)|)^2.
  \notag
  \end{multline}
Estimate \eqref{e9.2.6} is obtained.
\refL{9.2.1} is proved.
\end{proof}

\begin{proof}[Proof of \refT{9.1.4}]

Define $\Gamma_x(t,s)=0$ for $t$ or $s$ greater than $x$.
Let $w:=\Gamma_x(t,s)-\Gamma(t,s)$.
Then \eqref{e9.1.1} and \eqref{e9.1.6} imply

\begin{equation}\label{e9.2.12}
  (I+H_x)w=\int^\infty_x H(t-u)\Gamma(u,s)du:=h_x(t,s).  
  \end{equation}
If condition \eqref{e9.1.4} holds, then
 equations \eqref{e9.1.6} and \eqref{e9.1.22} have solutions in $L^1(\R_+)$,
and, since $\sup_{t\in \R}|H(t)|<\infty$, it is clear that
 this solution belongs to $L^\infty(\R_+)$ and consequently
to $L^2(\R_+)$, because 
$||\varphi||_2\leq ||\varphi||_\infty||\varphi||_1$.
The proof of \refT{9.1.3} shows that such a solution is unique and
does exist. From \eqref{e9.2.4} one gets
\begin{equation}
  \sup_{x\geq 0} ||(I+H_x)^{-1}||_{L^2(0,x)} \leq c^{-1}. 
  \end{equation}
For any fixed $s>0$ one sees that $\sup_{x\geq y}||h_x(t,s)||\to
0$ as $y\to \infty$, where the norm here stands for any of the three
norms $L^p(0,x), p=1,2,\infty$.
Therefore \eqref{e9.2.12} and \eqref{e9.2.11} imply
\begin{align}\label{e9.2.14}
  ||w||^2_{L^2(0,x)} &\leq c^{-2} ||h_x||^2_{L^2(0,x)}\notag\\ 
   &\leq c^{-2}
  \left\| \int^\infty_x H(t-u) \Gamma(u,s)du\right\|_{L^1(0,x)}
  \left\| \int^\infty_x H(t-u) \Gamma(u,s)du\right\|_{L^\infty(0,x)}
  \notag\\
  &\leq \hbox{\ const\ }
  \left\| \Gamma(u,s)\right\|^2_{L^1(x,\infty)}\to 0
  \hbox{\ as\ }x\to\infty, 
  \end{align}
since $\Gamma(u,s)\in L^1(\R_+)$ for any fixed $s>0$ and 
$H(t)\in L^1(\R)$.

Also
\begin{align}\label{e9.2.15}
  \|w(t,s)\|^2_{L^\infty(0,x)} 
      &\leq 2(||h_x||^2_{L^\infty(0,x)} +||H_xw||^2_{L^\infty(0,x)})\\
  \leq c_1 ||\Gamma (u,s)||^2_{L^1(x,\infty)}
      &+  c_2 \sup_{t\in \R}||H(t-u)||^2_{L^2(0,x)} ||w||^2_{L^2(0,x)}, 
  \end{align}
where $c_j>0$ are some constants. 
Finally, by \eqref{e9.2.6}, one has;
\begin{equation}\label{e9.2.16}
  \| w(t,s)\|^2_{L^2(0,x)}
   \leq c_3 (\int^\infty_x  |\Gamma(u,s)|du)^2\to 0
  \hbox{\ as\ } x\to+\infty. 
  \end{equation}
From \eqref{e9.2.15} and \eqref{e9.2.16} relation \eqref{e9.1.11} follows.
\refT{9.1.4} is proved.
\end{proof}

Let us now prove \refT{9.1.2}.
We need several lemmas.

\begin{lemma}\label{L:9.2.2}
The function \eqref{e9.1.8} satisfies the equations
\begin{equation}\label{e9.2.17}
  E^\prime=ikE - a(x)E_-, \quad E(0,k)=1, \quad E_-:=E(x,-k),  
  \end{equation}
\begin{equation}\label{e9.2.18}
  E^\prime_-=-ikE_- -a(x)E, \quad E_-(0,k)=1, 
  \end{equation}
where $E^\prime =\frac{dE}{dx}$,
and $a(x)$ is defined in \eqref{e9.1.14}.
\end{lemma}

\begin{proof}
Differentiate \eqref{e9.1.8} and get
\begin{equation}\label{e9.2.19}
  E^\prime =ikE-e^{ikx} \left(
    2\Gamma_{2x} (2x,0)e^{-ik2x}+2\int^{2x}_0
    \frac{\partial \Gamma_{2x}(s,0)}{\partial (2x)} e^{-iks} ds \right). 
    \end{equation}
We will check below that
\begin{equation}\label{e9.2.20}
  \frac{\partial\
  \Gamma_{x}(t,s)}{\partial x} =-\Gamma_x(t,x)\Gamma_x(x,s), 
  \end{equation}
and
\begin{equation}\label{e9.2.21}
  \Gamma_x(t,s)=\Gamma_x(x-t, x-s).  
  \end{equation}
Thus, by \eqref{e9.2.20},
\begin{equation}\label{e9.2.22}
  \frac{\partial \Gamma_{2x}(s,0)}{\partial(2x)} =-\Gamma_{2x}
  (s,2x)\Gamma_{2x}(2x,0).  
  \end{equation}
Therefore \eqref{e9.2.19} can be written as
\begin{equation}\label{e9.2.23}
  E^\prime =ikE-e^{-ikx} a(x)+a(x)e^{ikx}
  \int^{2x}_0 \Gamma_{2x}(s,2x) e^{-iks}ds.  
  \end{equation}

By \eqref{e9.2.21} one gets
\begin{equation}\label{e9.2.24}
  \Gamma_{2x}(s,2x)=\Gamma_{2x}(2x-s,0).  
  \end{equation}
Thus
\begin{align}\label{e9.2.25}
  e^{ikx} \int^{2x}_0 \Gamma_{2x} (s,2x) e^{-iks} ds
  & =\int^{2x}_0\Gamma_{2x}(2x-s,0)e^{ik(x-s)} ds
  \notag\\
  & =e^{-ikx} \int^{2x}_0 \Gamma_{2x}(y,0) e^{iky} dy.  
  \end{align}
From \eqref{e9.2.23} and \eqref{e9.2.25} one gets \eqref{e9.2.17}.

Equation \eqref{e9.2.18} can be obtained from \eqref{e9.2.17} by changing $k$ to $-k$.
\refL{9.2.2} is proved if formulas \eqref{e9.2.20} -- \eqref{e9.2.21} are checked.

To check \eqref{e9.2.21}, use $H(-t)=H(t)$ and compare the equation for
$\Gamma_x(x-t,x-s):=\varphi$,
\begin{equation}\label{e9.2.26}
  \Gamma_x(x-t,x-s)+\int^x_0H(x-t-u)\Gamma_x(u,x-s)du
  =H(x-t-x+s)=H(t-s),  
  \end{equation}
with equation \eqref{e9.1.1}. Let $u=x-y$. Then \eqref{e9.2.26} can be written as
\begin{equation}\label{e9.2.27}
  \varphi+\int^x_0 H(t-y)\varphi\,dy =H(t-s), 
   \end{equation}
which is equation \eqref{e9.1.1} for $\varphi$. Since \eqref{e9.1.1} has at most one
solution, as we have proved above (\refT{9.1.3}), formula \eqref{e9.2.21} is proved.

To prove \eqref{e9.2.20}, differentiate \eqref{e9.1.1} with respect to $x$ and get:
\begin{equation}\label{e9.2.28}
  \Gamma^\prime_x(t,s)+\int^x_0 H(t-u) \Gamma^\prime_x(u,s)du
  =-H(t-x)\Gamma_x(x,s), \quad \Gamma^\prime_x:=\frac {\partial 
  \Gamma_x}{\partial x}.  
   \end{equation}
Set $s=x$ in \eqref{e9.1.1}, multiply \eqref{e9.1.1} by $-\Gamma_x(x,s)$,
compare with \eqref{e9.2.28} and use again the uniqueness of the solution
to \eqref{e9.1.1}. This yields \eqref{e9.2.20}.

\refL{9.2.2} is proved.
\end{proof}

\begin{lemma}\label{L:9.2.3}
Equation \eqref{e1.1.4} holds for $\psi$ defined in \eqref{e9.1.7}.
\end{lemma}

\begin{proof}
From \eqref{e9.1.7} and \eqref{e9.2.17} -- \eqref{e9.2.18} one gets
\begin{equation}\label{e9.2.29}
  \psi^{\prime\prime}=\frac{E^{\prime\prime}-E^{\prime\prime}_-}{2i}
 = \frac{(ikE-a(x)E_-)^\prime -(-ikE_- -a(x)E)^\prime}{2i}.  
   \end{equation}
Using \eqref{e9.2.17} -- \eqref{e9.2.18} again one gets
\begin{equation}\label{e9.2.30}
  \psi^{\prime\prime}=-k^2\psi+q(x)\psi,
  \quad q(x):=a^2(x)+a^\prime(x).  
  \end{equation}

\refL{9.2.3} is proved.
\end{proof}

\begin{proof}[Proof of \refT{9.1.2}]
The function $\psi$ defined in \eqref{e9.1.7} solves equation \eqref{e1.1.4} and satisfies
the conditions
\begin{equation}\label{e9.2.31}
  \psi(0,k)=0,\quad \psi^\prime(0,k)=k.  
  \end{equation}
The first condition is obvious (in \cite{K1} there is a 
{\it misprint:} it is written
that $\psi(0,k)=1$),
and the second condition follows from \eqref{e9.1.7} and \eqref{e9.2.15}:
$$
  \psi^\prime(0,k)=
  \frac{E^\prime(0,k)-E^\prime_-(0,k)}{2i}
  =\frac{ikE-aE_--(ikE_- -aE)}{2i} \bigg|_{x=0}
  =\frac{2ik}{2i}=k.
$$
Let $f(x,k)$ be the Jost solution.
Since $f(x,k)$ and $f(x,-k)$ are linearly independent, one has
$ \psi=c_1f(x,k)+c_2f(x,-k)$, 
where $c_1$, $c_2$ are some constants independent of $x$
but depending on $k$.

From \eqref{e9.2.31} one gets
$  c_1=\frac {f(-k)}{2i}, \quad c_2=\frac{-f(k)}{2i}; \quad f(k):=f(0,k)$.
Indeed, the choice of $c_1$ and $c_2$ guarantees that
the first condition \eqref{e9.2.31} is obviously satisfied, while
the second follows from the Wronskian formula:
$  f^\prime(0,k)f(-k)-f(k)f^\prime(0,-k)=2ik$.

Comparing this with \eqref{e9.1.12} yields the conclusion of \refT{9.1.2}.
\end{proof}

\section*{Invertibility of the steps of the inversion procedure and proof of Theorem 1.1}

Let us start with a discussion of the inversion
steps 1) -- 4) described in the introduction.

Then we discuss the uniqueness of the solution to ISP
and the consistency of the inversion method, that is, 
the fact that $q(x)$, reconstructed from $S(k)$ by steps 1) -- 4),
generates the original $S(k)$.

Let us go through steps 1) -- 4) of the reconstruction method and
prove their invertibility.
The consistency of the inversion method follows from the invertibility
of the steps of the inversion method.

\vskip.15in
\underbar{Step 1.}  \quad $S(k)\Rightarrow f(k)$.

Assume $S(k)$ satisfying \eqref{e9.1.18} -- \eqref{e9.1.20} is given. 
Then solve the Riemann
problem
\begin{equation}\label{e9.2.32}
  f(k)=S(-k)f(-k), \qquad k\in\R. 
 \end{equation}
Since $\ind_\R S(k)=0$, one has $\ind_\R S(-k)=0$.
Therefore the problem \eqref{e9.2.32} of finding an analytic function $f_+(k)$
in $\C_+:=\{k:\Im k>0\}$,  $f(k):=f_+(k)$ in $\C_+$, 
(and an analytic function $f_-(k):=f(-k)$ in
$\C_-:=\{k:\Im k<0\},$) from equation \eqref{e9.2.32} 
can be solved in closed form.
Namely, define
\begin{equation}\label{e9.2.33}
  f(k)= \exp
  \left\{ \frac{1}{2\pi i} \int^\infty_{-\infty}
     \frac{\ln S(-y)dy}{y-k}\right\}, \quad \Im k>0. 
   \end{equation}
Then $f(k)$ solves \eqref{e9.2.32}, $f_+(k)=f(k)$, $f_-(k)=f(-k)$.
Indeed,
\begin{equation}\label{e9.2.34}
  \ln\, f_+(k)-\ln\,f_-(k)=\ln\, S(-k),\quad k\in\R  
  \end{equation}
by the known jump formula for the Cauchy integral. Integral \eqref{e9.2.33}
converges absolutely at infinity, $\ln\,S(-y)$
is differentiable with respect to $y$ for 
$y\neq 0$, and is bounded on the real axis, so the Cauchy integral in 
\eqref{e9.2.33} is
well defined.

To justify the above claims, one uses the known properties of the
Jost function
\begin{equation}\label{e9.2.35}
  f(k)=1+\int^\infty_0 A(0,y) e^{iky}dy
  := 1+\int^\infty_0 A(y) e^{iky}dy, 
  \end{equation}
where estimates \eqref{e1.1.24} and \eqref{e1.1.25} hold
 and $A(y)$ is a real-valued function. Thus
\begin{equation}\label{e9.2.36}
  f(k)=1-\frac{A(0)}{ik} -\frac{1}{ik}
  \int^\infty_0 A^\prime(t) e^{ikt}dt,  
  \end{equation}
\begin{equation}\label{e9.2.37}
 S(-k) =\frac{f(k)}{f(-k)}
  =\frac{1-\frac{A(0)}{ik} -\frac{1}{ik} \widetilde{A^\prime}(k)}
   {1+\frac{A(0)}{ik} + \frac{1}{ik} \widetilde{A^\prime}(-k)}
   =1+O\left( \frac{1}{k}\right).   
\end{equation}
Therefore
\begin{equation}\label{e9.2.38}
  \ln S(-k)=O\left(\frac{1}{k}\right)
  \quad\hbox{as}\quad |k|\to\infty, \quad k\in\R.  
  \end{equation}
Also
\begin{equation}\label{e9.2.39}
  \dotf(k)=i \int^\infty_0 A(y)y e^{iky}dy,
  \quad \dotf:=\frac{\partial f}{\partial k}.  
  \end{equation}
Estimate \eqref{e1.1.24} implies
\begin{equation}\label{e9.2.40}
  \int^\infty_0 y|A(y)| dy\leq 2\int^\infty_0 t|q(t)|dt<\infty,\quad
  A(y)\in L^2(\R_+),  
  \end{equation}
so that $\dot f(k)$ is bounded for all $k\in\R$,
$f(k)-1\in L^2(\R),$ $S(-k)$
is differentiable for $k\neq 0$, and $\ln\,S(-y)$ is bounded on the real 
axis, as claimed. Note that
\begin{equation}\label{e9.2.41}
  f(-k)=\overline{f(k)}, \quad k\in\R. 
  \end{equation}
The converse step
$ f(k)\Rightarrow S(k) $
is trivial: $ S(k)=\frac{f(-k)}{f(k)}$.
If $\ind_\R S=0$ then $f(k)$ is analytic in $\C_+$,
$f(k)\not= 0$ in $\C_+$, $f(k)=1+O\left(\frac{1}{k}\right)$
as $|k|\to\infty$, $k\in\C_+$, and \eqref{e9.2.41} holds.

\vskip.15in
\underbar{Step 2.} \quad $f(k)\Rightarrow H(t)$.

This step is done by formula \eqref{e9.1.13}:
\begin{equation}\label{e9.2.42}
  H(t)=\frac{1}{2\pi} \int^\infty_{-\infty} e^{-ikt}
  \left( \frac{1}{|f(k)|^2} -1\right) dk.  
  \end{equation}
One has $H\in L^2(\R)$.
Indeed, it follows from \eqref{e9.2.43} that
\begin{equation}\label{e9.2.43}
  |f(k)|^2-1 =-\frac{2}{k} \int^\infty_0 A^\prime(t) \sin(kt)dt
  +O\left( \frac{1}{|k|^2}\right),  \quad |k|\to\infty,\quad k\in\R.  
  \end{equation}
The function
\begin{equation}\label{e9.2.44}
  w(k):=\frac{1}{k} \int^\infty_0 A^\prime(t) \sin(kt)dt  
  \end{equation}
is continuous because $ A^\prime(t)\in L^1(\R_+)$
by \eqref{e1.1.25}, and $w\in L^2(\R)$ since $w=o\left(\frac{1}{|k|}\right)$
as $|k|\to\infty$, $k\in\R$. Thus, $H\in L^2(\R)$.

Also, $H\in L^1(\R)$. Indeed, integrating by parts, 
one gets from \eqref{e9.2.42} 
the relation: $2\pi H(t)=\frac i t 
\int_{-\infty}^{\infty}e^{-ikt}[\dot f(k)f(-k)-\dot f(-k)f(k)]\frac 
{dk}{|f(k)|^4}:=\frac i t g(t)$, and $g\in L^2(\R)$, therefore $H\in 
L^1(\R)$. To check that $g\in L^2(\R)$, one uses 
\eqref{e9.2.35}, \eqref{e1.1.24} -- \eqref{e1.1.25},
and \eqref{e9.2.39} -- \eqref{e9.2.40}, 
to conclude that $[\dot f(k)f(-k)-\dot f(-k)f(k)]\in L^2(\R)$, and,
since $f(k)\neq 0$ on $\R$ and $f(\infty)=1$, it follows that
$g\in L^2(\R)$. The inclusion $[\dot f(k)f(-k)-\dot f(-k)f(k)]\in L^2(\R)$
follows from 
\eqref{e9.2.35}, \eqref{e1.1.24} -- \eqref{e1.1.25},
and \eqref{e9.2.39} -- \eqref{e9.2.40}.  

By (9.2.43), the function $H'(t)$ is the Fourier transform
of $-ik(1-|f(k)|^2)|f(k)|^{-2}$, and, by (9.2.44),
$k(|f(k)|^2-1)= -2 \int^\infty_0 A^\prime(t) \sin(kt)dt
  +O\left( \frac{1}{|k|}\right),$  as $ |k|\to\infty,\quad k\in\R.$   
Thus, $H'(t)$ behaves, essentially, as $A'(t)$ plus a function,
whose Fourier transform is $O\left( \frac{1}{|k|}\right)$. 
Estimate (1.2.27) shows how  $A'(t)$ behaves. Equation (9.1.1)
shows that $\Gamma_{x}(t,0)$ is as smooth as $H(t)$, so that
formula (9.1.17) for $q(x)$ shows that $q$ is essentially
is as smooth as  $A'(t)$.

The converse step
\begin{equation}\label{e9.2.45}
  H(t)\Rightarrow f(k)  
  \end{equation}
is also done by formula \eqref{e9.1.13}: Fourier inversion
gives $|f(k)|^2=f(k)f(-k),$ and factorization yields the unique $f(k)$,
since $f(k)$ does not vanish in $\C_+$ and tends to $1$ at infinity.

\vskip.15in
\underbar{Step 3.} \quad $H\Rightarrow \Gamma_x(s,0)\Rightarrow
 \Gamma_{2x}(2x,0).$

This step is done by solving equation \eqref{e9.1.1}. By \refT{9.1.3}
equation \eqref{e9.1.1} is uniquely solvable since condition \eqref{e9.1.4} is 
assumed.
Formula \eqref{e9.1.13} holds and the known properties of the Jost function
are used:
$f(k)\to 1$ as $k\to\pm\infty$, $f(k)\not=0$ for $k\not= 0$, $k\in\R$,
$f(0)\not= 0$ since $\ind_\R S(k)=0$.

The converse step $\Gamma_x(s,0)\Rightarrow H(t)$ is done
by formula \eqref{e9.1.3}. The converse step
\begin{equation}\label{e9.2.46}
  \Gamma_{2x}(2x,0)\Rightarrow \Gamma_x(s,0)  
  \end{equation}
constitutes the essence of the inversion method.

This step is done as follows:
\begin{equation}\label{e9.2.47}
  \Gamma_{2x}(2x,0)\operatorname*\Rightarrow^{\eqref{e9.1.14}}
  a(x) \operatorname*\Rightarrow^{\eqref{e9.2.17} -- \eqref{e9.2.18}}
  E(x,k) \operatorname*\Rightarrow^{\eqref{e9.1.8}}
  \Gamma_x(s,0).   
  \end{equation}
Given $a(x)$, system \eqref{e9.2.17} -- \eqref{e9.2.18} is uniquely solvable for $E(x,k)$.

Note that the step $q(x)\Rightarrow f(k)$ can be done by solving the
uniquely solvable integral equation \eqref{e1.1.5}:
with $q\in L_{1,1}(\R_+)$, and then calculating
$f(k)=f(0,k)$.

\vskip.15in
\underbar{Step 4.} \quad $a(x):=2\Gamma_{2x}(2x,0)\Rightarrow q(x).$

This step is done by formula \eqref{e9.1.16}. The converse step
$$
q(x)\Rightarrow a(x)
$$
can be done by solving the Riccati problem \eqref{e9.1.16} for $a(x)$
given $q(x)$ and the initial condition $2H(0)$. 
Given  $q(x)$, one can find $2H(0)$ as follows: 
one finds $f(x,k)$ by solving equation \eqref{e1.1.5},
which is uniquely solvable if $q \in L_{1,1}(\R_+)$, then
one gets $f(k):=f(0,k)$, and then calculates 
$2H(0)$ using formula \eqref{e9.2.42} with $t=0$:
$$
2H(0)= \frac{1}{\pi} \int^\infty_{-\infty} 
  \left( \frac{1}{|f(k)|^2} -1\right) dk.
$$

\begin{proof}[Proof of \refT{9.1.1}.]
If \eqref{e9.1.18} -- \eqref{e9.1.20} hold, then, 
as has been proved in \refS{5.4}, there is a unique
$q(x)\in L_{1,1}(\R_+)$ which generates the given
$S$-matrix $S(k)$.

{\it It is not proved in \cite{K1} that $q(x)$ defined in (1.19)
(and obtained as a final result of steps 1) -- 4))
generates the scattering matrix $S(k)$ with which we started
the inversion.} 

Let us now prove this.
We have already discussed the following diagram:
\begin{equation}\label{e9.2.49}
  S(k)\operatorname*\Leftrightarrow^{\eqref{e9.2.33}}
  f(k)\operatorname*\Leftrightarrow^{\eqref{e9.1.13}}
  H(t)\operatorname*\Leftrightarrow^{\eqref{e9.1.1}}
  \Gamma_x(s,0)\operatorname*\Rightarrow\Gamma_{2x}(2x,0)
\operatorname*\Leftrightarrow^{\eqref{e9.1.14}}
  a(x)\operatorname*\Leftrightarrow^{\eqref{e9.1.16}} q(x).  
   \end{equation}
To close this diagram and therefore establish the basic one-to-one
correspondence $  S(k)\Leftrightarrow q(x),$
one needs to prove
$\Gamma_{2x}(2x,0)\Rightarrow\Gamma_x(s,0)$.
This is done by the scheme \eqref{e9.2.47}.

Note that the step
$q(x)\Rightarrow a(x)$ requires solving Riccati equation \eqref{e9.1.16}
with the boundary condition $a(0)=2H(0)$.
Existence of the solution to this problem on all of $\R_+$ is
guaranteed by the assumptions \eqref{e9.1.18} -- \eqref{e9.1.20}. 
The fact that these assumptions imply $q(x)\in L_{1,1}(\R_+)$ is proved
in \refS{5.4}.
\refT{9.1.1} is proved.
\end{proof}

Uniqueness theorems for the inverse scattering problem are not given
in \cite{K1}. They can be found in \refS{5.4}

\begin{remark}\label{R:9.2.1}
From our analysis one gets the following result:
\end{remark}

\begin{proposition}\label{P:9.2.1}

If $q(x)\in L_{1,1}(\R_+)$ and has no bounds states and no resonance at
zero, then Riccati equation \eqref{e9.1.16} with the initial condition \eqref{e9.1.15}
has the solution $a(x)$ defined for all $x\in\R_+$.
\end{proposition}

\section{Numerical aspects of the Krein inversion procedure.}\label{S:9.3}
The main step in this procedure from the numerical viewpoint is to
solve equation \eqref{e9.1.1} for all $x>0$ and all $0<s<x$, which are the
parameters in equation \eqref{e9.1.1}.

Since equation \eqref{e9.1.1} is an equation with the convolution kernel, its
numerical solution involves inversion of a Toeplitz matrix,
which is a well developed area of numerical analysis.
Moreover, such an inversion requires much less computer 
memory and time than the inversion based on
the Gel'fand-Levitan or Marchenko methods.
This is the main advantage of Krein's inversion method.

This method may become even more attractive if it is combined with
the Marchenko method. In the Marchenko method the equation to be solved
is \eqref{e1.4.13}
where $F(x)$ is defined in \eqref{e1.4.11} and is known if $S(k)$ is known.
The kernel $A(x,y)$ is to be found from \eqref{e1.4.11} and
if $A(x,y)$ is found then the potential is recovered by the formula:
Equation \eqref{e1.4.11} can be written in operator form:
$(I+F_x)A=-F$.
The operator $F_x$ is a contraction mapping in the Banach space
$L^1(x,\infty)$ for $x>0.$
The operator $H_x$ in \eqref{e9.1.1} is a contraction mapping in
$L^\infty(0,x)$ for $0<x<x_0$, where $x_0$ is chosen to that
$  \int^{x_0}_0 |H(t-u)| du<1$.
 Therefore it seems reasonable from the numerical point of view to use
the following approach:

\begin{description}
\item[1.] Given $S(k)$, calculate $f(k)$ and $H(t)$ as explained
in Steps 1 and 2, and also $F(x)$ by formula \eqref{e1.4.11}.
\item[2.] Solve by iterations equation \eqref{e9.1.1} for
$0<x<x_0$, where $x_0$ is chosen so that the iteration method
for solving \eqref{e9.1.6} converges rapidly. Then find $q(x)$ as explained in
Step 4.
\item[3.] Solve equation (1.5.13) for $x>x_0$ by iterations.
Find $q(x)$ for $x>x_0$ by formula (1.5.12).
\end{description}

\section{Discussion of the ISP when the bound states are present.}\label{S:9.4}

If the given data are \eqref{e9.1.15}, then one defines
$   w(k)=\prod^J_{j=1}
  \frac{k-ik_j}{k+ik_j}\hbox{\quad if\quad }\ind_\R S(x)=-2J$
and
$  W(k)=\frac{k}{k+i\gamma} w(k)\hbox{\quad if\quad }\ind_R S(k)=-2J-1$,
where $\gamma>0$ is arbitrary, and is chosen so that $\gamma\not=k_j$,
$1\leq j\leq J$.

Then one defines
$   S_1(k):=S(k) w^2(k)\hbox{\quad if\quad }\ind_\R S=-2J$
or
$  S_1(k):= S(k) W^2(k)\hbox{\quad if\quad }\ind_\R S=-2J-1$.
Since $\ind_\R w^2(k)=2J$ and $\ind_\R W^2(k)=2J+1$, one has
$   \ind_\R S_1(k)=0$.
The theory of \refS{9.2} applies to $S_1(k)$ and yields $q_1(x)$.
From $q_1(x)$ one gets $q(x)$ by adding bound states $-k_j^2$ and
norming constants $s_j$ using the known procedure (e.g. see \cite{M}).

\section{Relation between Krein's and GL's methods.}\label{S:9.5}

The GL (Gel'fand-Levitan) method in the case of absence of bound states  
of the following steps (see \refC{4}, for example):

\vskip.15in
\underbar{Step 1.}
Given $f(k)$, the Jost function, find
\begin{align}
  L(x,y)& :=\frac{2}{\pi} \int^\infty_0 dk\, k^2
  \left( \frac{1}{|f(k)|^2}-1\right)
  \frac{\sin kx}{k}\frac{\sin ky}{k}
  \notag\\
  &=\frac{1}{\pi} \int^\infty_0 dk\left(|f(k)|^{-2}-1\right)
  \left( \cos[k(x-y)]-\cos[k(x+y)]\right)
  \notag\\
  &:=M(x-y)-M(x+y),
\notag 
 \end{align}
where
$  M(x):=\frac{1}{\pi} \int^\infty_0dk  \left(|f(k)|^{-2}-1\right) \cos(kx)$.

\vskip.15in
\underbar{Step 2.}
Solve the integral equation \eqref{e1.4.4} for $K(x,y)$:

\vskip.15in
\underbar{Step 3.}
Find $  q(x)=2\frac{d K(x,x)}{dx}$.
Krein's function, $H(t)$, see \eqref{e9.1.13}, can be written as follows:
\begin{equation}\label{e9.5.1}
  H(t)=\frac{1}{2\pi} \int^\infty_{-\infty}
  \left( |f(k)|^{-2}-1 \right) e^{-ikt}dk
  =\frac{1}{\pi} \int^\infty_0 \left(|f(k)|^{-2}-1\right)
  \cos(kt)dk.  
  \end{equation}  

{\it Thus, the relation between the two methods is given by the formula:}
\begin{equation}\label{e9.5.2}
  M(x)=H(x).  
  \end{equation}

In fact, the GL method deals with the inversion of the spectral
foundation $d\rho$ of the operator $-\frac{d^2}{dx^2}+q(x)$
defined in $L^2(\R_+)$ by the Dirichlet boundary condition
at $x=0$. However, if $\ind_\R S(k)=0$ (in this case there are no bound 
states and no
resonance at $k=0$), then (see \eqref{e1.1.19}):
$
  d\rho(\lambda)=\left\{
  \begin{array}{rl}
  \frac{2k^2dk}{\pi |f(k)|^2}, &\quad\lambda>0,\quad\lambda=k^2,\\
  0, & \quad \lambda<0,
 \end{array} \right. $
so $d\rho(\lambda)$ in this case is uniquely defined by $f(k)$, $k\geq 0$.


\chapter{Inverse problems for the heat and wave equations.}\label{C:10}

\section{Inverse problem for the heat equation}\label{S:10.1}
Consider problem \eqref{e1.4.25} -- \eqref{e1.4.28}. Assume 
\begin{equation}\label{e10.1.1}
  a(t)=0 \hbox{\ for\ } t>T, \quad  \int^T_0 a(t)dt<\infty, \quad 
a(t)\not\equiv 0.
  \end{equation}
One can also take $a(t)=\delta (t)$ where $\delta(t)$ is the 
delta-function.
We prove that {\it the inverse problem of finding $q(x)\in L^1[0,1]$, 
$q=\barq$,
from the conditions \eqref{e1.4.25} -- \eqref{e1.4.28} has at most one 
solution. 

If \eqref{e1.4.28} is replaced by the condition
\begin{equation}\label{e10.1.2}
  u_x(0,t)=b_0(t),
 \end{equation}
then $q(x)$, in general, is not uniquely defined by the conditions 
\eqref{e1.4.25},
\eqref{e1.4.26}, \eqref{e1.4.27} and \eqref{e10.1.2}, but $q$ is uniquely
defined by these data if, for example, $q(\frac{1}{2}-x)=q(\frac{1}{2}+x)$,
or if $q(x)$ is known on $[\frac{1}{2},1]$.}

Let us take the Laplace transform of \eqref{e1.4.25} -- \eqref{e1.4.28} 
and put $v(x,\lambda):=\int^\infty_0u(x,t) e^{-\lambda t}dt$, 
$A(\lambda):=v(1,\lambda)$, 
$B(\lambda):=v_x(1,\lambda)$, $B_0(\lambda):=v_x(0,\lambda)$.
Then \eqref{e1.4.25} -- \eqref{e1.4.28} can be written as
\begin{equation}\label{e10.1.3}
  \ell v +\lambda v:=-v''+q(x)v+\lambda v=0,
  \quad 0\leq x\leq 1, \quad v(0,\lambda)=0, \quad v(1,\lambda)=A(\lambda)
  \end{equation}
\begin{equation}\label{e10.1.4}
  v'(1,\lambda)=B(\lambda)
  \end{equation}
and \eqref{e10.1.2} takes the form
\begin{equation}\label{e10.1.5}
  v'(0,\lambda)=B_0(\lambda).
  \end{equation}

\begin{theorem}\label{T:10.1.1}
The data $\{A(\lambda),B(\lambda)\}$, known on a set of 
$\lambda\in(0,\infty)$, which 
has a finite positive limit point, determine $q$ uniquely.
\end{theorem}

\begin{proof}
Since $A(\lambda)$ and $B(\lambda)$ are analytic in 
$\prod_+:=\{\lambda:\Re \lambda>0\}$, one can assume 
that $A(\lambda)$ and $B(\lambda)$ are known for all $\lambda>0$.
If $k=i\lambda^{\frac{1}{2}}$ and $\varphi$ is defined in \eqref{e1.1.2}
then $v(x,\lambda)=c(k)\varphi(x,k)$, $c(k)\not=0$,
$A(\lambda)=c(k)\varphi(1,k)$, $B(\lambda)=c(k)\varphi'(1,k)$, so
\begin{equation}\label{e10.1.6}
  \frac{B(\lambda)}{A(\lambda)} = \frac{\varphi'(1,k)}{\varphi(1,k)}.
  \end{equation}
Thus the function $\frac{B(\lambda)}{A(\lambda)}$ is meromorphic in $\C$,
its zeros on the axis $k\geq 0$ are the eigenvalues of $\ell=-\frac{d^2}{dx^2}+q(x)$,
corresponding to the boundary conditions $u(0)=u'(1)=0$
and its poles on the axis $k\geq 0$ are the eigenvalues of $\ell$ 
corresponding to $u(0)=u(1)=0$. The knowledge of two spectra determines 
$q$ 
uniquely (\refS{7.1}). $\Box$

{\it An alternative proof of \refT{10.1.1},} based on property 
$C_\varphi$, is:
assume that $q_1$ and $q_2$ generate the same data,
$p:=q_1-q_2$, $w:=v_1-v_2$, where $v_j$, $j=1,2$, solves 
\eqref{e10.1.3} -- \eqref{e10.1.4} with $q=q$, and get $(\ast)\ \ell_1w=pv_2$,
 $w(0,\lambda)=w(1,\lambda)=w'(1,\lambda)=0$.
Multiply $(\ast)$ by $\varphi_1$, $\ell_1\varphi_1+\lambda\varphi_1=0$, 
$\varphi_1(0,\lambda)=0$, $\varphi'(_1(0,\lambda)=1$, and integrate over 
$[0,1]$ to get
\begin{equation}\label{e10.1.7}
 \int^1_0 p v_2 \varphi_1 dx=0 \qquad  \forall \lambda>0.
 \end{equation}
By property $C_\varphi$ it follows from \eqref{e10.1.7} that $p=0$.
\refT{10.1.1} is proved.
\end{proof}

\begin{theorem}\label{T:10.1.2}
Data \eqref{e10.1.3}, \eqref{e10.1.5} does not determine $q$ uniquely in general.
It does if $q(x)$ is known on $[\frac{1}{2},1]$, or if  
$q(x+\frac{1}{2})=q(\frac{1}{2}-x)$.
\end{theorem}

\begin{proof}
Arguing as in the first proof of \refT{10.1.1}, one concluded that the 
data
\eqref{e10.1.3}, \eqref{e10.1.5} yields only one (Dirichlet) spectrum of $\ell$,
since $\varphi'(0,k)=1$. One spectrum determines $q$ only on ``a half of the interval'', 
$b=\frac{1}{2}$, see \refS{7.1}.
\refT{10.1.2} is proved.
\end{proof}

\section{What are the ``correct'' measurements?}\label{S:10.2}
From \refT{10.1.1} and \refT{10.1.2} it follows that the measurements 
$\{u_x(1,t)\}_{\forall t>0}$ are much more informative 
than $\{u_x(0,t)\}_{\forall t>0}$
for the problem \eqref{e1.4.25} -- \eqref{e1.4.27}.
In this section we state a similar result for the problem
\begin{equation}\label{e10.2.1}
 u_t=(a(x)u')', \quad 0\leq x\leq 1, \quad t>0; \quad u(x,0)=0,
 \quad u(0,t)=0,
 \end{equation}
\begin{equation}\label{e10.2.2}
  u(1,t)=f(t).
  \end{equation}
The extra data, that is, measurements, are
\begin{equation}\label{e10.2.3}
 a(1)u'(1,t)=g(t),
 \end{equation}
which is the flux. Assume:
\begin{equation}\label{e10.2.4}
 f\not\equiv 0, \quad f\in L^1(0,1), \quad a(x)\in W^{2,1}(0,1), \quad a(x)\geq c>0,
 \end{equation}
$W^{\ell, p}$ is the Sobolev space. Physically, $a(x)$ is the 
conductivity, $u$ is the 
temperature. We also consider in place of \eqref{e10.2.3} the following data:
\begin{equation}\label{e10.2.5}
 a(0) u'(0,t)=h(t).
 \end{equation}
Our results are similar to those in \refS{10.1}: 

{\it data 
$\{f(t),g(t)\}_{\forall t>0}$ determine $q(x)$ uniquely, while data
$\{f(t),h(t)\}_{\forall t>0}$ do not, in general, determine $a(x)$ 
uniquely.}

Therefore, the measurements $\{g(t)\}_{\forall t>0}$ are much more
informative than the measurements $\{h(t)\}_{\forall t>0}$.
We refer the reader to \cite{R9}.

\section{Inverse problem for the wave equation}\label{S:10.3}
Consider inverse problem \eqref{e1.4.20} -- \eqref{e1.4.24}.
Our result is

\begin{theorem}\label{T:10.3.1}
The above inverse problem has at most one solution.
\end{theorem}

\begin{proof}
Take the Fourier transform of \eqref{e1.4.20} -- \eqref{e1.4.24}
and get:
\begin{equation}\label{e10.3.1}
 \ell v-k^2v=0, \quad x\geq 0, \quad v(x,k)=\int^\infty_0 e^{ikt} u(x,t) 
dt,
 \end{equation}
\begin{equation}\label{e10.3.2}
 v(0,k)=1, \quad v(1,k)=A(k)=\int^\infty_0 a(t) e^{ikt}dt.
 \end{equation}
From \eqref{e10.3.1} one gets $v(x,k)=c(k)f(x,k)$, where $f(x,k)$ is the Jost
solution, and from \eqref{e10.3.2} one gets 
$v(x,k)=\frac{f(x,k)}{f(k)}$ and
$A(k)=\frac{f(1,k)}{f(k)}=\frac{e^{ik}}{f(k)}$, because $q=0$ for $x>1$.
Thus $f(k)=\frac{e^{ik}}{A(k)}$ is known.
By \refT{7.2.1} $q$ is uniquely determined. 
\refT{10.3.1} is proved.
\end{proof}

\begin{remark}\label{R:10.3.2}
The above method allows one to consider other boundary conditions at $x=0$,
such as $u'(0,t)=0$ or $u'(0,t)=hu(0,t)$, 
$h=\const>0$, and different data at $x=1$, for example, $u'(1,t)=b(t)$.
\end{remark}

\chapter{Inverse problem for an inhomogeneous Schr\"odinger equation}
\label{C:11}
\setcounter{section}{1}


In this chapter an inverse problem is studied for an inhomogeneous
Schr\"odinger equation. Most, if not all, of the
earlier studies dealt with inverse problems for homogeneous
equations.  Let
\begin{equation}\label{e11.1.1} 
\ell u-k^2u:=-u''+q(x)u-k^2u=\delta(x),\quad x\in\R^1,\quad
\frac{\partial u}{\partial|x|} -iku\to 0,\quad |x|\to\infty.  
\end{equation}
Assume that $q(x)$ is a real-valued function,
$ q(x)=0\ \hbox{for}\ |x|>1,\quad q\in L^\infty[-1,1].$
Suppose that the data
$ \{u(-1,k), u(1,k)\},\quad \forall k>0$
are given. 

{\it The inverse problem is: 

(IP) Given the data, find $q(x)$.}

This problem is of practical interest:
think about finding the properties of an
inhomogeneous slab (the governing equation is plasma equation)
from the boundary measurements of the field,
generated by a point source inside the slab.
Assume that the self-adjoint operator
$\ell=-\frac{d^2}{dx^2}+q(x)$ in $L^2(\R)$
has no negative eigenvalues (this is the case
when $q(x) \geq 0$, for example).
The operator $\ell$ is the closure in $L^2(\R)$
of the symmetric operator $\ell_0$ defined on $C^\infty_0(\R^1)$
by the formula $\ell_0u=-u''+q(x)u$.
Our result is:

\begin{theorem}\label{T:11.1}
Under the above assumptions IP has at most one solution.
\end{theorem}

\begin{proof}[Proof of Theorem 11.1:]
The solution to \eqref{e11.1.1} is
\begin{equation}\label{e11.1.2}
u=\left\{
\begin{array}{ll}\frac{g(k)}{[f,g]} f(x,k),& x>0, \\ \frac{f(k)}{[f,g]} g(x,k),& x<0.\end{array} 
 \right.
\end{equation}
Here $f(x,k)$ and $g(x,k)$ solve homogeneous version
of equation \eqref{e11.1.1} and have the following asymptotics:
\begin{equation}\label{e11.1.3}
 f(x,k)\sim e^{ikx},\quad x\to +\infty,\quad g(x,k)\sim e^{-ikx},
 \quad x\to -\infty,
\end{equation}
\begin{equation}\label{e11.1.4}
f(k):=f(0,k), \quad g(k):=g(0,k),
\end{equation}
\begin{equation} \label{e11.1.5}
 [f,g]:=fg'-f'g=-2ika(k),
\end{equation}
where the prime denotes differentiation with
respect to $x$-variable,
and $a(k)$ is defined by the equation
\begin{equation}\label{e11.1.6}
 f(x,k)=b(k)g(x,k)+a(k)g(x,-k).
\end{equation}
It is known that
\begin{equation}\label{e11.1.7}
 g(x,k)=-b(-k)f(x,k)+a(k)f(x,-k),
 \end{equation}
\begin{equation}\label{e11.1.8}
 a(-k)=\overline{a(k)},\quad b(-k)=\overline{b(k)},
   \quad |a(k)|^2=1+|b(k)|^2, \quad k\in\R,
\end{equation}
\begin{equation}\label{e11.1.9}
 a(k)=1+O(\frac 1k),\quad k\to\infty,\quad k\in\C_+;
   \quad b(k)=O(\frac 1k),\quad |k|\to\infty,\quad k\in\R,
\end{equation}
\begin{equation}\label{e11.1.10}
 [f(x,k),g(x,-k)]=2ikb(k),\quad [f(x,k),g(x,k)]=-2ika(k),
\end{equation}
$a(k)$ in analytic in $\C_+$, $b(k)$ in general does not admit analytic
continuation from $\R$, but if $q(x)$ is compactly supported,
then $a(k)$ and $b(k)$ are analytic functions of $k\in\C\setminus 0$.

The functions
\begin{equation}\label{e11.1.11}
A_1(k):=\frac{g(k)f(1,k)}{-2ika(k)},
   \quad A_2(k):=\frac{f(k) g(-1,k)}{-2ika(k)}
\end{equation}
are the data,
they are known for all $k>0$. Therefore one can assume the functions
\begin{equation}\label{e11.1.12}
 h_1(k):=\frac{g(k)}{a(k)},\quad h_2(k):=\frac{f(k)}{a(k)}
\end{equation}
to be known for all $k>0$ because
\begin{equation}\label{e11.1.13}
 f(1,k)=e^{ik},\quad g(-1,k)=e^{ik},
\end{equation}
as follows from the assumption $q=0$ if $|x|>1$, and from \eqref{e11.1.3}.

From \eqref{e11.1.12}, \eqref{e11.1.7} and \eqref{e11.1.6} it follows that
\begin{equation}\label{e11.1.14}
a(k) h_1(k)=-b(-k)f(k)+a(k)f(-k)
   =-b(-k)h_2(k)a(k) +h_2(-k)a(-k)a(k),
   \end{equation}
\begin{equation}\label{e11.1.15}
a(k)h_2(k) =b(k)a(k)h_1(k) +a(k)h_1(-k)a(-k).
   \end{equation}
From \eqref{e11.1.14} and \eqref{e11.1.15} it follows:
\begin{equation}\label{e11.1.16}
-b(-k)h_2(k) +h_2(-k)a(-k) =h_1(k),
   \end{equation}
\begin{equation}\label{e11.1.17}
b(k)h_1(k) +a(-k)h_1(-k) =h_2(k).
   \end{equation}
Eliminating $b(-k)$ from \eqref{e11.1.16} and \eqref{e11.1.17}, one gets:
\begin{equation}\label{e11.1.18}
a(k)h_1(k)h_2(k) +a(-k)h_1(-k)h_2(-k)
   =h_1(k)h_1(-k) +h_2(-k)h_2(k),
   \end{equation}
or
\begin{equation}\label{e11.1.19}
a(k)=m(k)a(-k) +n(k),\quad k\in\R
   \end{equation}
where
\begin{equation}\label{e11.1.20}
m(k):=-\frac{h_1(-k)h_2(-k)}{h_1(k)h_2(k)},
   \quad n(k):=\frac{h_1(-k)}{h_2(k)} + \frac{h_2(-k)}{h_1(k)}.
   \end{equation}

Problem \eqref{e11.1.19} is a Riemann problem 
 for the pair $\{a(k),a(-k)\}$,
the function $a(k)$ is analytic in $\C_+:=\{k:k\in\C,Imk>0\}$ and
$a(-k)$ is analytic in $\C_-$. The functions
$a(k)$ and $a(-k)$ tend to one as $k$ tends to infinity
in $\C_+$ and, respectively, in $\C_-$, see equation \eqref{e11.1.9}.

The function $a(k)$ has finitely many simple zeros at the points
$ik_j,1\leq j\leq J$, $k_j>0$, where $-k^2_j$ are the negative
eigenvalues of the operator $\ell$ defined by the differential
expression $\ell u=-u''+q(x)u$ in $L^2(\R)$.

The zeros $ik_j$ are the only zeros of $a(k)$ in the upper half-plane $k$.

Define
\begin{equation}\label{e11.1.21}
ind\, a(k):=\frac{1}{2\pi i}\int^\infty_{-\infty} d\,\ln\,a(k).
   \end{equation}
One has
\begin{equation}\label{e11.1.22}
ind\, a=J,
   \end{equation}
where $J$ is the number of negative eigenvalues of the
operator $\ell$, and, using \eqref{e11.1.12}, \eqref{e11.1.22}
and \eqref{e11.1.20}, one gets
\begin{equation}\label{e11.1.23}
ind\,m(k)=-2[ind\,h_1(k)+ind\, h_2(k)]
   =-2[ind\,g(k)+ind\, f(k)-2J].
   \end{equation}
Since $\ell$ has no negative eigenvalues, it follows that $J=0$.

In this case $ind\,f(k)=ind\,g(k)=0$ (see Lemma 1 below),
so $ind\,m(k)=0$, and $a(k)$ is uniquely recovered from the data
as the solution of \eqref{e11.1.19} which tends to 
one at infinity, see equation \eqref{e11.1.9}. If $a(k)$ is found,
then $b(k)$ is uniquely determined by equation \eqref{e11.1.17} and so
the reflection coefficient $r(k):=\frac{b(k)}{a(k)}$
is found. The reflection coefficient determines a compactly
supported $q(x)$ uniquely \cite{R9}, but we give a new proof.
If $q(x)$ is compactly supported, then the reflection coefficient
$r(k):=\frac {b(k)}{a(k)}$ is meromorphic. Therefore, its values
for all $k>0$ determine uniquely $r(k)$ in the whole
complex $k$-plane as a meromorphic function. The poles
of this function in the upper half-plane are the numbers
$ik_j, j=1,2,...,J$. They determine uniquely the numbers $k_j, 1\leq j
\leq J,$ which are a part of the standard scattering data
$\{r(k), k_j, s_j, 1\leq j \leq J\}$, where $s_j$ are the norming
constants.
Note that if $a(ik_j)=0$ then $b(ik_j)\neq 0$: otherwise
equation \eqref{e11.1.6} would imply $f(x,ik_j)\equiv 0$ in contradiction
to the first relation \eqref{e11.1.3}.
 If $r(k)$ is meromorphic, then the norming constants can
be calculated by the formula $s_j=-i\frac {b(ik_j)}{\dot a(ik_j)}=
-i Res_{k=ik_j} r(k)$, where the dot denotes differentiation
with respect to $k$, and $Res$ denotes the residue. So,
for compactly supported potential the values of $r(k)$ for
all $k>0$ determine uniquely the standard scattering data,
that is, the reflection coefficient, the bound states $-k_j^2$
and the norming constants $s_j$, $1\leq j \leq J.$
These data determine the potential uniquely.
\refT{11.1} is proved.
\end{proof}

\begin{lemma}\label{L:11.1.2}
If $J=0$ then $ind\,f=ind\,g=0$.
\end{lemma}

\begin{proof}
We prove $ind\,f=0$. The proof of the equation $ind\,g=0$ is similar.
Since $ind\,f(k)$ equals to the number of zeros of $f(k)$ in $\C_+$,
we have to prove that $f(k)$ does not vanish in $\C_+$.
If $f(z)=0$, $z\in\C_+$, then 
$z=ik$, $k>0$, and $-k^2$ is an eigenvalue of the
operator $\ell$ in $L^2(0,\infty)$ with the boundary condition
$u(0)=0$.

From the variational principle one
can find the negative eigenvalues of the operator $\ell$
in $L^2(\R_+)$ with the Dirichlet condition at $x=0$ as consequitive
minima of the quadratic functional. The minimal eigenvalue is:
\begin{equation}\label{e11.1.24}
-k^2=inf\int^\infty_0
   \left[ u^{\prime 2}+q(x)u^2  \right]    dx:=  \kappa_0,
   \quad u\in  \oH1 (\R_+),     \quad ||u||_{L^2(\R_+)}=1,
   \end{equation}
where $\oH1(\R_+)$ is the Sobolev space of
$H^1(\R_+)$-functions
satisfying the condition $u(0)=0$.

On the other hand, if $J=0$, then
\begin{equation}\label{e11.1.25}
0\leq inf\,\int^\infty_{-\infty} [u^{\prime 2}+q(x)u^2]\,dx:=
  \kappa_1,
   \quad u\in H^1(\R), \quad ||u||_{L^2(\R)}=1.
   \end{equation}
Since any element $u$ of $\oH1(\R_+)$ can be considered as
an element of $H^1(\R)$ if one extends $u$ to the whole axis by setting
$u=0$ for $x<0$, it follows from the variational definitions \eqref{e11.1.24}
and \eqref{e11.1.25} that
$\kappa_1\leq \kappa_0$. Therefore, if  $J=0$, 
then $\kappa_1\geq 0$ and therefore $\kappa_0 \geq 0$.
This means that the operator $\ell$ on $L^2(\R_+)$ with the Dirichlet 
condition
at $x=0$ has no negative eigenvalues.
This means that $f(k)$ does not have zeros in $\C_+$, if $J=0$. 
Thus $J=0$ implies $ind\,f(k)=0$.

\refL{11.1.2} is proved.
\end{proof}

\begin{remark}\label{R:11.1.3}
The above argument shows that in general
\begin{equation}\label{e11.1.26}
ind\, f\leq J \quad\hbox{and}\quad ind\,g\leq J,
   \end{equation}
so that \eqref{e11.1.23} implies
\begin{equation}\label{e11.1.27}
ind\,m(k)\geq 0.
   \end{equation}
Therefore the Riemann problem \eqref{e11.1.19} is always solvable.
\end{remark}


\end{document}